\documentstyle[preprint,fancybox,epsf,amssymb,amsbsy,amstext,subeqnarray,amsfonts,aps]{revtex}
\draft
\tighten
\begin{document}

\def\dsdt{$\frac{d\sigma}{dt}$}
\def\beqn{\begin{eqnarray}}
\def\eeqn{\end{eqnarray}}
\def\barr{\begin{array}}
\def\earr{\end{array}}
\def\btab{\begin{tabular}}
\def\etab{\end{tabular}}
\def\bite{\begin{itemize}}
\def\eite{\end{itemize}}
\def\bcen{\begin{center}}
\def\ecen{\end{center}}

\def\eq{\begin{equation}}
\def\ee{\end{equation}}
\def\eqa{\begin{eqnarray}}
\def\eea{\end{eqnarray}}

\def\rmqp{{ \hbox{\rm q}'} }
\def\rmq{{\hbox{\rm q}}}
\def\qt0{\tilde{q}_0}
\def\gqt{\tilde{Q}}
\def\calm{{\cal M}}
\def\dcalm{\Delta{\cal M}}
\def\psmu{P^{\prime \mu}}
\def\psnu{P^{\prime \nu}}
\def\ksmu{K^{\prime \mu}}
\def\pss{P^{\prime \hspace{0.05cm}2}}
\def\psf{P^{\prime \hspace{0.05cm}4}}
\def\kdagger{K\hspace{-0.3cm}/}
\def\ndagger{N\hspace{-0.3cm}/}
\def\qqs{q\!\cdot\!q'}
\def\lls{l\!\cdot\!l'}
\def\lp{l\!\cdot\!p}
\def\lps{l\!\cdot\!p'}
\def\lsp{l'\!\cdot\!p}
\def\lsps{l'\!\cdot\!p'}
\def\lqs{l\!\cdot\!q'}
\def\pps{p\!\cdot\!p'}
\def\psqs{p'\!\cdot\!q'}
\def\epsp{\varepsilon'\!\cdot\!p}
\def\epsps{\varepsilon'\!\cdot\!p'}
\def\epsl{\varepsilon'\!\cdot\!l}
\def\epsls{\varepsilon'\!\cdot\!l'}

\title{Dispersion relation formalism for virtual Compton 
scattering off the proton}
\normalsize
\author{B. Pasquini$^1$, M. Gorchtein$^2$, D. Drechsel$^2$, A. Metz$^3$, 
and M. Vanderhaeghen$^2$}
\address{$^1$ ECT*, European Center for Theoretical Studies 
in Nuclear Physics and Related Areas, I-38050 Villazzano (Trento), 
{\rm and} INFN, Trento, Italy}
\address{$^2$ Institut f\"ur Kernphysik, Johannes-Gutenberg-Universit\"at, D-55099 Mainz, Germany}
\address{$^3$ Division of Physics and Astronomy, Faculty of Science,
  Vrije Universiteit, 1081 HV Amsterdam, The Netherlands}
\date{\today}
\maketitle

\begin{abstract}
We present in detail a dispersion relation formalism for virtual
Compton scattering (VCS) off the proton from threshold into the 
$\Delta(1232)$-resonance region. 
Such a formalism can be used
as a tool to extract the generalized polarizabilities of the proton from
both unpolarized and polarized VCS observables over a larger energy
range. We present calculations for existing and forthcoming VCS
experiments and demonstrate that the VCS observables in the energy region
between pion production threshold and the $\Delta(1232)$-resonance
show an enhanced sensitivity to the generalized polarizabilities. 
\end{abstract}

\section{Introduction}

The field of virtual Compton scattering (VCS) has been opened up
experimentally in recent years by the new high precision electron
accelerator facilities. 
On the theoretical side, an important activity has emerged
over the last years around the VCS process in different kinematical regimes
(see e.g. \cite{GuiVdh,Vdh00} for reviews). 
\newline
\indent
In VCS off a nucleon target, a virtual photon interacts with the nucleon 
and a real photon is emitted in the process.
At low energy of the outgoing real photon, 
the VCS reaction amounts to a generalization 
of real Compton scattering (RCS) in which both energy and 
momentum of the virtual photon can be varied independently, which
allows us to extract response functions, parametrized by 
the so-called generalized polarizabilities (GPs) of the nucleon \cite{Gui95}. 
On the other side, VCS has also a close relation to elastic electron
scattering. More precisely this means, that the 
physics addressed with VCS is the same as if one would 
perform an elastic electron scattering experiment on a target 
placed between the plates of a capacitor or between the poles of a magnet.
In this way one studies the spatial distributions of the 
polarization densities of the target, by means of the GPs, which are functions 
of the square of the four-momentum, $Q^2$, transferred by the
electron. The GPs teach us about the interplay between nucleon-core
excitations and pion-cloud effects, and their measurement provides 
therefore a new test of our understanding of the nucleon structure. 
\newline
\indent
A first dedicated VCS experiment was performed at the MAMI accelerator,
and two combinations of the proton GPs have been measured \cite{Roc00}. 
Further experimental programs are underway 
at the intermediate energy electron accelerators 
(JLab \cite{JLab}, MIT-Bates \cite{Bates}, MAMI \cite{mamipol}) 
to measure the VCS observables. 
\newline
\indent
At present, VCS experiments at low outgoing photon energies 
are analyzed in terms of low-energy expansions (LEXs). 
In the LEX, only the leading term (in the energy of the real photon) 
of the response to the quasi-constant electromagnetic field, 
due to the internal structure of the system, is taken into account. 
This leading term depends linearly on the GPs.  
As the sensitivity of the VCS cross sections to the GPs 
grows with the photon energy, it is
advantageous to go to higher photon energies, provided one can keep the
theoretical uncertainties under control when approaching and crossing the pion
threshold. The situation can be compared to RCS, 
for which one uses a dispersion relation formalism \cite{lvov97,Dre99}
to extract the polarizabilities at energies 
above pion threshold, with generally larger effects on the observables. 
\newline
\indent
It is the aim of the present paper to present in detail such a
dispersion relation (DR) formalism for VCS on a proton target, 
which can be used as a tool to
extract the GPs from VCS observables over a larger energy range, into the
$\Delta(1232)$-resonance region. In Ref.~\cite{Pas00}, we have
given a first account of the DR predictions for
the GPs. In this paper we present the formalism in detail and show the
results for the VCS observables. 
\newline
\indent
In Sec.~\ref{invampl}, we start by specifying the kinematics 
and the invariant amplitudes of the VCS process.
\newline
\indent
In Sec.~\ref{disp}, we set up the DR formalism for
the VCS invariant amplitudes and show that for 10 of the 12 VCS
invariant amplitudes unsubtracted DRs hold.
\newline
\indent
In Sec.~\ref{dispgp}, it is shown that the DR formalism provides 
predictions for 4 of the 6 GPs of the proton. 
\newline
\indent
In Sec.~\ref{schannel}, it is discussed 
how the $s$-channel dispersion integrals, which correspond to the
excitation of $\pi N$, $\pi \pi N$,... intermediate states, are
calculated. In the numerical evaluation of the dispersion integrals, 
only the contribution of $\pi N$ states are taken into account. 
\newline
\indent
In Sec.~\ref{asympt}, we show how to deal with the two VCS invariant
amplitudes for which one cannot write down an unsubtracted DR. 
Our DR formalism involves two free parameters, 
being directly related to two GPs, and which are to be extracted
from a fit to experiment. 
\newline
\indent
In Sec.~\ref{results}, we show the results in the DR formalism for
both unpolarized and polarized VCS observables below and above pion
threshold. We compare with existing data and present predictions for
planned and forthcoming experiments. 
\newline
\indent
Finally, we present our conclusions in Sec.~\ref{conclusions}. 
\newline
\indent
Several technical details on VCS invariant amplitudes and helicity
amplitudes are collected in three appendices.


\section{Kinematics and invariant amplitudes for VCS}
\label{invampl}

\begin{figure}[ht]
\vspace{5cm}
\epsfxsize=14cm
\centerline{\epsffile{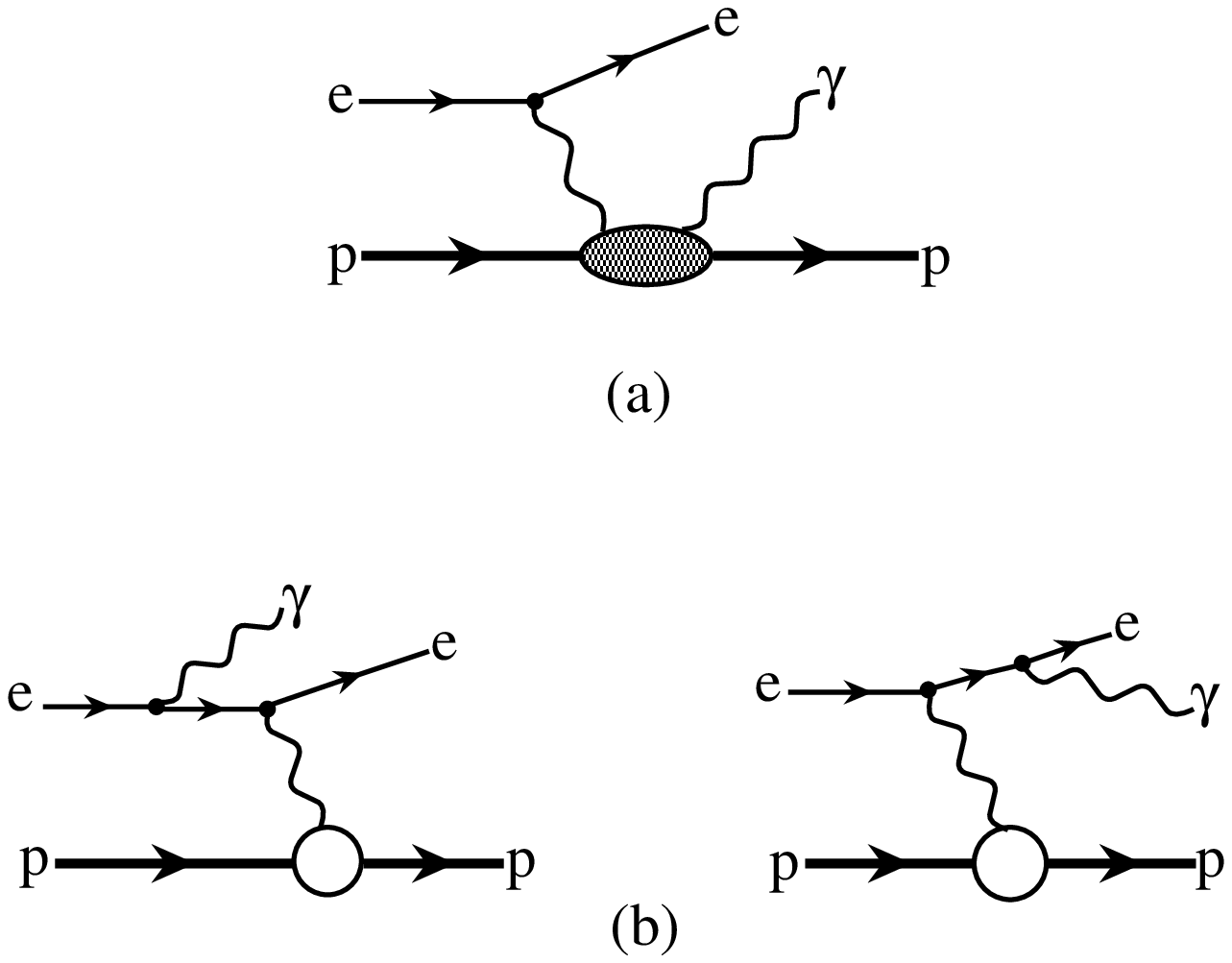}}
\vspace{-.5cm}
\caption[]{(a) FVCS process, (b) BH process.}
\label{fig:diagrams}
\end{figure}

\begin{figure}[ht]
\epsfxsize=11cm
\centerline{\epsffile{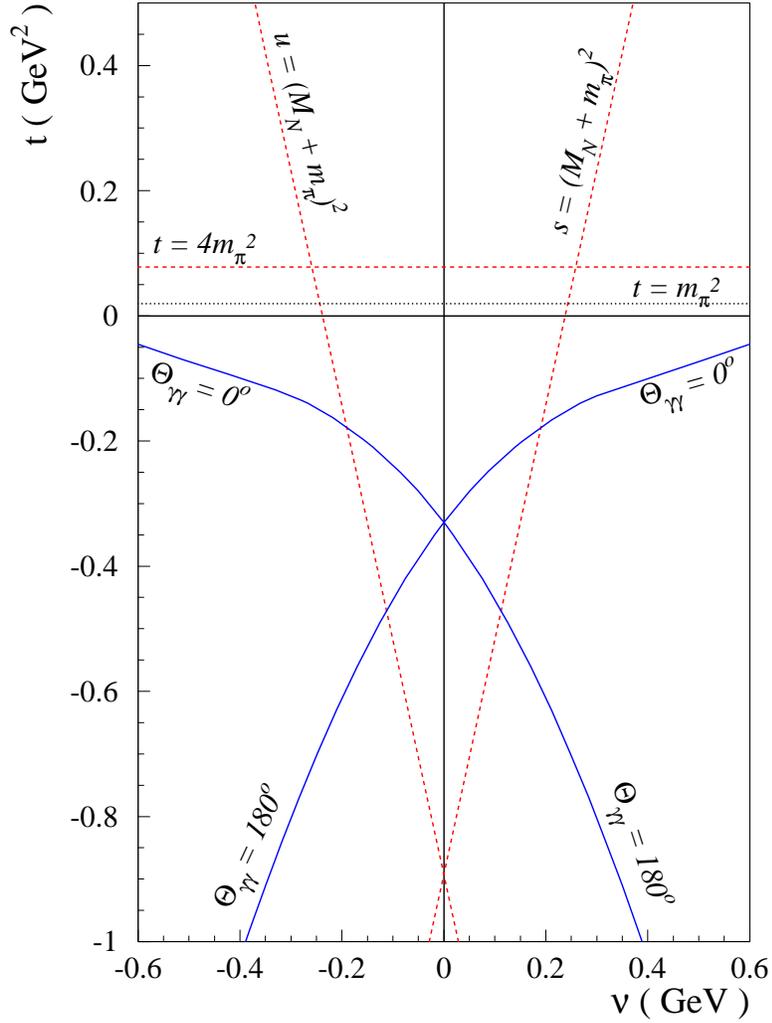}}
\vspace{.5cm}
\caption[]{The Mandelstam plane for virtual Compton scattering at 
  $Q^2$ = 0.33 GeV$^2$. The boundaries of the physical $s$-channel
  region are $\Theta_{\gamma \gamma} = 0^o$ and 
$\Theta_{\gamma \gamma} = 180^o$ for $\nu > 0$, the $u$-channel region
  is obtained by crossing, $\nu \to - \nu$. 
The curves for $\Theta_{\gamma \gamma} = 0^o$ and 
$\Theta_{\gamma \gamma} = 180^o$ intersect at $\nu = 0$, $t = - Q^2$, 
which is the point where the generalized polarizabilities are defined.}
\label{fig:mandelstam}
\end{figure}

In this section, we start by briefly recalling how the VCS process 
on the proton is accessed through the $e p \to e p \gamma$ reaction. 
In this process, the final photon can be emitted
either by the proton, which is referred to as the fully virtual
Compton scattering (FVCS) process, or by
the lepton, which is referred to as the Bethe-Heitler (BH) process.
This is shown graphically in Fig.~\ref{fig:diagrams},  
leading to the amplitude $T^{ee'\gamma}$ of the $e p \to e p \gamma$
reaction as the coherent sum of the BH and the FVCS process~:
\begin{equation}
T^{ee'\gamma}=T^{BH}+T^{FVCS}.
\end{equation}
The BH amplitude $T^{BH}$ is exactly calculable from QED if one knows 
the nucleon electromagnetic form factors. The FVCS amplitude
$T^{FVCS}$ contains, in the one-photon exchange approximation, the VCS
subprocess $\gamma^* p \to \gamma p$. We refer to Ref.~\cite{GuiVdh}
where the explicit expression of the BH amplitude is given, 
and where the construction of the
FVCS amplitude from the $\gamma^* p \to \gamma p$ process is discussed. 
In this paper, we present the details how to
construct the amplitude for the 
$\gamma^* p \to \gamma p$ VCS subprocess, in a DR formalism. 
\newline
\indent
We characterize the four-vectors of the virtual (real) photon in the VCS
process $\gamma^* p \to \gamma p$ 
by $q$ ($q'$) respectively, and the four-momenta of initial (final)
nucleons by $p$ ($p'$) respectively. 
In the VCS process, the initial photon is spacelike and we denote its
virtuality in the usual way by $q^2$ = - $Q^2$. 
Besides $Q^2$, the VCS process can be described by the 
Mandelstam invariants 
\begin{equation}
s = (q + p)^2, \hspace{.5cm} 
t = (q - q')^2, \hspace{.5cm} 
u = (q - p')^2\;,
\end{equation}
with the constraint 
\begin{equation}
s + t + u = 2 M^2 - Q^2 \;,
\end{equation}
where $M$ denotes the nucleon mass.
We furthermore introduce the variable $\nu$, which changes sign under 
$s \leftrightarrow u$ crossing : 
\begin{equation}
\nu = {{s - u} \over {4 M}} \;,
\end{equation}   
and which can be expressed in terms of the virtual photon
energy in the {\it lab} frame ($E_\gamma^{lab}$) as  
\begin{equation}
\nu \,=\, E_\gamma^{lab} \,+\, {1 \over {4 M}} \left( t - Q^2 \right) \;.
\end{equation}
In the following, we choose $Q^2$, $\nu$ and $t$ as the independent
variables to describe the VCS process. In Fig.~\ref{fig:mandelstam},
we show the Mandelstam plane for the VCS process at a fixed value 
of $Q^2$ = 0.33 GeV$^2$, at which the experiment of \cite{Roc00} was performed.
\newline
\indent
The VCS helicity amplitudes can be written as
\begin{equation}
T_{\lambda' \lambda_N'; \, \lambda \lambda_N} \;=\; 
- e^2 \varepsilon_\mu(q, \lambda) \; 
\varepsilon^{'*}_\nu(q', \lambda') \;
\bar u(p', \lambda_N') \, {\mathcal M}^{\mu \nu} \,u(p, \lambda_N)\;,
\label{eq:matrixele}
\end{equation}
with $e$ the proton electric charge ($e^2/4 \pi = 1/137.036$). 
The polarization four-vectors of the virtual (real) photons are
denoted by $\varepsilon$ ($\varepsilon^{'}$), and their helicities by
$\lambda$ ($\lambda'$), with $\lambda = 0, \pm 1$ and $\lambda' = \pm 1$.
The nucleon helicities are $\lambda_N, \lambda_N' = \pm 1/2$, 
and $u, \bar u$ are the nucleon spinors
(as specified in appendix \ref{app:shel}). 
The VCS tensor ${\mathcal M}^{\mu\nu}$ in Eq.~(\ref{eq:matrixele}) 
can be decomposed into a Born (B) and a non-Born part (NB) : 
\begin{equation}
{\mathcal M}^{\mu \nu} \;=\; {\mathcal M}^{\mu \nu}_{B} \,+\,
{\mathcal M}^{\mu \nu}_{NB} \;.
\label{eq:vcsbnb}
\end{equation}
In the Born process, the virtual photon is
absorbed on a nucleon and the intermediate state remains a nucleon,
whereas the non-Born process contains all nucleon excitations 
and meson-loop contributions. 
The separation between Born and non-Born parts is performed in the
same way as described in Ref.~\cite{Gui95}, to which we refer for details. 
\newline 
One can proceed by parametrizing the VCS tensor of Eq.~(\ref{eq:vcsbnb}) 
in terms of 12 independent amplitudes.
In Ref.~\cite{Dre98}, a tensor basis was found so that the resulting
non-Born invariant amplitudes are free of kinematical singularities and
constraints, which is an important property when setting up a
dispersion relation formalism.  
In detail, we denote the tensor ${\mathcal M}^{\mu \nu}$ as \cite{Dre98}
\begin{equation}
{\mathcal M}^{\mu \nu} \;=\; \sum_{i = 1}^{12} 
\; f_i(Q^2, \nu, t) \, \rho^{\mu \nu}_i \;, 
\label{eq:nonborn}
\end{equation}
where the 12 independent tensors $\rho^{\mu \nu}_i$ 
are given in appendix \ref{app:tensor}.
The 12 independent invariant amplitudes $f_i$ are expressed 
in terms of the invariants $Q^2$, $\nu$ and $t$, 
but are otherwise identical with the amplitudes used in \cite{Dre98}. 
\newline
\indent
The tensor basis $\rho^{\mu \nu}_i$ of Eq.~(\ref{eq:vcstensors}) was chosen 
in \cite{Dre98} such that the resulting invariant amplitudes 
$f_i$ are either even or odd under crossing. Photon crossing leads to the 
symmetry relations among the $f_i$ at the real photon point :
\begin{eqnarray}
&&f_i \left( Q^2 = 0, \nu, t \right) \,=\, 
+ \, f_i \left( Q^2 = 0, - \nu, t \right)\;, 
\hspace{.5cm} (i = 1, 2, 6, 11) \;, \nonumber \\
&&f_i \left( Q^2 = 0, \nu, t \right) \,=\, 
- \, f_i \left( Q^2 = 0, - \nu, t \right)\;, 
\hspace{.5cm} (i = 4, 7, 9, 10) \;, 
\label{eq:phocross}
\end{eqnarray}
while the amplitudes $f_3$, $f_5$, $f_8$, $f_{12}$ 
do not contribute at the real photon point, because 
the corresponding tensors in Eq.~(\ref{eq:vcstensors}) 
vanish in the limit $Q^2 \rightarrow 0$. 
\newline
\indent
Nucleon crossing combined with charge conjugation provides the
following constraints on the $f_i$ at arbitrary virtuality
$Q^2$ :
\begin{eqnarray}
&&f_i \left( Q^2, \nu, t \right) \,=\, 
+ \, f_i \left( Q^2, - \nu, t \right)\;, 
\hspace{.5cm} (i = 1, 2, 5, 6, 7, 9, 11, 12) \;, \nonumber \\
&&f_i \left( Q^2, \nu, t \right) \,=\, 
- \, f_i \left( Q^2, - \nu, t \right)\;, 
\hspace{.5cm} (i = 3, 4, 8, 10) \; . 
\label{eq:nucross}
\end{eqnarray}
When using dispersion relations, it will be convenient to work with
12 amplitudes that are all even in $\nu$. Therefore, 
we define new amplitudes $F_i$ ($i$ = 1,...,12) as follows : 
\begin{eqnarray}
F_i \left( Q^2, \nu, t \right) \,&=&\, f_i \left( Q^2, \nu, t \right)
\;, \hspace{.5cm} (i = 1, 2, 5, 6, 7, 9, 11, 12)\;, \nonumber\\
F_i \left( Q^2, \nu, t \right) \,&=&\, 
{1 \over \nu} \, f_i \left( Q^2, \nu, t \right)
\;, \hspace{.5cm} (i = 3, 4, 8, 10)\;,  
\label{eq:fampl}
\end{eqnarray} 
satisfying $F_i \left( Q^2, -\nu, t \right) 
= F_i\left( Q^2, \nu, t \right)$ for
$i$ = 1,...,12.  
As the non-Born invariant amplitudes 
$f^{NB}_{3, 4, 8, 10} \thicksim \nu$ for $\nu \rightarrow 0$, 
the definition of Eq.~(\ref{eq:fampl}) ensures that also all the 
non-Born $F_i^{NB}$ ($i$ = 1,...,12) are free from kinematical singularities.  
The results for the Born amplitudes $F_i^B$ are listed in 
Appendix~\ref{app:born}.
\newline
\indent
From Eqs.~(\ref{eq:phocross}) and (\ref{eq:nucross}), one furthermore sees that
$F_7$ and $F_9$ vanish at the real photon point. 
Since 4 of the tensors vanish in the limit $Q^2 \rightarrow 0$, only 
the six amplitudes $F_1$, $F_2$, $F_4$, $F_6$, $F_{10}$
and $F_{11}$ enter in real Compton scattering (RCS). 
\newline
\indent
Dispersion relation formalisms for RCS were worked out in  
Refs.~\cite{lvov97,Dre99} in terms of another set of invariant
amplitudes, also free from kinematical singularities and constraints and
denoted as $A_i(\nu, t)$ ($i$ = 1,...,6) (see Appendix A of 
Ref.~\cite{lvov97} for definitions). It is therefore useful to
relate the amplitudes $F_{1, 2, 4, 6, 10, 11}(0, \nu, t)$ to 
the RCS amplitudes $A_i \left(\nu, t \right)$ $(i = 1,...,6)$. 
We find after some algebra the following relations at $Q^2 = 0$ :
\begin{eqnarray}
- e^2 \; F_1 \,&=&\, - A_1 \,-\, \left( {{t - 4 M^2} \over {4 M^2}} 
\right) \, A_3 \,+\, {{\nu^2} \over {M^2}} \, A_4 \,+\, A_6 \;, \nonumber\\
- e^2 \; F_2 \,&=&\, - {1 \over {2 M^2}} \, \left[ A_3 \,+\, A_6
  \,-\, {t \over {4 M^2}} \, A_4 \right] \;, \nonumber\\
- e^2 \; F_4 \,&=&\, {1 \over {2 M^2}} \, A_4 \;, \nonumber\\
- e^2 \; F_6 \,&=&\, {1 \over {4 M^2}} \, 
\left[ - \left( {{t - 4 M^2} \over {4 M^2}} \right) \, A_4 
\,+\, A_6 \right]  \;, \nonumber\\
- e^2 \; F_{10} \,&=&\, - {1 \over {2 M}} \, \left[ A_5 - A_6
\right]\;, \nonumber\\
- e^2 \; F_{11} \,&=&\, - {1 \over 4 M} \, \left[ A_2 \,-\, 
{{t - 4 M^2 + 4 \nu^2} \over {4 M^2}} \, A_4 \,+\, A_6 \right] \;, 
\label{eq:fversusa}
\end{eqnarray}
where the charge factor $- e^2$ 
appears explicitely on the {\it lhs} of Eq.~(\ref{eq:fversusa}), 
because this factor is included in the usual definition of the
$A_i$. The values of the RCS invariant amplitudes $A_i$ ($i$ = 1,...,6)
at $\nu = t = 0$ 
can be expressed in terms of the scalar polarizabilities $\alpha$, $\beta$,
and the spin polarizabilities 
$\gamma_1, \gamma_2, \gamma_3, \gamma_4$, as specified in Ref.~\cite{lvov97}.


\section{Dispersion relations at fixed $t$ and fixed $Q^2$ for VCS}
\label{disp}

With the choice of the tensor basis of Eq.~(\ref{eq:vcstensors}), 
and taking account of the crossing relation Eq.~(\ref{eq:nucross}), 
the resulting non-Born VCS invariant amplitudes $F_i$ ($i$ = 1,...,12)
are free of all kinematical singularities and constraints 
and are all even in $\nu$, i.e. $F_i(Q^2,\nu, t) \,=\, F_i(Q^2,-\nu, t)$.
\newline
\indent
Assuming further analyticity and
an appropriate high-energy behavior, the amplitudes $F_i(Q^2, \nu, t)$ fulfill 
unsubtracted dispersion relations with respect to the variable $\nu$ at
fixed $t$ and fixed virtuality $Q^2$ :
\begin{equation}
{\mathrm Re} F_i^{NB}(Q^2, \nu, t) \;=\; 
{2 \over \pi} \; {\mathcal P} \int_{\nu_{thr}}^{+ \infty} d\nu' \; 
{{\nu' \; {\mathrm Im}_s F_i(Q^2, \nu',t)} \over {\nu'^2 - \nu^2}}\;,
\label{eq:unsub} 
\end{equation}
where we indicated explicitely that the {\it lhs} of Eq.~(\ref{eq:unsub})
represents the non-Born ($NB$) parts of the
amplitudes. Furthermore, in Eq.~(\ref{eq:unsub}), ${\mathrm
  Im}_s F_i$ are the discontinuities across the $s$-channel cuts of the
VCS process, starting at the pion production threshold, 
which is the first inelastic channel, 
i.e.  $\nu_{thr} = m_\pi + (m_\pi^2 + t/2 + Q^2/2)/(2 M)$, 
with $m_\pi$ the pion mass. 
\newline
\indent
Besides the absorptive singularities due to 
physical intermediate states which contribute
to the {\it rhs} of dispersion integrals as Eq.~(\ref{eq:unsub}),
one might wonder if other singularities exist giving rise to
imaginary parts. Such additional singularities could come from 
so-called anomalous thresholds \cite{Bj65,Pil79}, which  
arise when a hadron is a loosely bound system of other hadronic
constituents which can go on-shell (such as is the case of a nucleus 
in terms of its nucleon constituents), leading to so-called triangular
singularities. It was shown that in the case of strong
confinement within QCD, the quark-gluon structure of hadrons 
does not give rise to additional anomalous thresholds
\cite{Jaf92,Oeh95}, and the quark singularities are turned into hadron
singularities described through an effective field theory. 
Therefore, the only anomalous thresholds arise for those 
hadrons which are loosely bound composite systems of other 
hadrons (such as e.g. the $\Sigma$ particle in terms of $\Lambda$ and $\pi$).
For the nucleon case, such anomalous thresholds are
absent, and the imaginary parts entering the dispersion integrals as in
Eq.~(\ref{eq:unsub}) are calculated from 
absorptive singularities (due to $\pi N$, $\pi \pi N$,
... physical intermediate states). 
\newline
\indent
The assumption that unsubtracted dispersion relations 
as in Eq.~(\ref{eq:unsub}) hold, requires 
that at high energies ($\nu \rightarrow \infty$ 
at fixed $t$ and fixed $Q^2$) the
amplitudes ${\mathrm Im}_s F_i(Q^2,\nu,t)$ ($i$ = 1,...,12) 
drop fast enough so that the
integrals of Eq.~(\ref{eq:unsub}) are convergent and the contribution from
the semi-circle at infinity can be neglected. 
\newline
\indent
For the RCS invariant amplitudes $A_1$,...,$A_6$ which appear on the
{\it rhs} of Eq.~(\ref{eq:fversusa}), Regge theory 
leads to the following high-energy behavior
for $\nu \rightarrow \infty$ and fixed $t$~:
\begin{eqnarray}
A_{1}, A_{2} \;&\sim&\; \nu^{\alpha_M(t)} \;, 
\label{eq:rcsregge1}\\
\left(A_3 + A_6 \right) \;&\sim&\; \nu^{\alpha_P(t) - 2} \;, 
\label{eq:rcsregge2}\\
A_{3}, A_{5} \;&\sim&\; \nu^{\alpha_M(t) - 2} \; 
\label{eq:rcsregge3}\\ 
A_{4} \;&\sim&\; \nu^{\alpha_M(t) - 3} \;,
\label{eq:rcsregge4}
\end{eqnarray}
where $\alpha_M(t) \lesssim 0.5$ (for $t \leq$ 0) 
is a meson Regge trajectory, and where $\alpha_P(t)$ is the Pomeron
trajectory which has an intercept $\alpha_P(0) \approx $ 1.08. Note that the
Pomeron dominates the high energy behavior of the combination of $A_3 + A_6$.
From the asymptotic behavior of Eqs.~(\ref{eq:rcsregge1} -
\ref{eq:rcsregge4}), it follows that for RCS unsubtracted
dispersion relations do not exist for the amplitudes $A_1$ and $A_2$. 
The reason for the divergence of the unsubtracted integrals 
is essentially given by fixed poles in the $t$-channel, 
notably the exchange of the neutral pion (for $A_2$) 
and of a somewhat fictitious $\sigma$-meson (for $A_1$) with a mass of about
600~MeV and a large width, which models the two-pion continuum with
the quantum numbers $I=J=0$. 
\newline
\indent
We consider next the VCS amplitudes $F_1,...,F_{12}$, 
in the Regge limit ($\nu \rightarrow \infty$ at fixed $t$ and fixed
$Q^2$) to determine for which of the amplitudes unsubtracted
dispersion relations as in Eq.~(\ref{eq:unsub}) exist. 
The high-energy behavior of the amplitudes $F_i$ is deduced 
from the high-energy behavior of the VCS helicity amplitudes that are
defined and calculated in Appendix~\ref{app:shel}. 
This leads, after some algebra, to the following behavior 
in the Regge limit ($\nu \rightarrow \infty$, at fixed $t$ and fixed $Q^2$) 
\footnote{We note that some of the $F_i$ in 
Eqs.~(\ref{eq:reggevcs1} - \ref{eq:reggevcs7}) decrease faster with
increasing $\nu$ than reported in Ref.~\cite{Pas00}. This is because a more
detailed calculation has shown a cancellation in
the highest power of $\nu$ for some of the $F_i$, which leads to the
behavior of Eqs.~(\ref{eq:reggevcs1} - \ref{eq:reggevcs7}). 
However, this does not change the conclusion obtained 
in Ref.~\cite{Pas00} that unsubtracted DR only exist 
for 10 of the 12 $F_i$. The asymptotic
behavior of Eqs.~(\ref{eq:reggevcs1} - \ref{eq:reggevcs7}) only shows
that for some of those 10 amplitudes, the dispersion integrals
converge even faster than anticipated earlier \cite{Pas00}. }~: 
\begin{eqnarray}
F_{1}, \; F_{5} \;&\sim&\; \nu^{\alpha_P(t) - 2} \;,\hspace{.5cm}
\nu^{\alpha_M(t)} \; , 
\label{eq:reggevcs1}\\
F_{5} + 4 F_{11} 
\;&\sim&\; \nu^{\alpha_P(t) - 2} \;,\hspace{.5cm} 
\nu^{\alpha_M(t) - 1} \;,
\label{eq:reggevcs2}\\
F_{2}, \; F_{6}, \; F_{10}  
\;&\sim&\; \nu^{\alpha_P(t) - 2} \;,\hspace{.5cm} \nu^{\alpha_M(t) - 2} \; , 
\label{eq:reggevcs3} \\
F_{7}  
\;&\sim&\; \nu^{\alpha_P(t) - 3} \;,\hspace{.5cm} \nu^{\alpha_M(t) - 1} \; , 
\label{eq:reggevcs4}\\
F_{3}, \; F_{8}  
\;&\sim&\; \nu^{\alpha_P(t) - 3} \;,\hspace{.5cm} \nu^{\alpha_M(t) - 2} \; , 
\label{eq:reggevcs5} \\
F_{9}, \; F_{12}  
\;&\sim&\; \nu^{\alpha_P(t) - 4} \;,\hspace{.5cm} \nu^{\alpha_M(t) - 2} \; , 
\label{eq:reggevcs6} \\
F_{4}  
\;&\sim&\; \nu^{\alpha_P(t) - 4} \;,\hspace{.5cm} \nu^{\alpha_M(t) - 3} \; . 
\label{eq:reggevcs7}
\end{eqnarray}
In Eqs.~(\ref{eq:reggevcs1} - \ref{eq:reggevcs7}), 
we have indicated the high energy behavior from the Pomeron ($\alpha_P$) 
and from the meson ($\alpha_M$) contributions separately. 
It then follows that for the two amplitudes $F_1$ and
$F_5$, an unsubtracted dispersion integral as in Eq.~(\ref{eq:unsub})
does not exist, whereas the other ten amplitudes 
on the {\it lhs} of Eqs.~(\ref{eq:reggevcs2} - \ref{eq:reggevcs7}) can
be evaluated through unsubtracted dispersion integrals as in
Eq.~(\ref{eq:unsub}). 
\newline
\indent 
Having specified the VCS invariant amplitudes and their high energy
behavior, we are now ready to set up the DR formalism. 
First, we will show in Sec.~\ref{dispgp} that 4 of the 6 GPs of the 
nucleon can be evaluated using unsubtracted DR. 
We will then discuss in Sec.~\ref{schannel} how the $s$-channel 
dispersion integrals of Eq.~(\ref{eq:unsub}) are evaluated. 
In particular, unitarity will allow us to express the imaginary parts 
of the VCS amplitudes in terms of 
$\pi N$, $\pi \pi N$,... intermediate states. 
Finally, we will show in Sec.~\ref{asympt} how to deal with the 
remaining two VCS invariant amplitudes for which one cannot write 
unsubtracted DRs.

\section{Dispersion relations for the generalized polarizabilities}
\label{dispgp}

The behavior of the non-Born VCS tensor  
${\mathcal M}^{\mu \nu}_{NB}$ of Eq.~(\ref{eq:nonborn})
at low energy ($\rmqp \equiv |\vec q^{\; '}| \to 0$) 
but at arbitrary three-momentum $\rmq \,\equiv\, | \vec q \,|$ 
of the virtual photon, can be parametrized by six generalized
polarizabilities (GPs), which are functions of $\rmq$ and which are 
denoted by $P^{(\rho' \, L', \rho \,L)S}(\rmq)$  
\cite{Gui95,Dre97,Dre98}. 
In this notation, $\rho$ ($\rho'$) refers to the
electric (2), magnetic (1) or longitudinal (0) nature of the initial 
(final) photon, $L$ ($L' = 1$) represents the angular momentum of the
initial (final) photon, and $S$ differentiates between the 
spin-flip ($S=1$) and non spin-flip ($S=0$) 
character of the transition at the nucleon side.   
A convenient choice for the 6 GPs has been proposed in \cite{GuiVdh}~:
\begin{eqnarray}
&&P^{(01,01)0}(\rmq),\; P^{(11,11)0}(\rmq), \;
\label{eq:defgpunpol} \\
&&P^{(01,01)1}(\rmq),\; P^{(11,11)1}(\rmq),\;
P^{(11,02)1}(\rmq),\; P^{(01,12)1}(\rmq). 
\label{eq:defgppol}
\end{eqnarray}
\indent
In the limit $\rmq \to 0$ for the GPs, one finds the following  
relations with the polarizabilities (in gaussian units) of RCS \cite{Dre98}~:
\begin{eqnarray}
&&P^{(01,01)0}(0)=-{{4 \pi} \over {e^2}}\, \sqrt{\frac{2}{3}} \,
\alpha \;, \nonumber\\
&&P^{(11,11)0}(0)=-{{4 \pi} \over {e^2}}\, \sqrt{\frac{8}{3}} \, 
\beta \;, \nonumber \\
&&P^{(01,12)1}(0)=-{{4 \pi} \over {e^2}}\, \frac{\sqrt{2}}{3} \,
\gamma_3 \;,  \nonumber\\ 
&&P^{(11,02)1}(0)=-{{4 \pi} \over {e^2}}\, \frac{2 \sqrt{2}}{3 \sqrt{3}} \,
\left( \gamma_2 + \gamma_4 \right) \;, \nonumber \\
&&P^{(01,01)1}(0)= 0 \;, \nonumber\\
&&P^{(11,11)1}(0)= 0 \;, 
\label{eq_3_47}
\end{eqnarray}
\indent
In terms of invariants, the limit $\rmqp \to 0$ at finite
three-momentum $\rmq$ of the virtual photon corresponds to 
$\nu \to 0$ and $t \to -Q^2$ at finite $Q^2$. 
One can therefore express the GPs in terms of
the VCS invariant amplitudes $F_i$ at the point $\nu = 0, t = -Q^2$ for
finite $Q^2$, for which we introduce the shorthand~:
\begin{equation}
\bar F_i(Q^2) \;\equiv\; F_i^{NB} \left(Q^2, \nu = 0, t = - Q^2 \right) \;.
\label{eq:fbardef}
\end{equation}
The relations between the GPs and the $\bar F_i(Q^2)$ can be found in
\cite{Dre98}. 
\newline
\indent 
The present work aims at evaluating the GPs through unsubtracted
DRs of the type of Eq.~(\ref{eq:unsub}). 
We have seen from the high-energy behavior 
that the unsubtracted DRs do not exist for the amplitudes 
$F_1$ and $F_5$, but can be written down for the other amplitudes. 
Therefore, unsubtracted DRs for the GPs will hold for those GPs which do not  
depend on the two amplitudes $F_1$ and $F_5$. 
However, the amplitude $F_5$ can appear in the form $F_5 + 4 \, F_{11}$, 
because this combination has a high-energy behavior 
(Eq.~(\ref{eq:reggevcs3})) leading to a convergent integral. 
Among the six GPs we find four combinations which do not depend on
$F_1$ and $F_5$~:
\begin{eqnarray}
&&P^{\left(0 1, 0 1\right)0} + {1 \over 2}  
P^{\left(1 1, 1 1\right)0} = 
{{-2} \over {\sqrt{3}}}\,\left( {{E + M} \over
    E}\right)^{1/2} M \,\tilde q_0\, 
\left\{ {{\rmq^2} \over {\tilde q_0^2}}\, \bar F_2 + 
\left( 2 \, \bar F_6 + \bar F_9 \right) - \bar F_{12} \right\}, 
\label{eq:gpdisp1} \\
&&P^{\left(0 1, 0 1\right)1} = 
{1 \over {3 \sqrt{2}}}\,\left( {{E + M} \over E}\right)^{1/2} 
\,\tilde q_0\, 
\left\{ \left( \bar F_5 + \bar F_7 + 4\, \bar F_{11} \right) 
+ 4 \, M \, \bar F_{12} \right\}, 
\label{eq:gpdisp2} \\
&&P^{\left(0 1, 1 2\right)1} - {1 \over {\sqrt{2} \, \tilde q_0}}  
P^{\left(1 1, 1 1\right)1} = 
{1 \over {3}} \left( {{E + M} \over E}\right)^{1/2} 
{{M \, \tilde q_0} \over {\rmq^2}} \nonumber\\
&&\hspace{4.cm} \times
\left\{ \left( \bar F_5 + \bar F_7 + 4\, \bar F_{11} \right) 
+ 4 \, M \left( 2 \, \bar F_6 + \bar F_9 \right) \right\}, 
\label{eq:gpdisp3} \\
&&P^{\left(0 1, 1 2\right)1} +  
{{\sqrt{3}} \over {2}}  P^{\left(1 1, 0 2\right)1} =
{1 \over {6}} \left( {{E + M} \over E}\right)^{1/2} \, 
{{\tilde q_0} \over {\rmq^2}} \nonumber \\
&&\hspace{4cm} \times 
\left\{ \tilde q_0 \left( \bar F_5 + \bar F_7 + 4\, \bar F_{11} \right) 
+ 8 \, M^2 \left( 2 \, \bar F_6 + \bar F_9 \right) \right\}, 
\label{eq:gpdisp4} 
\end{eqnarray}
where $E = \sqrt{\rmq^2 + M^2}$ denotes the initial proton c.m. energy and  
$\tilde q_0 = M - E$ the virtual photon c.m. energy in the limit 
$\rmqp$ = 0. For small values of $\rmq$, 
we observe the relation $\tilde q_0 \approx - \, \rmq^2 / (2 M)$. 
Furthermore, in the limit $\rmqp$ = 0, the value of $Q^2$ is
always understood as being $\tilde Q^2 \equiv \rmq^2 - \tilde q_0^2$,
which we denote by $Q^2$ for simplicity of the notation.
\newline
\indent
The four combinations of GPs on the {\it lhs} of 
Eqs.~(\ref{eq:gpdisp1} - \ref{eq:gpdisp4}) can then be evaluated in a 
framework of unsubtracted DRs through the following integrals
for the corresponding $\bar F_i(Q^2)$~: 
\begin{equation}
\bar F_i(Q^2) \;=\; 
{2 \over \pi} \; \int_{\nu_{thr}}^{+ \infty} d\nu' \; 
{{{\mathrm Im}_s F_i(Q^2, \nu',t = - Q^2)} \over {\nu'}}\;.
\label{eq:sumrule} 
\end{equation}

\section{s-channel dispersion integrals}
\label{schannel}

The imaginary parts of the amplitudes $F_i$ in Eq.~(\ref{eq:unsub}) are 
obtained through the imaginary part of the VCS helicity amplitudes 
defined in Eq.~(\ref{eq:matrixele}).
The latter are determined by using unitarity. 
Denoting the VCS helicity amplitudes by $T_{fi}$, the
unitarity relation takes the generic form
\begin{equation}
2 \, {\mathrm Im}_s \, T_{fi} = \sum_{X} (2\pi)^4 \delta^4 (P_X - P_i)
T_{Xf}^{\dagger} T_{Xi}\,,
\label{eq:schunit}
\end{equation}
where the sum runs over all possible intermediate states $X$.
In this work, we are mainly interested in VCS through the
$\Delta(1232)$-resonance region.  
Therefore, we restrict ourselves to the dominant contribution by
only taking account of the $\pi N$ intermediate states. 
The influence of additional channels, like the $\pi \pi N$ 
intermediate states which are indispensable when extending the
dispersion formalism to higher energies, 
will be investigated in a future work.
\newline
\indent
The VCS helicity amplitudes can be expressed by the $F_i$ 
in a straightforward manner, even though the calculation is cumbersome.
The main difficulty, however, is the inversion of the relation 
between the two sets of amplitudes,
i.e., to express the twelve amplitudes $F_i$ 
in terms of the twelve independent helicity amplitudes.
To solve this problem we proceeded in two different ways. First, 
the inversion was performed numerically by applying different 
algorithms. Second, we succeeded in 
obtaining an analytical inversion using a two-step procedure. 
To this end we used an additional set of amplitudes,
called $B_i$, which were introduced by Berg and Lindner
\cite{Ber58} and which are defined in Appendix~{\ref{app:berglindner}}.
Both the relations between the $B_i$ and the $F_i$ on the one hand, 
and between the helicity amplitudes and the $B_i$ on the other hand 
can be inverted analytically. 
The expressions of the $F_i$ amplitudes in terms of the $B_i$
amplitudes are given in Appendix~\ref{app:berglindner}, and the
expressions of the $B_i$ amplitudes in terms of the VCS helicity
amplitudes are given in Appendix~\ref{app:relblhel} (for the
definition of the VCS helicity amplitudes, see 
Appendices~\ref{app:sheldef} and ~\ref{app:reduced_ampl}).
In our calculations, we checked that 
the two methods to express the $F_i$ amplitudes in terms of the VCS
helicity amplitudes lead numerically to the same results.
\newline
\indent
The imaginary parts of the s-channel VCS helicity amplitudes 
are calculated through unitarity 
taking into account the contribution from $\pi N$ intermediate states.
They are expressed in terms of pion photo- and electroproduction
multipoles as specified in Appendix~\ref{app:pionprod}. 
For the calculation of the pion photo- and electroproduction
multipoles, we use the phenomenological MAID analysis \cite{maid00}, 
which contains both resonant and non-resonant pion production mechanisms.

\section{Asymptotic parts and dispersive contributions beyond $\pi N$}
\label{asympt}

To evaluate the VCS amplitudes $F_1$ and $F_5$ in an unsubtracted 
DR framework, we proceed as 
in the case of RCS \cite{lvov97}. This amounts to perform the  
unsubtracted dispersion integrals~(\ref{eq:unsub}) for $F_1$ and $F_5$  
along the real $\nu$-axis only in the range 
$-\nu_{max}\leq\nu\leq+\nu_{max}$,
and to close the contour by a semi-circle with radius $\nu_{max}$ 
in the upper half of the complex $\nu$-plane, with the result
\begin{equation}
{\mathrm Re} F_{i}^{NB}(Q^2, \nu, t) \;=\; 
F_{i}^{int}(Q^2, \nu, t) \;+\; F_{i}^{as}(Q^2, \nu, t) \;,
\hspace{1cm} ({\rm for} \;\; i = 1, 5) \;, 
\label{eq:unsub2} 
\end{equation}
where the integral contributions $F_{i}^{int}$ (for $i = 1, 5$) are given by
\begin{equation}
F_{i}^{int}(Q^2, \nu, t) \;=\; 
{2 \over \pi} \; {\mathcal P} \int_{\nu_{thr}}^{\nu_{max}} d\nu' \; 
{{\nu' \; {\mathrm Im}_s F_{i}(Q^2, \nu',t)} \over {\nu'^2 - \nu^2}}\;, 
\hspace{1cm} ({\rm for} \;\; i = 1, 5) \;,
\label{eq:unsub3} 
\end{equation}
and with the contributions of the semi-circle of radius $\nu_{max}$ 
identified with the asymptotic contributions 
($F_1^{as}$, $F_5^{as}$). 
\newline
\indent
Evidently, the separation between asymptotic and integral contributions 
in Eq.~(\ref{eq:unsub2}) is specified by the value of $\nu_{max}$. 
The total result for $F_i^{NB}$ is formally independent 
of the specific value of $\nu_{max}$. 
In practice, however, $\nu_{max}$ is chosen to be not too large so
that one can evaluate the dispersive integrals of
Eq.~(\ref{eq:unsub3}) from threshold up to $\nu_{max}$ sufficiently
accurate. On the other hand, $\nu_{max}$ should also be large enough
so that one can approximate the asymptotic contribution $F_i^{as}$ 
by some energy-independent (i.e. $\nu$-independent) function. 
In the calculations, we therefore choose some intermediate value 
$\nu_{max} \approx 1.5$~GeV, and parametrize the asymptotic contributions
$F_i^{as}$ by $t$-channel poles, which will be discussed next for   
the cases of $F_5^{as}$ and $F_1^{as}$.

\subsection{The asymptotic contribution $F_5^{as}$}

\begin{figure}[ht]
\vspace{7cm}
\epsfxsize=12cm
\centerline{\epsffile{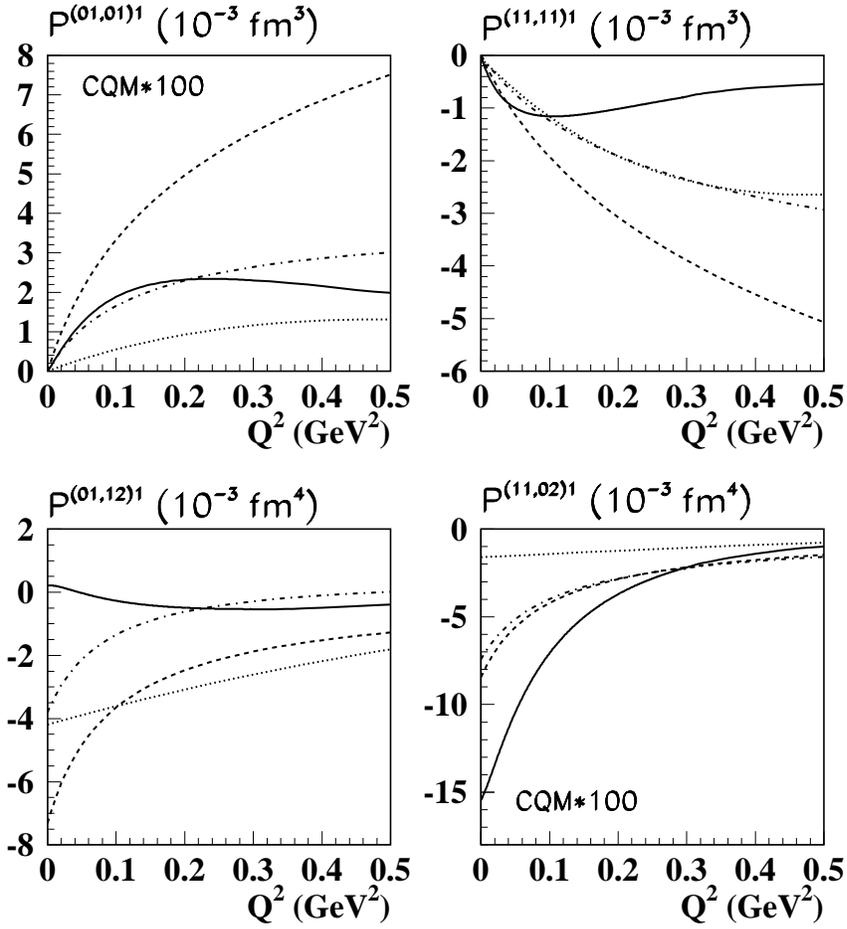}}
\caption[]{Results for the spin-flip GPs excluding the $\pi^0$-pole
  contribution in different model calculations, 
as function of the squared momentum transfer.
The full curves correspond to the dispersive $\pi N$ contribution.
The dashed curves show the results of ${\cal O}(p^3)$ HBChPT~\cite{Hemmert00},
the dashed-dotted curves correspond to the predictions of the 
linear $\sigma$-model~\cite{Metz96}, and the dotted curves are the results of 
the nonrelativistic constituent quark model~\cite{PSD00}.
Note that the constituent quark model (CQM) results for $P^{(01,01)1}$ and
$P^{(11,02)1}$ are multiplied (for visibility) by a factor 100.} 
\label{fig:polarizab_comp}
\end{figure}

\begin{figure}[ht]
\vspace{7cm}
\epsfxsize=12cm
\centerline{\epsffile{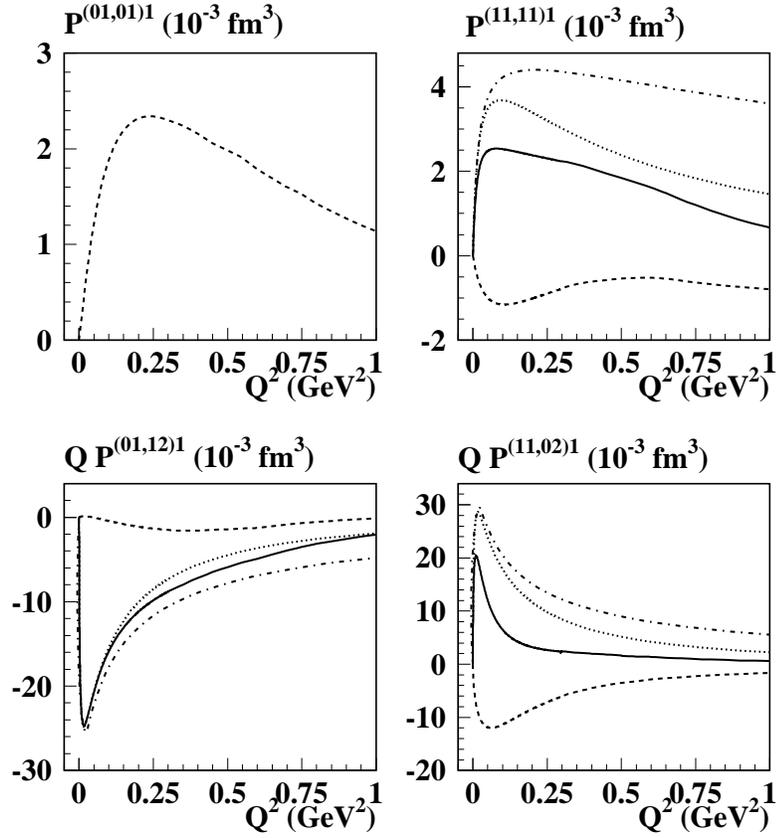}}
\caption[]{Results for the spin-flip GPs 
as function of the squared momentum transfer.
The dashed curves correspond to the dispersive $\pi N$ contribution,
the dotted curves show the $\pi^0$-pole contribution, and the 
full curves are the sum of the dispersive and $\pi^0$-pole
contributions. 
For comparison, we also show the $\pi^0$-pole contribution when setting
the $\pi^0 \gamma^* \gamma$ form factor equal to 1 (dashed-dotted curves).  
Note that $P^{(01,01)1}$ has no $\pi^0$-pole contribution.} 
\label{fig:polarizab_anom}
\end{figure}

The asymptotic contribution to the amplitude $F_5$ predominantly
results from the $t$-channel $\pi^0$-exchange,
\begin{equation}
F^{as}_{5}(Q^2, \nu, t) \,\approx\, F_{5}^{\pi^0}(Q^2, t) 
\,=\, -4 \, F_{11}^{\pi^0}(Q^2, t) \,=\, 
 {1 \over {M}} \, 
{ g_{\pi NN} \; {F_{\pi^0 \gamma \gamma}\left( Q^2 \right)} 
\over {t - m_\pi^2}} \, . 
\label{eq:piopole}
\end{equation}
As mentioned before, the $\pi^0$-pole only contributes to the amplitudes 
$F_5$ and $F_{11}$, but drops out in the combination 
$( F_5 + 4 \, F_{11} )$, which therefore has a different high-energy
behavior as expressed in Eq.~(\ref{eq:reggevcs2}). 
In Eq.~(\ref{eq:piopole}), the $\pi NN$ coupling $g_{\pi NN}$ 
is taken from Ref.~\cite{vpi99}~: 
$g^2_{\pi NN}/(4 \pi)$ = 13.73.  
Furthermore, in~(\ref{eq:piopole}), $F_{\pi^0 \gamma \gamma}\left( Q^2 \right)$
represents the $\pi^0 \gamma^* \gamma$ form factor. Its value 
at $Q^2 = 0$ is fixed by the axial anomaly~: 
$F_{\pi^0 \gamma \gamma}\left( 0 \right)$ =   
$1 / (4 \, \pi^2 \, f_\pi)$ = 0.274~GeV$^{-1}$, where $f_\pi$ =
0.0924 GeV is the pion decay constant.
For the $Q^2$-dependence of $F_{\pi^0 \gamma \gamma} \left( Q^2 \right)$, 
we use the interpolation formula proposed by
Brodsky-Lepage \cite{bl81}~: 
\begin{equation}
F_{\pi^0 \gamma \gamma} \left( Q^2 \right) \; = \; 
{{1 / (4 \, \pi^2 \, f_\pi)} 
\over {1 + Q^2/ (8 \, \pi^2 \, f_\pi^2)}} \, ,
\label{eq:piogagaff}
\end{equation}
which provides a rather good parametrization of the 
$\pi^0 \gamma^* \gamma$ form factor data over the whole $Q^2$ range, and 
which leads to the asymptotic prediction at large $Q^2$ : 
$F_{\pi^0 \gamma \gamma} \left( Q^2 \gg \right) \,\rightarrow\, 
2 \, f_\pi / Q^2$.
\newline
\indent
When fixing the asymptotic contribution $F_5^{as}$ through its
$\pi^0$-pole contribution as in Eq.~(\ref{eq:piopole}), 
one can determine one more GP of the nucleon, in addition to the
four combinations of Eqs.~(\ref{eq:gpdisp1} - \ref{eq:gpdisp4}). 
In particular, the GP $P^{\left(1 1, 1 1\right)1}$ can be expressed by~:
\begin{equation}
P^{\left(1 1, 1 1\right)1} \left( Q^2 \right) \,=\, -
{{\sqrt{2}} \over {3}}\,\left( {{E + M} \over E}\right)^{1/2} 
{{M \, \tilde q_0^2} \over {\rmq^2}} 
\left\{ \bar F_5 (Q^2) \,+\, \tilde q_0 \, \bar F_{12} (Q^2) \right\}. 
\label{eq:gp1111} 
\end{equation}
In Fig.~\ref{fig:polarizab_comp}, we show the results of the dispersive
contribution to the four spin GPs, and compare them to the results of the 
${\cal O}(p^3)$ 
heavy-baryon chiral perturbation theory (HBChPT)~\cite{Hemmert00},
the linear $\sigma$-model~\cite{Metz96}, 
and the nonrelativistic constituent quark model~\cite{PSD00}.
It is obvious that the DR calculations show more structure in $Q^2$ than
the different model calculations. 
\newline
\indent
The ${\cal O}(p^3)$ HBChPT results predict for the GPs 
$P^{\left(0 1, 0 1\right)1}$ and $P^{\left(1 1, 1 1\right)1}$ a
rather strong increase with $Q^2$, which would have to be checked by a 
${\cal O}(p^4)$ calculation. 
\newline
\indent
The constituent quark model calculation gives negligibly small
contributions for the GPs $P^{\left(0 1, 0 1\right)1}$ and 
$P^{\left(1 1, 0 2\right)1}$, whereas the GPs 
$P^{\left(1 1, 1 1\right)1}$ and $P^{\left(0 1, 1 2\right)1}$ 
receive their dominant contribution from the excitation of the
$\Delta (1232)$ ($M1 \to M1$ transition) and $N^*$ and $\Delta^*$ resonances 
($E1 \to M2$ transition) respectively. 
\newline
\indent
The linear $\sigma$-model, which takes account of part of the higher order
terms of a consistent chiral expansion, in general results in smaller values
for the GPs than the corresponding calculations to leading order in HBChPT.
\newline
\indent
The comparison in Fig.~\ref{fig:polarizab_comp} clearly indicates 
that a satisfying theoretical description of the GPs over a 
larger range in $Q^2$ is a challenging task.
\newline
\indent
In Fig.~\ref{fig:polarizab_anom}, we show the dispersive and $\pi^0$-pole
contributions to the 4 spin GPs as well as their sum. 
For the presentation, we multiply in Fig.~\ref{fig:polarizab_anom} 
the GPs $P^{\left(0 1, 1 2\right)1}$ and $P^{\left(1 1, 0 2\right)1}$
with $Q$, in order to better compare the $Q^2$ dependence when including
the $\pi^0$-pole contribution, which itself drops very fast with
$Q^2$. The $\pi^0$-pole does
not contribute to the GP $P^{\left(0 1, 0 1\right)1}$, but 
is seen to dominate the other three spin GPs. 
It is however possible to find, besides the
GP $P^{\left(0 1, 0 1\right)1}$, the two combinations 
given by Eqs.~(\ref{eq:gpdisp3},\ref{eq:gpdisp4}) 
of the remaining three spin GPs,  
for which the $\pi^0$-pole contribution drops out \cite{Pas00}.

\subsection{The asymptotic part and dispersive contributions
  beyond $\pi N$ to $F_1$}
\label{sec:f1}

\begin{figure}[ht]
\vspace{7cm}
\epsfxsize=12cm
\centerline{\epsffile{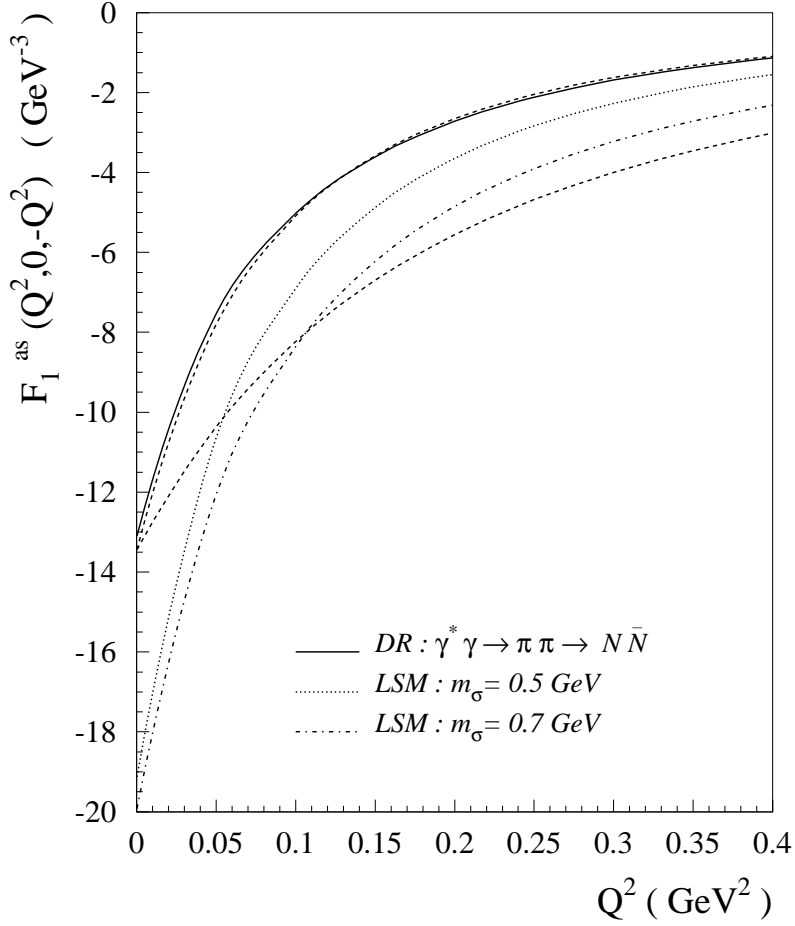}}
\caption[]{Theoretical estimates of the asymptotic contribution
  $F_1^{as}$ : DR calculation \cite{Dre99}
of the $\gamma^* \gamma \to \pi \pi \to N \bar N$ process
as described in the text in Sec.~\ref{sec:f1} 
(full curve); linear $\sigma$-model (LSM) 
calculation \cite{Metz96} with $m_\sigma$ = 0.5 GeV 
(dotted curve) and $m_\sigma$ = 0.7 GeV (dashed-dotted curve). 
The dashed curves are dipole parametrizations 
according to Eq.~(\ref{eq:gpbetaparam}), which are fixed 
to the phenomenological value at $Q^2$ = 0 and are shown for
two values of the mass-scale, $\Lambda_\beta$ = 0.4 GeV 
(upper dashed curve, nearly coinciding with full curve) 
and $\Lambda_\beta$ = 0.6 GeV (lower dashed curve).}
\label{fig:f1_asymp}
\end{figure}

We next turn to the high-energy contribution to $F_1$.
As we are mainly interested in a description of VCS up to
$\Delta(1232)$-resonance energies, 
we saturate the dispersion integrals by their $\pi N$ contribution. 
Furthermore, we will estimate the remainder by an energy-independent
function, which parametrizes the asymptotic contribution (i.e. the
contour with radius $\nu_{max}$ in the complex $\nu$-plane), and  
all dispersive contributions beyond the $\pi N$ channel 
up to the value $\nu_{max} = 1.5$ GeV.
\newline
\indent
Before turning to the case of VCS, we briefly outline the parametrization 
of the asymptotic part of $F_1$ in the case of RCS, 
and how one expresses it in terms of
a polarizability, which is then extracted from a fit to experiment.
\newline
\indent 
The asymptotic contribution to the amplitude $F_1$ 
originates predominantly from the $t$-channel $\pi \pi$ intermediate 
states, and will be calculated explicitly in two model calculations. 
In the phenomenological analysis, this continuum 
is parametrized through the exchange of a scalar-isoscalar particle in the
$t$-channel, i.e. an effective ``$\sigma$''-meson, as suggested in
Ref.~\cite{lvov97}. 
For RCS, this leads to the parametrization of the 
difference of $F_1^{NB}$ and its $\pi N$ contribution, as
an energy-independent function~:
\begin{equation}
F_1^{NB}(Q^2 = 0, \nu, t) - F_1^{\pi N}(Q^2 = 0, \nu, t) \, \approx \,
\left[ F_1^{NB}(0, 0, 0) - F_1^{\pi N}(0, 0, 0) \right] 
\, {{1} \over {1 - t/m_\sigma^2}} \;,  
\label{eq:f1asrcs}
\end{equation}
where $F_1^{\pi N}$ on the $lhs$ and $rhs$ are evaluated through a
dispersive integral as discussed in section \ref{schannel}. 
In Eq.~(\ref{eq:f1asrcs}), the effective ``$\sigma$''-meson mass  
$m_\sigma$ is a free parameter in the RCS dispersion analysis, 
which is obtained from a fit to the $t$-dependence of RCS data, and  
turns out to be around $m_\sigma \approx$ 0.6 GeV \cite{lvov97}.  
The value $F_1^{NB}(0, 0, 0)$ is then considered as a remaining gobal 
fit parameter to be extracted from experiment. 
It can be expressed physically in terms of the magnetic polarizability
$\beta$~:
\begin{equation}
F_1^{NB}(0, 0, 0) \,=\, {{4 \pi} \over {e^2}} \; \beta  \;.
\label{eq:f1beta}
\end{equation}
In RCS, one usually takes $(\alpha - \beta)$ as fit parameter instead of
$\beta$ because the sum $(\alpha + \beta)$ at the real photon point 
can be determined independently, and rather accurately, 
through Baldin's sum rule, 
which leads for the proton to the phenomenological value \cite{Bab98}~:
\begin{equation} 
\alpha + \beta \;=\; \left(\, 13.69 \,\pm\, 0.14 \, \right) \,
\times\,10^{-4}\,{\rm fm}^3.
\label{eq:baldinexp}
\end{equation}
Using a dispersive formalism as outlined above, the most recent global
fit to RCS data for the proton yields the following values 
for the electric and magnetic polarizabilities of the proton \cite{Wiss00}~:
\begin{eqnarray}
\alpha \;&=&\; \left(\, 12.1 \,\pm\, 0.3 \, {\mathrm (stat.)} \, \mp \, 
0.4 \, {\mathrm (syst.)} \pm \, 0.3 \, {\mathrm (model)} \, \right) 
\,\times\,10^{-4}\,{\rm fm}^3  \, , 
\label{eq:alpharcs}\\
\beta \;&=&\; \left( \,1.6 \,\pm\, 0.4 \, {\mathrm (stat.)} \, \pm \, 
0.4 \,  {\mathrm (syst.)} \pm \, 0.4 \, {\mathrm (model)} \, \right)  
\,\times\,10^{-4}\,{\rm fm}^3  \, .
\label{eq:betarcs}
\end{eqnarray}
From Eqs.~(\ref{eq:alpharcs}, \ref{eq:betarcs}), one then obtains 
for the difference ($\alpha - \beta$), the following 
global average \cite{Wiss00}~:
\begin{equation}
\alpha - \beta \;=\; \left( \, 10.5 \,\pm\, 0.9 \, 
{\mathrm (stat. + syst.)} \,\pm\, 0.7 \, {(\mathrm model)} \, \right) 
\,\times\,10^{-4}\,{\rm fm}^3  \, .
\label{eq:ambexp}
\end{equation}
\indent
The term $F_1^{\pi N}(0, 0, 0)$ in Eq.~(\ref{eq:f1asrcs}),
when calculated through a dispersion integral, has the value~:
\begin{equation}
F_1^{\pi N}(0, 0, 0) \,=\, {{4 \pi} \over {e^2}} \; \beta^{\pi N} 
\, = \,  {{4 \pi} \over {e^2}} \; 
\left( 9.1 \,\times\, 10^{-4}\,{\rm fm}^3 \right) \;.
\label{eq:betapin}
\end{equation}
From the $\pi N$ contribution $\beta^{\pi N}$ of Eq.~(\ref{eq:betapin}), 
and the phenomenological value $\beta$ of Eq.~(\ref{eq:betarcs}), 
one obtains the difference~:
\begin{equation}
(\beta - \beta^{\pi N}) = 
- 7.5 \,\times\,10^{-4}\,{\rm fm}^3 \; , 
\label{eq:bmbpin}
\end{equation}
which enters in the {\it rhs} of Eq.~(\ref{eq:f1asrcs}).
By comparing the value of Eq.~(\ref{eq:bmbpin}) 
with the total value for $\beta$
(Eq.(\ref{eq:betarcs})), one sees that the small experimental value of the 
magnetic polarizability comes about by a near cancellation between a large
(positive) paramagnetic contribution ($\beta^{\pi N}$) and a 
large (negative) diamagnetic contribution ($\beta - \beta^{\pi N}$),  
i.e. the asymptotic part of $F_1$ parametrizes the diamagnetism.
\newline
\indent
Turning next to VCS, we proceed analogously by parametrizing the
non-Born term $F_1^{NB}(Q^2, \nu, t)$ 
beyond its $\pi N$ dispersive contribution, by 
an energy-independent $t$-channel pole of the form~:
\begin{equation}
F_1^{NB}(Q^2, \nu, t) \, - \, F_1^{\pi N}(Q^2, \nu, t) \, \approx \,  
{{f(Q^2)} \over {1 - t/m_\sigma^2}} \;,    
\label{eq:f1asvcs1}
\end{equation}
where the parameter $m_\sigma$ is taken as for RCS : 
$m_\sigma \approx$ 0.6 GeV.
The function $f(Q^2)$ in Eq.~(\ref{eq:f1asvcs1}) can be obtained by
evaluating the $lhs$ of Eq.~(\ref{eq:f1asvcs1}) at the point where the
GPs are defined, i.e. $\nu = 0$ and $t = - Q^2$, at finite $Q^2$. 
This leads to~:
\begin{equation}
f(Q^2) \, = \, 
\left[ \bar F_1(Q^2) - \bar F_1^{\pi N}(Q^2) \right] \; 
\left(1 + Q^2/m_\sigma^2 \right) \;,   
\label{eq:f1asvcs2}
\end{equation}
where we introduced the shorthand $\bar F_1(Q^2)$ 
as defined in Eq.~(\ref{eq:fbardef}).
$\bar F_1(Q^2)$ can be expressed in terms of 
the generalized magnetic polarizability $P^{(1 1, 1 1)0} (Q^2)$ 
of Eq.~(\ref{eq:defgpunpol}) as \cite{Dre98}~: 
\begin{eqnarray}
\bar F_1(Q^2) \,&=&\,
- \sqrt{{3 \over 8}}\,\left( {{2 E} \over {E + M}}\right)^{1/2} \, 
P^{\left(1 1, 1 1\right)0}(Q^2) \\
\,&\equiv&\, {{4 \pi} \over {e^2}} \, 
\left( {{2 E} \over {E + M}}\right)^{1/2} \, \beta(Q^2) \, ,
\label{eq:betaq} 
\end{eqnarray}
where $\beta(Q^2$) is the generalized magnetic polarizability, 
which reduces at $Q^2$ = 0 to the polarizability $\beta$ of RCS.
\newline
\indent
Eqs.(\ref{eq:f1asvcs1}, \ref{eq:f1asvcs2}) then lead  
to the following expression for the VCS amplitude $F_1^{NB}$~:
\begin{equation}
F_1^{NB}(Q^2, \nu, t) \approx F_1^{\pi N}(Q^2, \nu, t) \;+\; 
\left[ \bar F_1(Q^2) \,-\, \bar F_1^{\pi N}(Q^2) \right]\, 
{{1 + Q^2/m_\sigma^2} \over {1 - t/m_\sigma^2}} \;,  
\label{eq:f1asvcs}
\end{equation}
where the $\pi N$ contributions $F_1^{\pi N}(Q^2, \nu, t)$ and 
$\bar F_1^{\pi N}(Q^2)$ (or equivalently $\beta^{\pi N}(Q^2)$) 
are calculated through a dispersion integral as outlined above. 
Consequently, the only unknown quantity on the {\it rhs} of 
Eq.~(\ref{eq:f1asvcs}) is $\bar F_1(Q^2)$, which can be directly used as
a fit parameter at finite $Q^2$. This amounts to fit the generalized
magnetic polarizability $\beta(Q^2)$ from VCS observables. 
\newline
\indent
The parametrization of Eq.~(\ref{eq:f1asvcs}) for $F_1$ permits to 
extract $\beta(Q^2)$ from VCS observables at some finite $Q^2$ and over
a larger range of energies with as few model dependence as possible. 
In the following, we consider a convenient parametrization of the 
$Q^2$ dependence of $\beta(Q^2)$ in order to provide predictions 
for VCS observables.  
For this purpose we use a dipole form for the difference of $\beta(Q^2) -
\beta^{\pi N}(Q^2)$, which enters in the {\it rhs}
of Eq.~(\ref{eq:f1asvcs}) via Eq.~(\ref{eq:betaq}).
This leads to the form~:
\begin{eqnarray}
\beta(Q^2) - \beta^{\pi N}(Q^2) \,=\, {{ \left( \beta - \beta^{\pi N}\right) } 
\over {\left( 1 + Q^2 / \Lambda^2_{\beta} \right)^2 }} \, ,
\label{eq:gpbetaparam}
\end{eqnarray}
where the RCS value $(\beta - \beta^{\pi N})$ on the {\it rhs} is
given by Eq.~(\ref{eq:bmbpin}). The mass scale $\Lambda_{\beta}$ 
in Eq.~(\ref{eq:gpbetaparam}) determines the $Q^2$ dependence, 
and hence gives us the information how the diamagnetism 
is spatially distributed in the nucleon. Using the dipole
parametrization of Eq.~(\ref{eq:gpbetaparam}), one can extract 
$\Lambda_{\beta}$ from a fit to VCS data at different $Q^2$ values. 
\newline
\indent
To have some educated guess on the physical value of $\Lambda_\beta$, we next
discuss two microscopic calculations of the diamagnetic contribution
to the GP $\beta(Q^2)$. The diamagnetism of the nucleon is dominated
by the pion cloud surrounding the nucleon. Therefore,
we calculate the diamagnetic contribution through a
dispersion relation estimate of the $t$-channel $\pi \pi$ intermediate
state contribution to $F_1$. Such a dispersive estimate 
has been performed before in the case of RCS \cite{Hol94,Dre99},
where it was shown that the asymptotic part of $F_1$ can be related to
the $\gamma \gamma \to \pi \pi \to N \bar N$ process. The dominant
contribution is due to the $\pi \pi$ intermediate state with 
spin and isospin zero ($I = J = 0$). The generalization to VCS 
leads then to the identification of $F_1^{as}$ with 
the following unsubtracted DR in $t$ at fixed energy $\nu = 0$~:
\begin{eqnarray}
\bar F_1^{as} (Q^2)\,=\,
\frac{1}{\pi}\int^\infty_{4m_{\pi}^2} dt' 
\frac{{\mathrm Im}_t F_1 (Q^2,0, t')}{t'+Q^2}\,.
\label{eq:t-DR}
\end{eqnarray}
The evaluation of the imaginary part on the {\it rhs} of
Eq.~(\ref{eq:t-DR}), originating mainly from the  $\pi \pi$ intermediate
state contribution, requires information on the subprocesses
$\gamma^* \gamma\rightarrow \pi\pi$ and $\pi\pi\rightarrow
N{\bar{N}}$. For the latter we use the extrapolation 
of Ref.~\cite{Hohler} for the $\pi N$-scattering amplitude 
to the unphysical region of positive $t$. 
For the $\gamma^* \gamma\rightarrow \pi\pi$ amplitude, 
we use the unitarized Born amplitude, following Ref.~\cite{Dre99}. 
At the pion electromagnetic vertex, the pion electromagnetic form
factor is included. 
At $Q^2$ = 0, it was found \cite{Dre99} that the 
unitarization procedure enhances the $\gamma \gamma \rightarrow
\pi \pi$ cross section in the threshold region, compared to the Born
result, which is required to get agreement with the data. 
This becomes obvious from the DR of Eq.~(\ref{eq:t-DR}), 
where the imaginary part of $F_1$ is weighted by 1/$t$, 
so that the threshold contribution dominates the dispersion integral. 
The dispersive evaluation of Eq.~(\ref{eq:t-DR}) contains no free parameters 
as it uses as input the $\gamma \gamma \to \pi \pi$ and $\pi \pi \to N
\bar N$ processes, and therefore provides a more microscopic model for 
the phenomenological ``$\sigma$''-exchange. 
For RCS, the dispersion integral Eq.~(\ref{eq:t-DR}) yields the value 
$\beta^{as} \approx - 7.3 \,\times\,10^{-4}\,{\rm fm}^3$.   
However, the unsubtracted dispersion integral can only be evaluated up to 
$-t\,=\,0.778$ GeV$^2$, because the $\pi\pi\rightarrow N{\bar{N}}$
amplitudes of Ref.~\cite{Hohler} were only determined up to this
value, and the dispersion integral of Eq.~(\ref{eq:t-DR}) 
may not have fully converged at this value.
Therefore, one should consider the near perfect agreement between the
value of $\beta^{as}$ from this calculation 
with the phenomenological value of~(\ref{eq:bmbpin}) as a
coincidence. However, our estimate indicates that the dispersive estimate
through $\pi \pi$ $t$-channel intermediate states provides the
dominant physical contribution to the diamagnetism, and 
that it can be used to give a first guess of the distribution of
diamagnetism in the nucleon. With this model we show the $Q^2$
dependence of $\bar F_1^{as}$ in Fig.~\ref{fig:f1_asymp}.    
\newline
\indent
To have a second microscopic calculation for comparison, we also show in 
Fig.~\ref{fig:f1_asymp} an evaluation of $\bar F_1^{as}(Q^2)$ 
in the linear $\sigma$-model (LSM) of Ref.~\cite{Metz96}.
The LSM calculation overestimates the value of 
$\bar F_1^{as}(0)$ (or equivalently $\beta_{as}$) by 
about 30\% at any realistic value of $m_\sigma$ (which is a free
parameter in this calculation). However, as for the dispersive
calculation, it also shows a steep $Q^2$ dependence. 
\newline
\indent 
Furthermore, we compare in Fig.~\ref{fig:f1_asymp} 
the two model calculations discussed above with the dipole 
parametrization for $\beta(Q^2) - \beta^{\pi N}(Q^2)$ 
of Eq.~(\ref{eq:gpbetaparam}) for the two values~: $\Lambda_\beta$ = 0.4 GeV 
and $\Lambda_\beta$ = 0.6 GeV. It is seen that these values are
compatible with the microscopic estimates discussed before. 
In particular, the result for $\Lambda_\beta$ = 0.4 GeV is nearly
equivalent to the dispersive estimate of $\pi \pi$
exchange in the $t$-channel. The value of the mass scale
$\Lambda_\beta$ is small compared to the typical scale of 
$\Lambda_D \approx$ 0.84 GeV appearing in the nucleon 
magnetic (dipole) form factor.
This reflects the fact that diamagnetism has its physical 
origin in the pion cloud, i.e. is situated in the surface region of
the nucleon.

\subsection{Dispersive contributions beyond $\pi N$ to $F_2$}

Though we can write down unsubtracted DRs 
for all invariant amplitudes (or combinations of invariant
amplitudes) except for $F_1$ and $F_5$, 
one might wonder about the quality of our approximation to saturate the
unsubtracted dispersion integrals by $\pi N$ intermediate states only. 
We shall show that this question is particularly relevant for the
amplitude $F_2$, for which we next investigate the size of 
dispersive contributions beyond the $\pi N$ channel.  
We start with the case of RCS, where one can
quantify the higher dispersive corrections to $F_2$, because the value of 
$F_2^{NB}$ at the real photon point can be expressed exactly 
(see Eqs.~(\ref{eq_3_47}, \ref{eq:gpdisp1})) in terms of the scalar
polarizability sum $(\alpha + \beta)$ as~: 
\begin{equation}
F_2^{NB}(0, 0, 0) \,=\, - {{4 \pi} \over {e^2}} \, {1 \over {(2 M)^2}} \,
\left(\alpha + \beta \right) \, .
\label{eq:f2rcs}
\end{equation}
The $\pi N$ dispersive contribution to $(\alpha + \beta)$ 
provides the value~: 
\begin{equation}
(\alpha+\beta)^{\pi N} = 11.6 \times 10^{-4} \, {\mathrm fm}^3 \, , 
\label{eq:apbpin}
\end{equation}
which falls short by about 15 \% compared to the sum rule value 
of Eq.~(\ref{eq:baldinexp}). The remaining part originates from higher
dispersive contributions ($\pi \pi N$, ...) to $F_2$. 
These higher dispersive contributions could be calculated through
unitarity, by use of Eq.~(\ref{eq:schunit}), similarly to the 
$\pi N$ contribution. However, the present data for the production of
those intermediate states (e.g. $\gamma^* N \to \pi \pi N$) are still
too scarce to evaluate the imaginary parts of the VCS amplitude $F_2$
directly. Therefore, we estimate the dispersive contributions 
beyond $\pi N$ by an energy-independent constant, 
which is fixed to its phenomenological value at $\nu = t = 0$. This yields~: 
\begin{equation}
F_2^{NB}(Q^2 = 0, \nu, t) \approx F_2^{\pi N}(Q^2 = 0, \nu, t) \;-\; 
{{4 \pi} \over {e^2}} \, {1 \over {(2 M)^2}} \,
\left[ \, \left(\alpha + \beta \right) \,-\, 
\left(\alpha + \beta \right)^{\pi N} \, \right] \, ,
\label{eq:f2approx}
\end{equation}
which is an exact relation at $\nu = t = 0$, the point where 
the polarizabilities are defined. 
\newline
\indent
The approximation of Eq.~(\ref{eq:f2approx}) to replace 
the dispersive contributions beyond $\pi N$ by a
constant can only be valid if one stays below the thresholds for those
higher contributions. Since the next threshold beyond $\pi N$ is $\pi \pi N$,
the approximation of Eq.~(\ref{eq:f2approx}) restricts us in practice
to energies below the $\Delta(1232)$-resonance.  
If one wanted to extend the DR formalism to energies above two-pion
production threshold, one could proceed in an analogous way by 
replacing Eq.~(\ref{eq:f2approx}) as follows~:
\begin{eqnarray}
F_2^{NB}(Q^2 = 0, \nu, t) &\approx& F_2^{\pi N}(Q^2 = 0, \nu, t) 
\;+\; F_2^{\pi \pi N}(Q^2 = 0, \nu, t) \nonumber \\ 
&-&\; {{4 \pi} \over {e^2}} \, {1 \over {(2 M)^2}} \,
\left[ \, \left(\alpha + \beta \right) \,-\, 
\left(\alpha + \beta \right)^{\pi N} \, \,-\,
\left(\alpha + \beta \right)^{\pi \pi N} \, \right] \, ,
\label{eq:f2approx2pi}
\end{eqnarray}
i.e. the energy-dependence associated with $\pi N$ and $\pi \pi
N$ dispersive contributions would have to be calculated explicitly 
and the remainder be parametrized by
an energy-independent constant fixed to the phenomenological
value of $(\alpha + \beta)$. Eq.~(\ref{eq:f2approx2pi}), and 
Eq.~(\ref{eq:f1asrcs}) for $F_1^{NB}$ modified in an analogous way to
include the $\pi \pi N$ dispersive contributions,
would then allow an extension of the DR formalism to energies into the
second resonance region. 
Such an extension remains to be investigated in a future work, 
but because of the present lack of experimental input for the 
$\pi \pi N$ channel, we restrict ourselves in the present work 
to energies up to the $\Delta (1232)$-resonance region. 
\newline
\indent
We next consider the extension to VCS, and focus our efforts to 
describe VCS into the $\Delta(1232)$-resonance region. 
Analogously to Eq.~(\ref{eq:f2approx}) for RCS, the 
dispersive contributions beyond $\pi N$ are approximated by an
energy-independent constant. This constant is fixed 
at arbitrary $Q^2$, $\nu = 0$, and $t = - Q^2$, which is the point 
where the GPs are defined. One thus obtains for $F_2^{NB}$~:
\begin{equation}
F_2^{NB}(Q^2, \nu, t) \approx F_2^{\pi N}(Q^2, \nu, t) \;+\; 
\left[ \bar F_2(Q^2) \,-\, \bar F_2^{\pi N}(Q^2) \right] \, , 
\label{eq:f2asvcs}
\end{equation}
where $\bar F_2(Q^2)$ is defined as in Eq.~(\ref{eq:fbardef}), and can be
expressed in terms of GPs. 
In this paper, we saturate the three combinations of spin
GPs of Eqs.~(\ref{eq:gpdisp2} - \ref{eq:gpdisp4}) by their $\pi N$
contribution, and calculate   
the fourth spin GP of Eq.~(\ref{eq:gp1111}) through
its $\pi N$ contributions plus the $\pi^0$-pole contribution 
as shown in Fig.~\ref{fig:polarizab_anom}. 
Therefore, we only consider dispersive contributions beyond the $\pi N$
intermediate states for the two scalar GPs, which are then
two fit quantities that enter our DR formalism for
VCS. In this way, and by using Eq.~(\ref{eq:gpdisp1}), 
one can write the difference $\bar F_2(Q^2) - \bar
F_2^{\pi N}(Q^2)$ entering in the {\it rhs} of Eq.~(\ref{eq:f2asvcs}) 
as follows~:
\begin{eqnarray}
\bar F_2(Q^2) - \bar F_2^{\pi N}(Q^2) \, &\approx & \,
{{4 \pi} \over {e^2}} \, 
\left( {{2 E} \over {E + M}}\right)^{1/2} 
\, {{\tilde q_0} \over {\rmq^2}} \, {1 \over {2 M}} 
\nonumber \\
&& \times \left\{\;  \left[\alpha(Q^2) \,-\, \alpha^{\pi N}(Q^2)\right] 
\; + \;  \left[\beta(Q^2) \,-\, \beta^{\pi N}(Q^2)\right] \; \right\} \, ,
\label{eq:alphaq} 
\end{eqnarray}
where $\beta(Q^2)$ is the generalized magnetic polarizability 
of Eq.~(\ref{eq:betaq}). Furthermore, 
$\alpha(Q^2$) is the generalized electric polarizability
which reduces at $Q^2$ = 0 to the electric polarizability $\alpha$ of RCS, and
which is related to the GP $P^{(01,01)0}(Q^2)$ of
Eq.~(\ref{eq:defgpunpol}) by~:
\begin{eqnarray}
P^{(01,01)0}(Q^2) \,\equiv\,-{{4 \pi} \over {e^2}}\, \sqrt{\frac{2}{3}} \,
\alpha(Q^2) \;. 
\end{eqnarray}
\newline
\indent
We stress that Eqs.~(\ref{eq:f1asvcs}) and (\ref{eq:f2asvcs}) 
are intended to extract the two GPs $\alpha(Q^2)$ and $\beta(Q^2)$  
from VCS observables minimizing the model dependence as much as possible. 
As in the previous case for $\beta(Q^2)$, we next consider 
a convenient parametrization of the $Q^2$ dependence 
of $\alpha(Q^2)$ in order to provide predictions for VCS observables.  
Again we propose a dipole form for the difference  
$\alpha(Q^2) - \alpha^{\pi N}(Q^2)$ which enters in the {\it rhs} of 
Eq.~(\ref{eq:alphaq}), 
\begin{eqnarray}
\alpha(Q^2) - \alpha^{\pi N}(Q^2) \,=\, 
{{\left( \alpha - \alpha^{\pi N} \right)} 
\over {\left( 1 + Q^2 / \Lambda^2_{\alpha} \right)^2 }} \, ,
\label{eq:gpalphaparam}
\end{eqnarray}
where the $Q^2$ dependence is governed by the mass-scale
$\Lambda_\alpha$, again a free parameter. 
In Eq.~(\ref{eq:gpalphaparam}), the RCS value 
\begin{equation}
(\alpha - \alpha^{\pi N}) \,=\,  9.6 \,\times\,10^{-4}\,{\rm fm}^3 \, ,
\end{equation} 
is obtained from the phenomenological value of Eq.~(\ref{eq:alpharcs}) for
$\alpha$, and from the calculated $\pi N$ contribution~: 
$\alpha^{\pi N} =  2.5 \times 10^{-4} \, {\rm fm}^3$.
Using the dipole parametrization of~(\ref{eq:gpalphaparam}), 
one can then extract the free parameter $\Lambda_{\alpha}$ from 
a fit to VCS data at different $Q^2$ values.

\section{Results for $e p \to e p \gamma$ observables and discussion}
\label{results}

\begin{figure}[ht]
\epsfxsize=13cm
\centerline{\epsffile{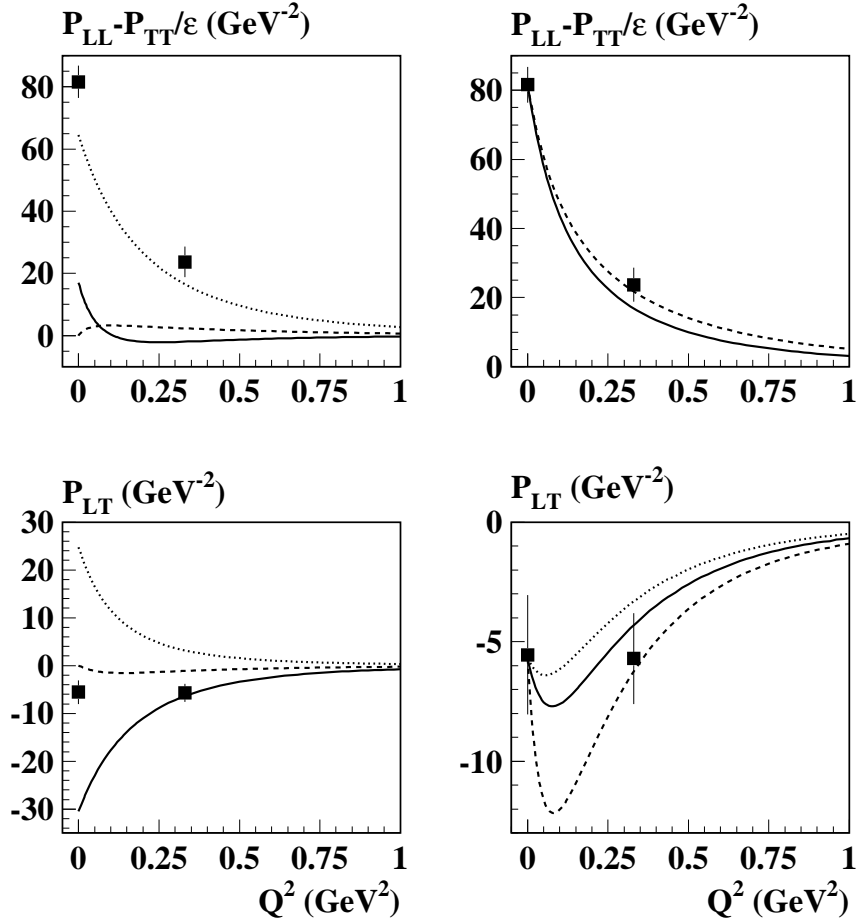}}
\caption[]{Results for the unpolarized structure functions
$P_{LL} - P_{TT}/\varepsilon$ (upper panels),
and $P_{LT}$ (lower panels), for $\varepsilon$ = 0.62.
In the upper left panel, a comparison is shown between 
the dispersive $\pi N$ contribution of the GP $\alpha$ (full curve), 
the dispersive $\pi N$ contribution of the spin-flip GPs (dashed curve), and 
the asymptotic contribution of $\alpha$ according to 
Eq.~(\ref{eq:gpalphaparam}) with $\Lambda_\alpha=1$ GeV (dotted curve).
The upper right panel displays the total result for 
$P_{LL} - P_{TT}/\varepsilon $ 
(sum of the three contributions on the upper left panel) 
for $\Lambda_\alpha=1$ GeV (full curve) and 
$\Lambda_\alpha=1.4$ GeV (dashed curve).
In the lower left panel, a comparison is shown between the dispersive 
$\pi N$ contribution of the GP $\beta$ (full curve), the contribution 
of the spin-flip GPs (dashed curve), and the asymptotic contribution 
of $\beta$ according to Eq.~(\ref{eq:gpbetaparam}) 
with $\Lambda_\beta=0.6$ GeV (dotted curve). 
The lower right panel displays the total result for $P_{LT}$,   
for $\Lambda_\beta=0.7$ GeV (dotted curve), 
$\Lambda_\beta=0.6$ GeV (full curve), 
and $\Lambda_\beta=0.4$ GeV (dashed curve).
The RCS data are from Ref.~\cite{Wiss00}, and the VCS data at $Q^2=0.33
\mbox{\, GeV}^2$ from Ref.~\cite{Roc00}.}
\label{fig:meaning_pol}
\end{figure}

\begin{figure}[ht]
\epsfxsize=11cm
\centerline{\epsffile{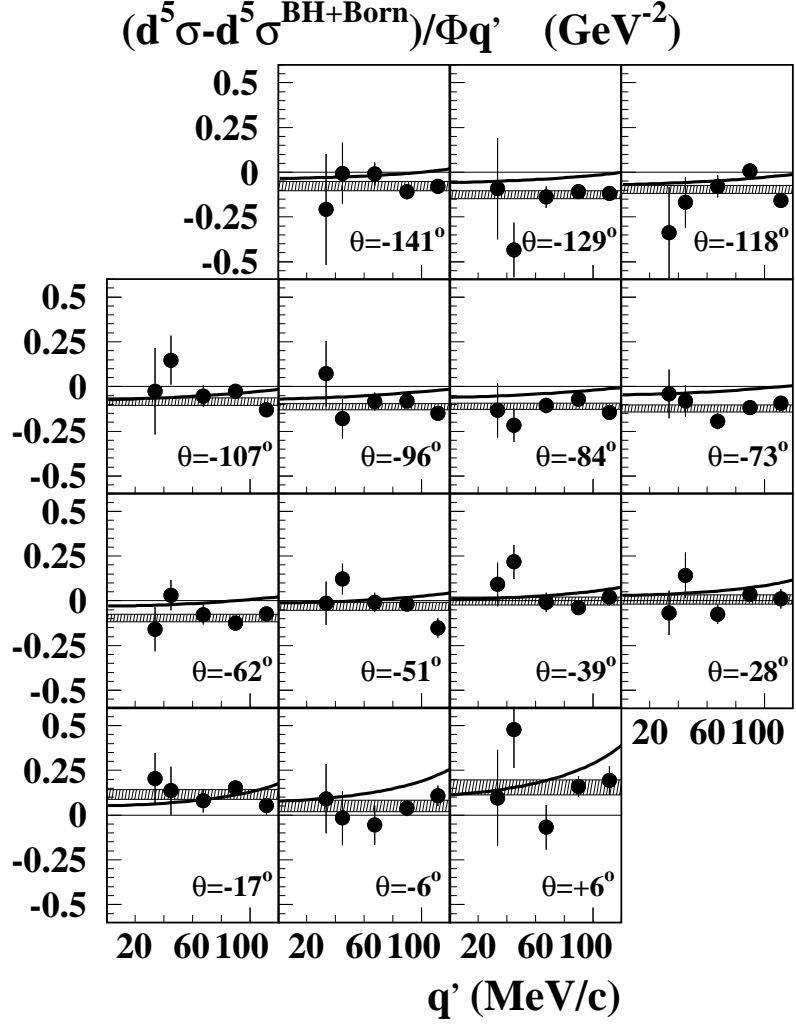}}
\vspace{-1cm}
\caption[]{ $(d^5\sigma-d^5\sigma^{{\rm BH+Born}})/\Phi \rmqp$ 
for the $e p \to e p \gamma$ reaction as function 
of the outgoing-photon energy $\rmqp$ in MAMI kinematics : $\varepsilon
=0.62$, $\rmq = 0.6$ GeV, and for different photon c.m. angles 
$\Theta_{\gamma \gamma}^{c.m.}$. 
The data and the shaded bands, representing the best fit to 
the data within the LEX formalism, are from Ref.~\cite{Roc00}. 
The solid curves are the DR results taking into 
account the full $\rmqp$ dependence of the non-Born contribution to the cross 
section. The asymptotic contributions are calculated according to  
Eqs.~(\ref{eq:gpbetaparam}, \ref{eq:gpalphaparam}), 
with $\Lambda_\beta=0.6$ GeV and $\Lambda_\alpha = 1$ GeV, respectively.}
\label{fig:vcs_mami_spectrum}
\end{figure}

\begin{figure}[ht]
\epsfxsize=10.5cm
\centerline{\epsffile{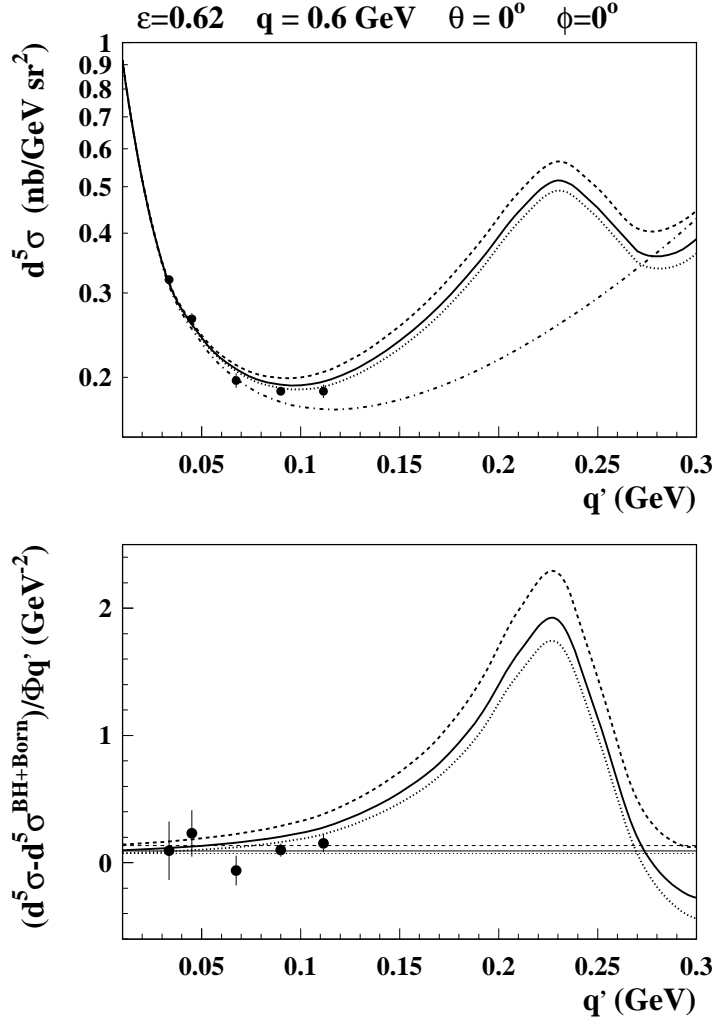}}
\caption[]{
Upper panel: The differential cross section for the reaction 
$e p \to e p \gamma$ as function of the outgoing-photon energy $\rmqp$ 
in MAMI kinematics : $\varepsilon =0.62$, $\rmq = 0.6$ GeV, 
and for $\Theta_{\gamma \gamma}^{c.m.} = 0^{\rm o}$, 
in plane ($\phi=0^{\rm o}$).
The BH + Born contribution is given by the dashed-dotted curve.
The total DR results are obtained with the
asymptotic parts of Eqs.~(\ref{eq:gpbetaparam}, \ref{eq:gpalphaparam}),
using a fixed value of $\Lambda_\alpha = 1 $ GeV and 
for the three values of $\Lambda_\beta$ as displayed in the lower
right plot of Fig.~\ref{fig:meaning_pol}, i.e. 
$\Lambda_\beta = 0.7$ GeV (dotted curve), 
$\Lambda_\beta = 0.6$ GeV (solid curve), and 
$\Lambda_\beta = 0.4$ GeV (dashed curve).
Lower panel: 
Results for $(d^5\sigma-d^5\sigma^{{\rm BH+Born}})/\Phi \rmqp$ 
as function of $\rmqp$ .
The DR calculation taking into account the full energy dependence 
of the non-Born contribution (thick curves) are compared to the 
corresponding results within the LEX formalism (thin horizontal curves).
The curves in the lower panel correspond to the same
values of $\Lambda_\alpha$ and $\Lambda_\beta$ as in the upper panel. 
The data are from Ref.~\cite{Roc00}.}
\label{fig:vcs_mami_thetap0_qp_dep}
\end{figure}

\begin{figure}[ht]
\epsfxsize=11cm
\centerline{\epsffile{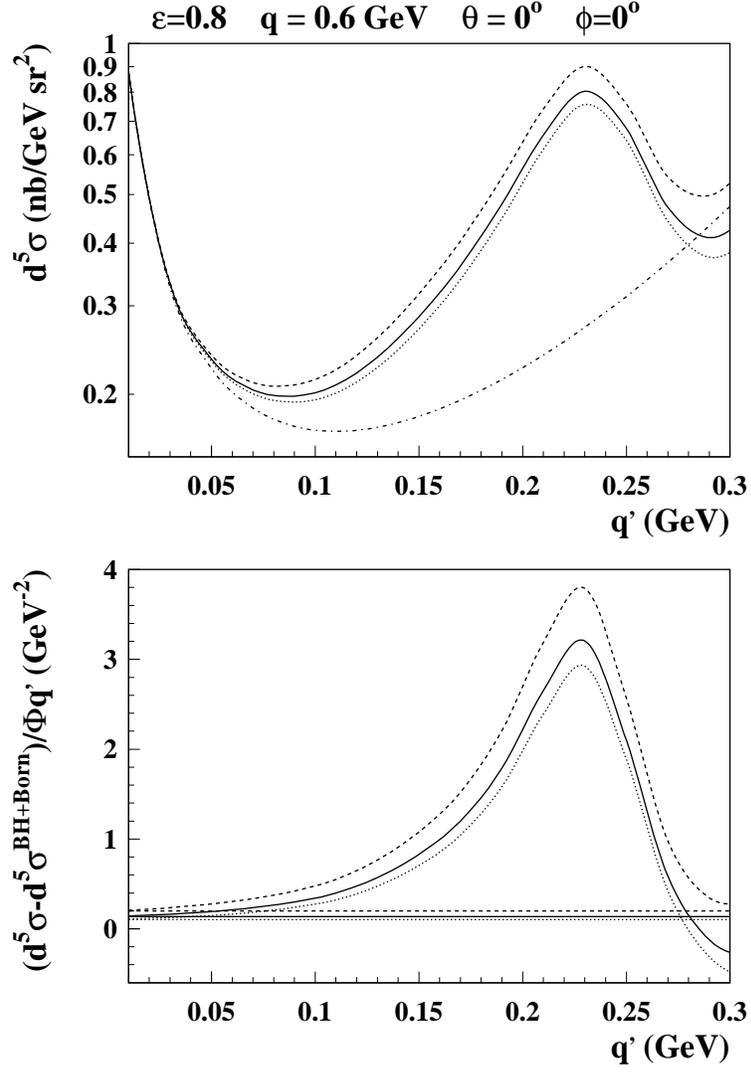}}
\caption[]{Same as Fig.~\ref{fig:vcs_mami_thetap0_qp_dep} but for
  $\varepsilon$ = 0.8.}
\label{fig:vcs_mami_thetap0_qp_dep_eps0p8}
\end{figure}

\begin{figure}[ht]
\epsfxsize=15cm
\centerline{\epsffile{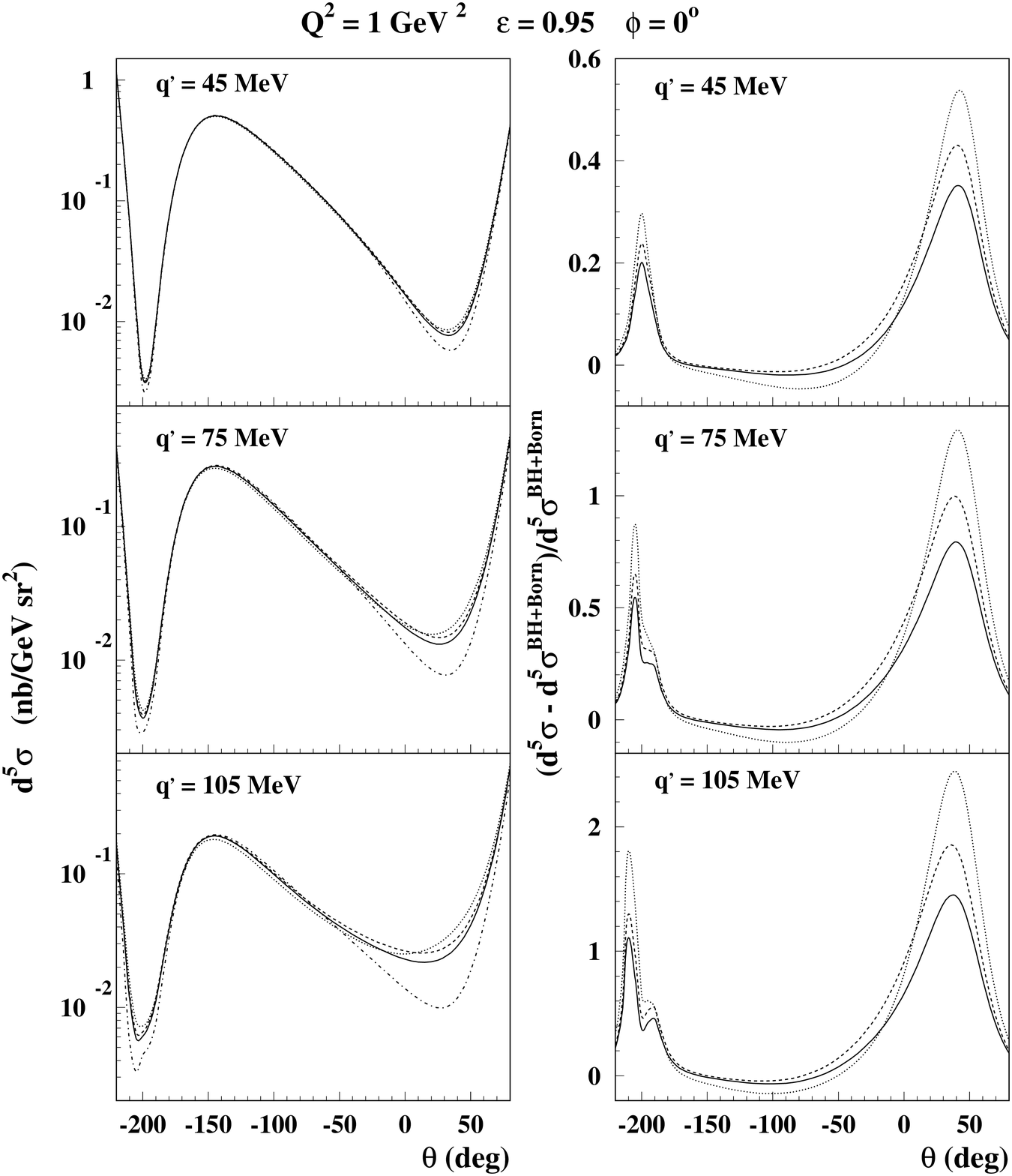}}
\caption[]{Left panels : 
The differential cross section for the reaction $e p \to e p \gamma$ 
as function of the photon scattering angle and at 
different values of the outgoing-photon 
energy $\rmqp$ in JLab kinematics. Right panels : ratio of cross
sections $(d \sigma - d \sigma^{BH + Born}) / d \sigma^{BH + Born}$. 
Dashed-dotted curves on the left panels : BH+Born contribution.
The DR results are displayed (on both left and right panels) 
with the asymptotic terms parametrized as in 
Eqs.~(\ref{eq:gpalphaparam}, \ref{eq:gpbetaparam}), using the
values : 
$\Lambda_\alpha$ = 1~GeV and $\Lambda_\beta$ = 0.6~GeV (full curves), 
$\Lambda_\alpha$ = 1~GeV and $\Lambda_\beta$ = 0.4~GeV (dashed curves),
$\Lambda_\alpha$ = 1.4~GeV and $\Lambda_\beta$ = 0.6~GeV (dotted curves).}
\label{fig:vcs_jlab_5fold_q2_1p0}
\end{figure}

\begin{figure}[ht]
\epsfxsize=10cm
\centerline{\epsffile{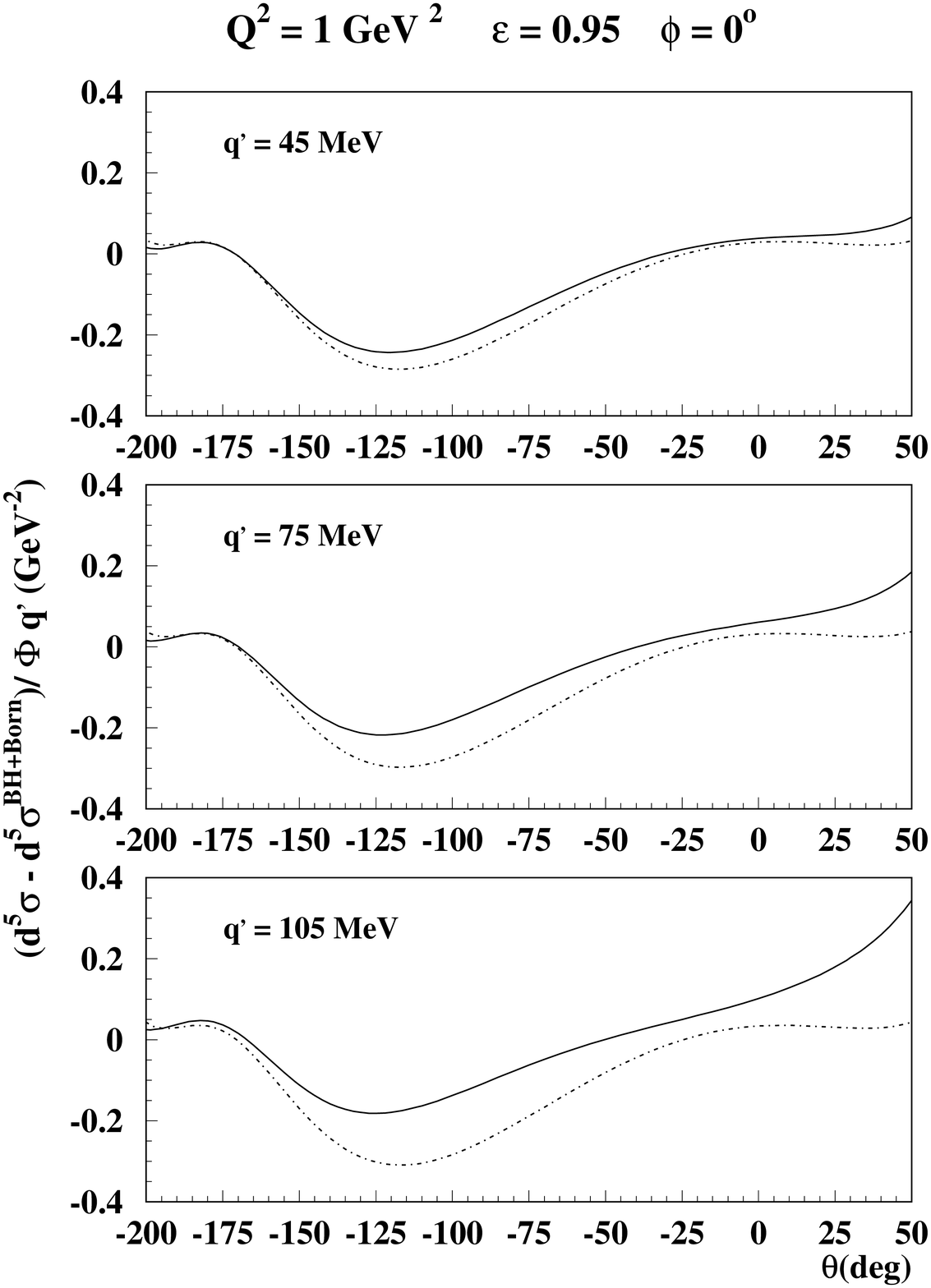}}
\vspace{-.5cm}
\caption[]{$(d^5\sigma-d^5\sigma^{{\rm BH+Born}})/\Phi \rmqp$ for the 
$e p \to e p \gamma$ reaction as  function
of the scattering angle and at different values of the outgoing-photon energy
$\rmqp$ in JLab kinematics.
The solid curves correspond to the DR calculation with the full energy
dependence of the non-Born contribution to the cross section. 
The dashed-dotted curves are the corresponding results obtained from
the LEX.
The asymptotic contributions have been calculated with the parametrizations 
in Eqs.~(\ref{eq:gpalphaparam}, \ref{eq:gpbetaparam}), using  
$\Lambda_\alpha=1.4$ GeV 
and $\Lambda_\beta=0.6$ GeV 
.}
\label{fig:vcs_jlab_5fold_diff}
\end{figure}

\begin{figure}[ht]
\epsfxsize=11cm
\centerline{\epsffile{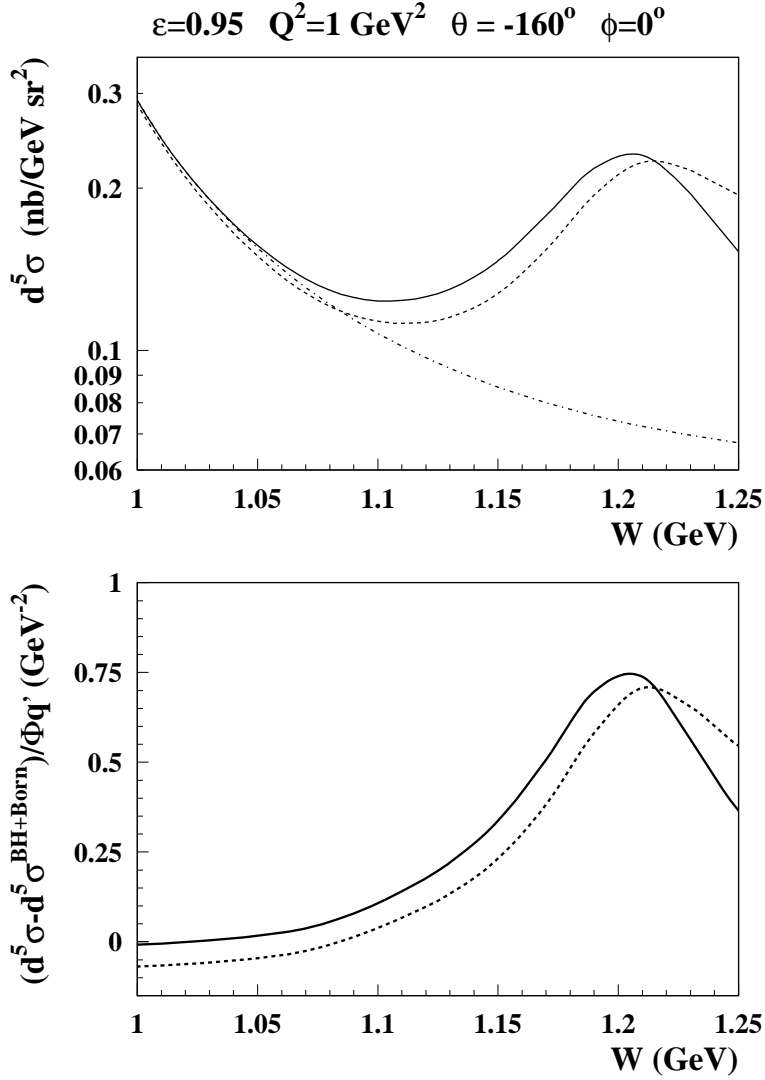}}
\caption[]{
Upper panel: The differential cross sections for the $e p \to e p
\gamma$ reaction as function of the c.m. energy W in JLab kinematics : 
$\varepsilon =0.95$, $Q^2=1$ GeV$^2$, and for fixed scattering angle 
$\Theta_{\gamma \gamma}^{c.m.} = -160 ^{\rm o}$, 
in plane ($\phi = 0^{\rm o}$). 
The BH + Born contribution is given by the dashed-dotted curve.
The total result including the non-Born contribution is shown 
for $\Lambda_\beta =0.6$ GeV and for the two values~: 
$\Lambda_\alpha =1 $ GeV (full curve), 
and $\Lambda_\alpha = 1.4 $ GeV (dashed curve).
Lower panel: results for $(d^5\sigma-d^5\sigma^{{\rm BH+Born}})/\Phi \rmqp$ 
as function of W. The curves in the lower panel correspond to the same
values of $\Lambda_\alpha$ and $\Lambda_\beta$ as in the upper panel.}
\label{fig:vcs_jlab_thetam160_spectrum}
\end{figure}

\begin{figure}[ht]
\epsfxsize=13cm
\centerline{\epsffile{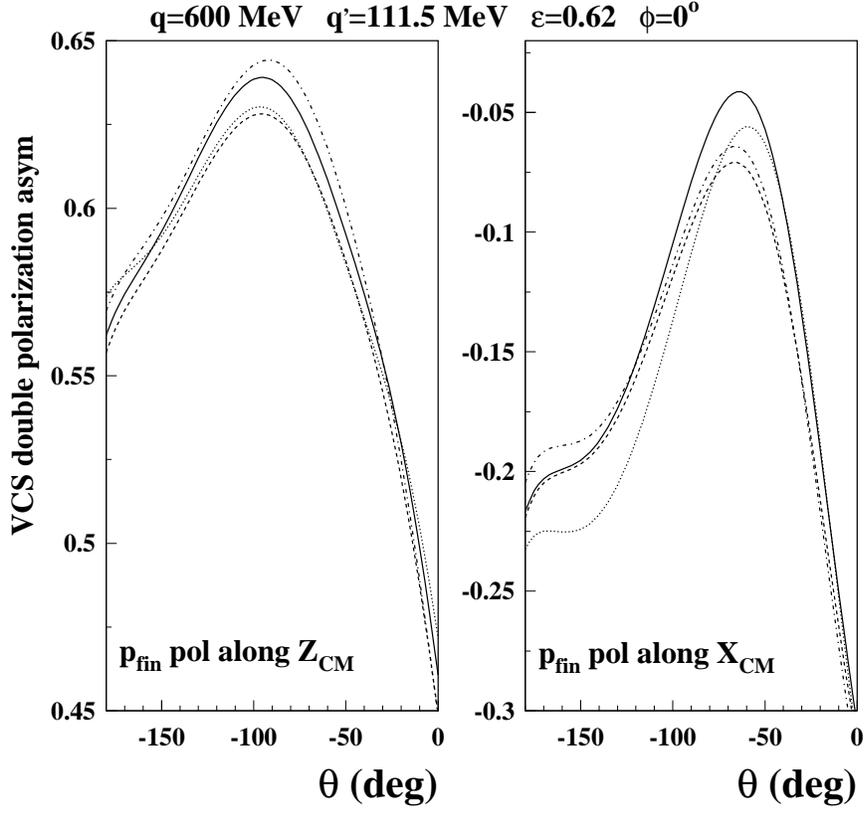}}
\caption[]{VCS double-polarization asymmetry (polarized electron,
recoil proton polarization along either the $z$- or $x$- directions 
in the c.m. frame) 
in MAMI kinematics as function of the photon scattering angle.
The dotted curves correspond to the BH+Born contribution.
The dispersion results for the total BH+Born+non-Born cross section
are shown for the values of the mass-scale 
$\Lambda_\alpha$ = 1~GeV, $\Lambda_\beta$ = 0.6~GeV (full curves)
and $\Lambda_\alpha$ = 1~GeV, $\Lambda_\beta$ = 0.4~GeV (dashed curves). 
To see the effect of the $\pi^0$-pole contribution, we also show the
results for the values $\Lambda_\alpha$ = 1~GeV, $\Lambda_\beta$
= 0.6~GeV, when turning off the $\pi^0$-pole 
contribution (dashed-dotted curves).}
\label{fig:vcs_mami_doublepol}
\end{figure}

\begin{figure}[ht]
\epsfxsize=8.5cm
\centerline{\epsffile{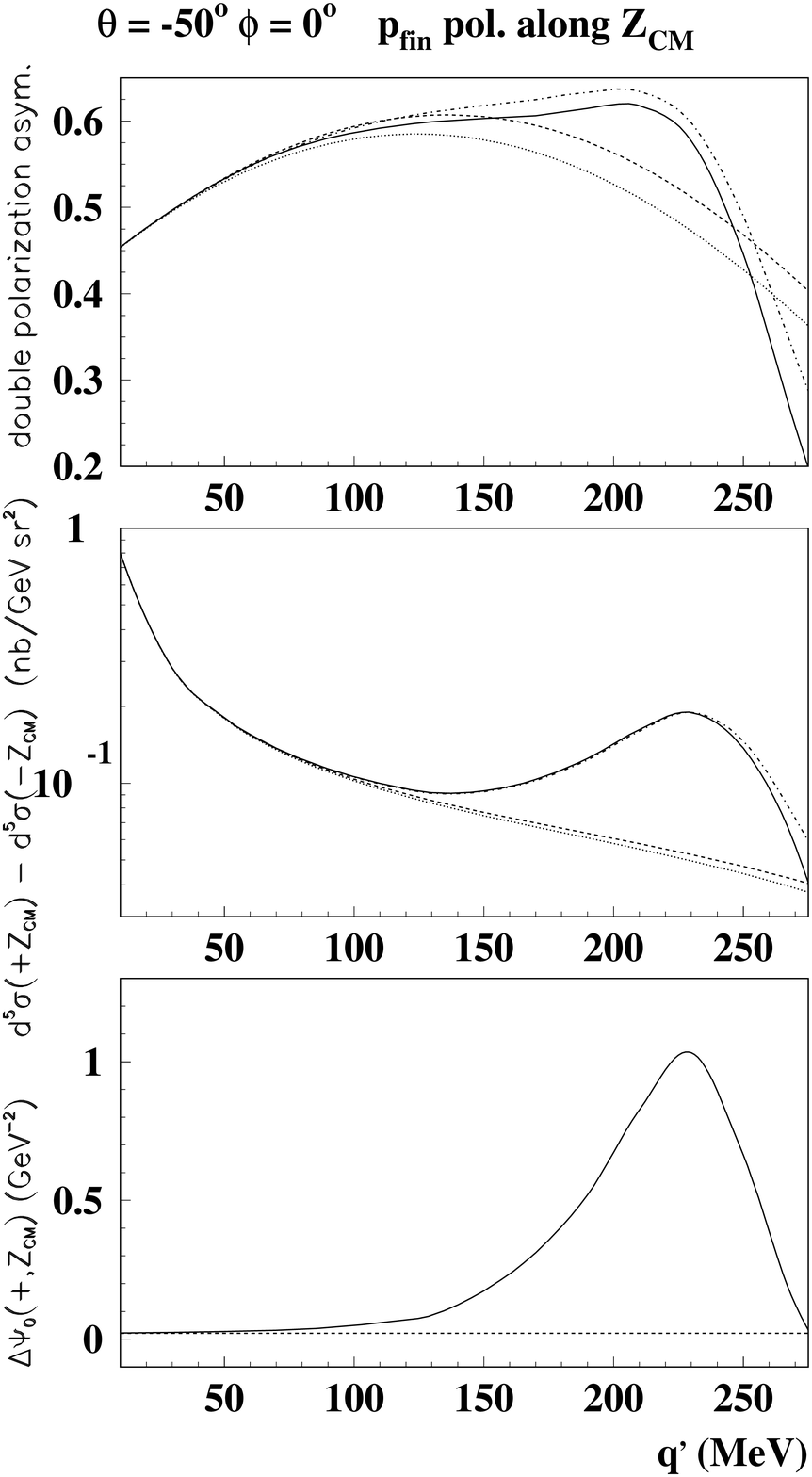}}
\caption[]{Upper panel : VCS double-polarization asymmetry (polarized electron,
recoil proton polarization along the $z$-direction in the c.m. frame) 
in MAMI kinematics (same value of $\rmq$ and $\varepsilon$ as in 
Fig.~\ref{fig:vcs_mami_doublepol}) as function of the 
outgoing-photon energy at a fixed 
photon scattering angle $\Theta_{\gamma \gamma}^{c.m.} = -50 ^{\rm o}$, 
in plane ($\phi = 0^{\rm o}$). 
The middle panel is the corresponding difference of polarized cross
sections and the lower panel is the non-Born contribution to the 
corresponding polarized squared matrix element (according to 
Eq.~(\ref{eq_3_50})).  
The dotted curves correspond to the BH+Born contribution.
The dispersion results for the total BH+Born+non-Born cross section
(full curves) are calculated using the values 
$\Lambda_\alpha$ = 1~GeV and $\Lambda_\beta$ = 0.6~GeV.
The dashed curves are the corresponding results obtained from the LEX.
To see the effect of the $\pi^0$-pole contribution, we also show the
results of the dispersion calculation, when turning off the $\pi^0$-pole 
contribution (dashed-dotted curves).} 
\label{fig:vcs_mami_doublepolz_spectrum_thm50}
\end{figure}

\begin{figure}[ht]
\epsfxsize=12cm
\centerline{\epsffile{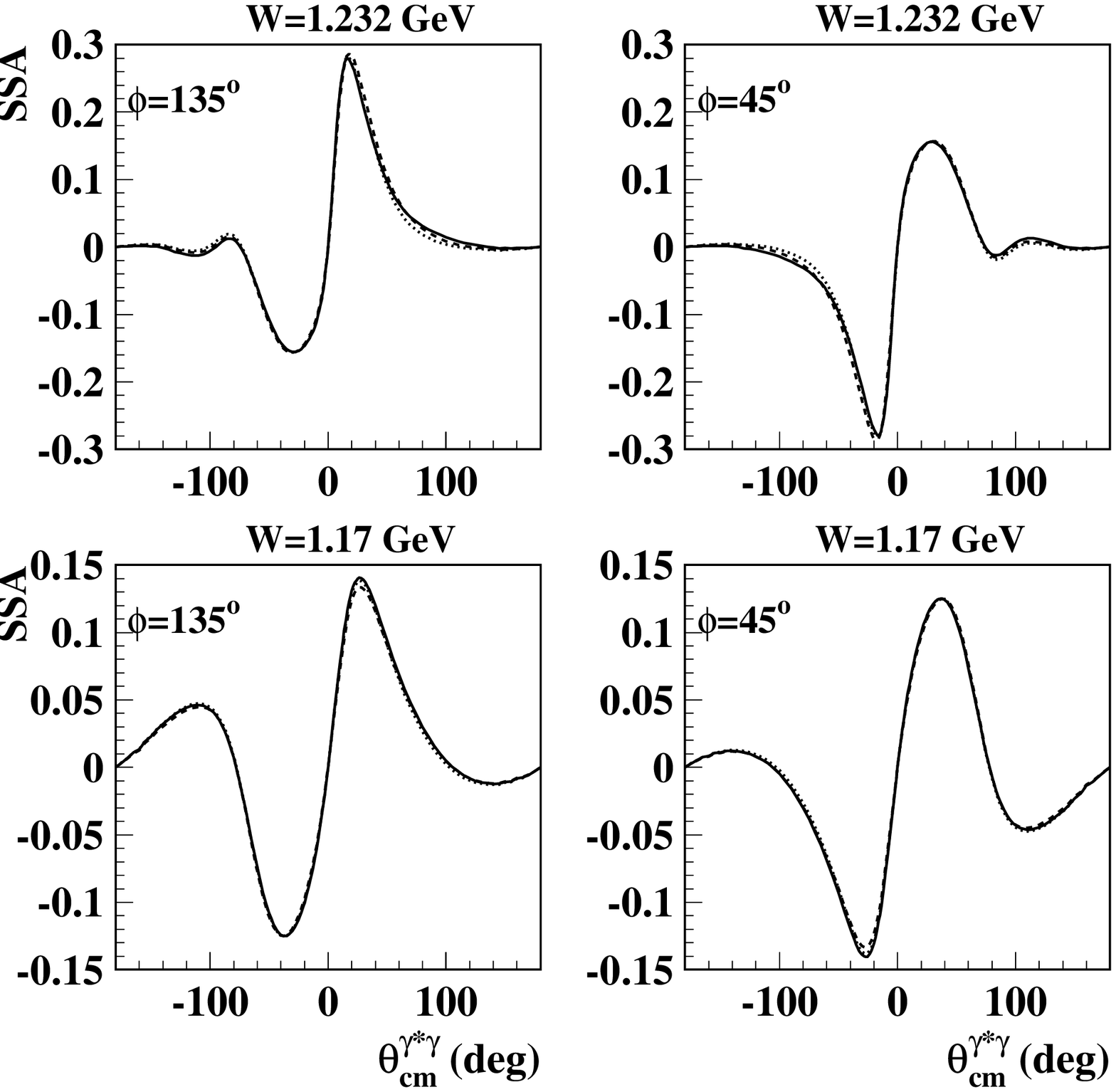}}
\caption[]{Electron single spin asymmetry (SSA) for VCS at $Q^2=0.12$ GeV$^2$, 
for two kinematics in the $\Delta(1232)$ region~:
W = 1.232 GeV, $\varepsilon =0.75$ (upper plots) and 
W = 1.17 GeV, $\varepsilon =0.81$ (lower plots).
In both cases the SSA is shown as function of the photon scattering angle 
for two out-of-plane angles $\Phi$, as accessible at MIT-Bates
\cite{Kalos97}. 
The full dispersion results are shown for the values~:
$\Lambda_\alpha$ = 1~GeV, $\Lambda_\beta$ = 0.6~GeV (full curves), 
$\Lambda_\alpha$ = 1~GeV, $\Lambda_\beta$ = 0.4~GeV (dashed curves),
and $\Lambda_\alpha$ = 1.4~GeV, $\Lambda_\beta$ = 0.6~GeV (dotted curves).}
\label{fig:vcs_ssa}
\end{figure}

Having set up the dispersion formalism for VCS, we now show the
predictions for the different $e p \to e p \gamma$ observables 
for energies up to the $\Delta(1232)$-resonance region. 
The aim of the experiments is to extract the 6 GPs of 
Eqs.~(\ref{eq:defgpunpol},\ref{eq:defgppol}) 
from both unpolarized and polarized
observables. We will compare the DR results, which take account of 
the full dependence of the $e p \to e p \gamma$ observables 
on the energy ($\rmqp$) of the emitted photon, 
with a low-energy expansion (LEX) in $\rmqp$. 
In the LEX of observables, only the first three terms of a Taylor
expansion in $\rmqp$ are taken into account. 
\newline
\indent
In such an expansion in $\rmqp$, the experimentally extracted 
VCS unpolarized squared amplitude 
$\calm^{\rm exp}$ takes the form \cite{Gui95}~:
\begin{eqnarray}
\calm^{\rm exp}=\frac{\calm^{\rm exp}_{-2}}{\rmqp^2}
+\frac{\calm^{\rm exp}_{-1}}{\rmqp}
+\calm^{\rm exp}_0+O(\rmqp) \, .
\label{eq:unpolsqramp}
\end{eqnarray}
Due to the low-energy theorem (LET), the threshold coefficients 
$\calm^{\rm exp}_{-2}$ and $\calm^{\rm exp}_{-1}$ are known (see
Ref.~\cite{Gui95} for details). 
The information on the GPs is contained in \( \calm^{\rm exp}_0\), 
which contains a part originating from the (BH+Born) amplitude 
and another one which is a linear combination of the GPs, with
coefficients determined by the kinematics. 
It was found in Ref.~\cite{Gui95} that 
the unpolarized observable $\calm^{\rm exp}_0$ 
can be expressed in terms of 3 structure functions
$P_{LL}(\rmq)$, $P_{TT}(\rmq)$, and $P_{LT}(\rmq)$ by~:
\begin{eqnarray}
\calm^{\rm exp}_0 -  \calm^{\rm BH+Born}_0 
\,=\, 2 K_2  \left\{ 
v_1 \left[ {\varepsilon  P_{LL}(\rmq)  -  P_{TT}}(\rmq)\right] 
\,+\, \left(v_2-\frac{\qt0}{\rmq}v_3\right)\sqrt {2\varepsilon \left( 
{1+\varepsilon }\right)} P_{LT}(\rmq) \right\},
\label{eq:vcsunpol}
\end{eqnarray}
where $K_2$ is a kinematical factor, $\varepsilon$ is the virtual
photon polarization (in the standard notation used in electron
scattering), and $v_1, v_2, v_3$ are kinematical
quantities depending on $\varepsilon$ and $\rmq$  
as well as on the polar and azimuthal angles
($\Theta_{\gamma \gamma}^{c.m.}$ and $\Phi$, respectively) of  the
produced real photon (for details see Ref.~\cite{GuiVdh}). 
\newline
\indent
After some algebra, one finds that the 3 unpolarized 
observables of Eq.~(\ref{eq:vcsunpol}) can be expressed in terms of the 6
GPs as \cite{Gui95,GuiVdh}~:
\begin{eqnarray}
&&P_{LL} \;=\; - 2\sqrt{6} \, M \, G_E \, P^{\left( {01,01} \right)0} \;, 
\label{eq:unpolobsgp1} \\
&&P_{TT} \;=\; - 3 \, G_M \, \frac{\rmq^2}{\qt0} 
\left( P^{(11,11)1}\,-\,\sqrt{2} \, \qt0 \, P^{(01,12)1} \right) \;, 
\label{eq:unpolobsgp2} \\
&&P_{LT} \;=\; \sqrt{\frac{3}{2}} \, \frac{M \, \rmq}{Q} \, G_E \, 
P^{(11,11)0} \,+\,\frac{3}{2} \, \frac{Q \, \rmq}{\qt0}\,G_M\, P^{(01,01)1} \;,
\label{eq:unpolobsgp3}
\end{eqnarray}
where $G_E$ and $G_M$ stand for the electric and magnetic nucleon form factors
$G_E(Q^2)$ and $G_M(Q^2)$, respectively. 
\newline
\indent
In Fig.~\ref{fig:meaning_pol}, we show the calculations of 
$P_{LL} - P_{TT}/\varepsilon$ and $P_{LT}$, which have been measured
at MAMI at $Q^2$ = 0.33 GeV$^2$ \cite{Roc00}. 
The virtual photon polarization $\varepsilon$ is fixed to the
experimental value ($\varepsilon$ = 0.62), 
and for the electromagnetic form factors
in Eqs.~(\ref{eq:unpolobsgp1} - \ref{eq:unpolobsgp3}) 
we use the H\"ohler parametrization \cite{Hoe76} 
as in the analysis of the MAMI experiment \cite{Roc00}. 
\newline
\indent
In the lower panel of Fig.~\ref{fig:meaning_pol}, the  
$Q^2$-dependence of the VCS response function $P_{LT}$ is displayed, 
which reduces to the magnetic polarizability $\beta$  
at the real photon point ($Q^2$ = 0). 
At finite $Q^2$, it contains both the scalar GP $\beta(Q^2)$ and  
the spin GP $P^{(0 1, 0 1)1}$, as seen from Eq.~(\ref{eq:unpolobsgp3}). 
It is obvious from Fig.~\ref{fig:meaning_pol} that 
the structure function $P_{LT}$ results from a large
dispersive $\pi N$ contribution and a large asymptotic contribution
(to $\beta$) with opposite sign, leading to a relatively small net result. 
At the real photon point, the small value of $\beta$ is indeed known
to result from the near cancellation of a large paramagnetic
contribution from the $\Delta$-resonance, and a large diamagnetic
contribution (asymptotic part). The latter is shown in 
Fig.~\ref{fig:meaning_pol} with the parametrization of 
Eq.~(\ref{eq:gpbetaparam}) for the values 
$\Lambda_\beta$ = 0.4 and $\Lambda_\beta$ = 0.6 GeV, 
which were also displayed in Fig.~\ref{fig:f1_asymp}.
Due to the large cancellation in $P_{LT}$, its $Q^2$ dependence is
a very sensitive observable to study the interplay of the two mechanisms. 
In particular, one expects a faster fall-off of the asymptotic
contribution with $Q^2$ in comparison to the 
$\pi N$ dispersive contribution, as
discussed before. This is already highlighted by the measured value 
of $P_{LT}$ at $Q^2$ = 0.33 GeV$^2$ \cite{Roc00}, 
which is comparable to the value of $P_{LT}$ at $Q^2$ = 0 \cite{Wiss00}. 
As seen from Fig.~\ref{fig:meaning_pol}, 
this points to an interesting structure in the  
$Q^2$ region around 0.1~GeV$^2$,
where forthcoming data are expected from an experiment at MIT-Bates
\cite{Bates}.  
\newline
\indent
In the upper panel of Fig.~\ref{fig:meaning_pol}, we show the 
$Q^2$-dependence of the VCS response function 
$P_{LL}$~-~$P_{TT}/\varepsilon$, which reduces 
at the real photon point ($Q^2$ = 0) to the electric polarizability 
$\alpha$. At non-zero $Q^2$, $P_{LL}$ 
is directly proportional to the scalar GP $\alpha(Q^2)$, as 
seen from Eq.~(\ref{eq:unpolobsgp1}), and the response function  
$P_{TT}$ of Eq.~(\ref{eq:unpolobsgp2}) contains only spin GPs.
As is shown by Fig.~\ref{fig:meaning_pol}, the $\pi N$
dispersive contribution to $\alpha$ and to the spin GPs 
are smaller than the asymptotic contribution to $\alpha$, which is
evaluated for $\Lambda_\alpha$ = 1 GeV. 
At $Q^2$ = 0, the $\pi N$ dispersive and asymptotic contributions to
$\alpha$ have the same sign, in contrast to $\beta$ 
where both contributions have opposite sign and largely cancel each
other in their sum. 
\newline
\indent
The response functions $P_{LT}$ and $P_{LL}$~-~$P_{TT}/\varepsilon$ 
were extracted in \cite{Roc00} by performing a LEX to VCS data,
according to Eq.~(\ref{eq:vcsunpol}). To test the validity of such a
LEX, we show in Fig.~\ref{fig:vcs_mami_spectrum} the DR predictions for
the full energy dependence of the non-Born part of the $e p \to e p
\gamma$ cross section in the kinematics of the MAMI experiment
\cite{Roc00}. This energy dependence is compared with the LEX, which
predicts a linear dependence in $\rmqp$ 
for the difference between the experimentally
measured cross section and its BH + Born contribution. 
The result of a best fit to the data in the framework of the LEX is
indicated by the horizontal bands in Fig.~\ref{fig:vcs_mami_spectrum} 
for the quantity $(d^5\sigma-d^5\sigma^{{\rm BH+Born}})/\Phi
\rmqp$, where $\Phi$ is a phase space factor defined in \cite{Gui95}. 
The fivefold differential cross section $d^5 \sigma$ is differential
with respect to the electron {\it lab} energy and {\it lab} angles and
the proton {\it c.m.} angles, and stands in all of the following for 
$d \sigma\,/\, dk^e_{lab} \, d \Omega^e_{lab} \,d \Omega^p_{c.m.}$. 
It is seen from Fig.~\ref{fig:vcs_mami_spectrum} that the DR results 
predict only a modest additional energy dependence 
up to $\rmqp \simeq$ 0.1 GeV/c and for most of the photon angles involved, 
and therefore seems to support the LEX analysis of \cite{Roc00}. Only for 
forward angles, $\Theta_{\gamma \gamma}^{c.m.} \approx 0$, 
which is the angular range from which the value of $P_{LT}$ is extracted, 
the DR calculation predicts a stronger energy dependence in the range 
up to $\rmqp \simeq$ 0.1 GeV/c, as compared to the LEX. 
It will be interesting to perform a best fit of the MAMI data using
the DR formalism, extract the two fit parameters 
$\alpha(Q^2)$ and $\beta(Q^2)$, and consequently the values of 
$P_{LL} - P_{TT}/\varepsilon$ and $P_{LT}$ respectively. 
Such a best fit using the DR formalism is planned in a future investigation.  
\newline
\indent
Increasing the energy, we show 
in Fig.~\ref{fig:vcs_mami_thetap0_qp_dep} the DR predictions
for photon energies in the $\Delta(1232)$-resonance region. 
It is seen that the $e p \to e p \gamma$ cross section 
rises strongly when crossing the
pion threshold. In the dispersion relation formalism, which is based
on unitarity and analyticity, the rise of the cross section with $\rmqp$
below pion threshold, due to virtual $\pi N$ intermediate states,
is connected to the strong rise of the cross section with $\rmqp$ 
when a real $\pi N$ intermediate state can be produced. 
It is furthermore seen from Fig.~\ref{fig:vcs_mami_thetap0_qp_dep}
(lower panel) that the region
between pion threshold and the $\Delta$-resonance peak displays 
an enhanced sensitivity to the GPs through the interference with the
rising Compton amplitude due to $\Delta$-resonance excitation. 
For example, at $\rmqp \simeq$ 0.2 GeV/c, the 
predictions for $P_{LT}$ in the lower right panel of
Fig.~\ref{fig:meaning_pol} for $\Lambda_\beta$ = 0.4
GeV and $\Lambda_\beta$ = 0.6 GeV give a difference of about 20 \%
in the non-Born squared amplitude. In contrast, the LEX prescription 
results in a relative effect for the same two values of $P_{LT}$ 
of about 10\% or less. 
This is similar to the situation in RCS, where the region between pion
threshold and the $\Delta$-resonance position 
also provides an enhanced sensitivity to the
polarizabilities and is used to extract those polarizabilities from
data \cite{lvov97,Dre99} using a DR formalism. 
Therefore, the energy region between pion threshold and the
$\Delta$-resonance seems promising to measure VCS observables 
with an increased sensitivity to the GPs. The presented DR
formalism can be used as a tool to extract the GPs from such data. 
\newline
\indent
When increasing the value of $\varepsilon$, the Born and non-Born parts of the
$e p \to e p \gamma$ cross section increase relative to the BH
contribution, due to the increasing virtual photon flux factor \cite{GuiVdh}.
This is seen by comparing the non-Born cross section 
in Fig.~\ref{fig:vcs_mami_thetap0_qp_dep} 
(corresponding to $\varepsilon$ = 0.62), with the result 
for $\varepsilon$ = 0.8 at the same value of $\rmq$ and 
$\Theta_{\gamma \gamma}^{c.m.}$, as is shown in 
Fig.~\ref{fig:vcs_mami_thetap0_qp_dep_eps0p8}.
Besides giving rise to higher non-Born cross sections, an experiment
at a higher value of $\varepsilon$ (keeping $\rmq$ fixed) also allows to
disentangle the unpolarized structure functions $P_{LL}(\rmq)$ and 
$P_{TT}(\rmq)$ in Eq.~(\ref{eq:vcsunpol}). 
This will provide a nice opportunity for the MAMI-C facility 
where such a higher $\varepsilon$ value 
(as compared to the value $\varepsilon$ = 0.62 of
the first VCS experiment of Ref.~\cite{Roc00}) will be reachable for the
same value of $\rmq$. 
\newline
\indent
Recently, VCS data have also been taken at JLab \cite{JLab} both below pion
threshold at $Q^2$ = 1 GeV$^2$ \cite{natalie}, 
and at $Q^2$ = 1.9 GeV$^2$ \cite{helene}, as well as in the resonance
region around $Q^2$ = 1 GeV$^2$ \cite{geraud}.
\newline
\indent
The extraction of GPs from VCS data at these higher values of $Q^2$,
requires an accurate knowledge of the nucleon 
electromagnetic form factors (FFs) in this region. 
For the proton magnetic FF $G_M^p(Q^2)$, we
use the Bosted parametrization \cite{Bost95}, which has an accuracy of
around 3 \% in the $Q^2$ region of 1 - 2 GeV$^2$. The ratio of the 
proton electric FF $G_E^p$ to the magnetic FF $G_M^p$ 
was recently measured with high accuracy in a polarization
experiment at JLab in the $Q^2$ range 0.4 - 3.5 GeV$^2$ \cite{Jon00}. 
It was found in \cite{Jon00} that $G_E^p$
drops considerably faster with $Q^2$ than $G_M^p$. 
In the region of interest here, i.e. $Q^2$ in the 
1 - 2~GeV$^2$ range, the JLab data of Ref.~\cite{Jon00} are well described by
the fit \cite{natalie}~:
\begin{eqnarray}
{{\mu_p \, G_E^p(Q^2)} \over {G_M^p(Q^2)}} 
\; \approx \; 1 \;-\; 0.13 \; (Q^2)^2 
\;+\; 0.028 \; (Q^2)^3 \; ,
\label{eq:jlabge}
\end{eqnarray}
where $\mu_p$ is the proton magnetic moment. 
In the following VCS calculations at $Q^2$ = 1 GeV$^2$, 
we use the parametrization of Eq.~(\ref{eq:jlabge}) to specify $G_E^p$
(with the Bosted parametrization for $G_M^p$). 
\newline
\indent
In Fig.~\ref{fig:vcs_jlab_5fold_q2_1p0}, we show the DR predictions for
the $e p \to e p \gamma$ reaction at $Q^2$ = 1 GeV$^2$, for three values
of the outgoing photon energy, below pion threshold. In these
kinematics, data have been taken at JLab and, at the time of writing 
this paper, 
preliminary results on VCS cross sections and GPs have been reported
in \cite{natalie}. For those kinematics, we show in 
Fig.~\ref{fig:vcs_jlab_5fold_q2_1p0} the differential cross sections
as well as the non-Born effect relative to the BH + Born
cross section. It is seen from 
Fig.~\ref{fig:vcs_jlab_5fold_q2_1p0} that the sensitivity to the GPs
is largest where the BH + Born cross section becomes small, in
particular in the angular region between 0$^o$ and 50$^o$. 
In Fig.~\ref{fig:vcs_jlab_5fold_q2_1p0}, we show the non-Born effect
for different values of the polarizabilities. For $P_{LL}$, 
the calculation for the $\pi N$ dispersive contribution 
at $Q^2 = 1$~GeV$^2$ gives~:  
\begin{equation}
P_{LL}^{\pi N} ( 1 \; {\mathrm GeV}^2) \,=\, -\,0.3 
\;\; {\mathrm GeV}^{-2} \, , 
\label{eq:pll1pin}
\end{equation}
leading to the total results for $P_{LL}$ within the DR formalism~: 
\begin{eqnarray}
\label{eq:pll1a}
P_{LL} ( 1 \; {\mathrm GeV}^2) \,&=&\, +\, 2.3 \;\; {\mathrm GeV}^{-2} ,
\hspace{.5cm} {\mathrm for} \; 
\left( \Lambda_{\alpha} \,=\, 1 \; {\mathrm GeV} \right)\, , \\ 
\label{eq:pll1b}
P_{LL} ( 1 \; {\mathrm GeV}^2) \,&=&\, +\, 4.2 \;\; {\mathrm GeV}^{-2} , 
\hspace{.5cm} {\mathrm for} \; 
\left( \Lambda_{\alpha} \,=\, 1.4 \; {\mathrm GeV} \right) \, .  
\end{eqnarray}
For $P_{LT}$, the calculation for the $\pi N$ dispersive contribution 
at $Q^2 = 1$~GeV$^2$ gives~:
\begin{equation}  
P_{LT}^{\pi N} ( 1 \; {\mathrm GeV}^2)\,=\, -\, 0.9 \;\; {\mathrm GeV}^{-2}\, ,
\label{eq:plt1pin}
\end{equation}
leading to the total results for $P_{LT}$ within the DR formalism~:
\begin{eqnarray}
P_{LT} ( 1 \; {\mathrm GeV}^2) \,&=&\, -\, 0.6 \;\; {\mathrm GeV}^{-2} ,
\hspace{.5cm} {\mathrm for} \; 
\left( \Lambda_{\beta} \,=\, 0.6 \; {\mathrm GeV} \right)\, , 
\label{eq:plt1b}\\ 
P_{LT} ( 1 \; {\mathrm GeV}^2) \,&=&\, -\, 0.9 \;\; {\mathrm GeV}^{-2} , 
\hspace{.5cm} {\mathrm for} \; 
\left( \Lambda_{\beta} \,=\, 0.4 \; {\mathrm GeV} \right) \, .  
\end{eqnarray}
It will be interesting to compare the sensitivity of the cross
sections to these values of
the GPs, as displayed in Fig.~\ref{fig:vcs_jlab_5fold_q2_1p0}, to the JLab
data which have been taken in this region \cite{natalie}. 
The deviation of the experimental values from the dispersive $\pi N$
values of~(\ref{eq:pll1pin}) for $P_{LL}$ and of~(\ref{eq:plt1pin}) for
$P_{LT}$ will provide us with interesting information, allowing to  
test our understanding of
the electric and magnetic polarizability at this large virtuality of
$Q^2$ = 1 GeV$^2$. 
\newline
\indent
For the same kinematics as in Fig.~\ref{fig:vcs_jlab_5fold_q2_1p0}, we
compare in Fig.~\ref{fig:vcs_jlab_5fold_diff}  
the DR calculation for the non-Born cross section with the
corresponding result using the LEX.
It is seen that the deviation of the DR results 
from the LEX becomes already noticeable for
$\rmqp$ = 75 MeV, over most of the photon angular range. Therefore, the
DR analysis seems already to be needed at those lower values of $\rmqp$
to extract GPs from the JLab data. 
\newline
\indent
In Fig.~\ref{fig:vcs_jlab_thetam160_spectrum}, we increase the energy
through the $\Delta(1232)$-resonance region, 
and show the results for the $e p \to e p \gamma$ reaction at $Q^2 =
1$~GeV$^2$ and at a backward angle. 
We display the calculations of the cross section
and of the non-Born effect for the values in~(\ref{eq:pll1a}) and 
(\ref{eq:pll1b}) for $P_{LL}$, and for the value 
in~(\ref{eq:plt1b}) for $P_{LT}$. One sees a sizeable sensitivity 
to $P_{LL}$ in this backward angle cross section, 
and it therefore seems very promising to extract information
on the electric polarizability from such anticipated data.  
\newline
\indent
Until now, we discussed only unpolarized VCS observables. 
An unpolarized VCS experiment gives access to only 3 combinations of
the 6 GPs, as given by Eqs.~(\ref{eq:unpolobsgp1}-\ref{eq:unpolobsgp3}). 
It was shown in Ref.~\cite{Vdh97a} that VCS double
polarization observables with polarized lepton and polarized target
(or recoil) nucleon,
will allow us to measure three more combinations of GPs. Therefore a
measurement of unpolarized VCS observables (at different values of 
$\varepsilon$) and of 3 double-polarization observables 
will give the possibility to disentangle all 6 GPs. 
The VCS double polarization observables, which are denoted by 
\(\Delta{\cal M} (h, i) \) for an electron of helicity $h$, are defined as 
the difference of the squared amplitudes for recoil (or target) proton
spin orientation in the direction and opposite to 
the axis $i$ ($i = x, y, z$) (see Ref.~\cite{Vdh97a} for details). 
In a LEX, this polarized squared amplitude yields~:   
\begin{eqnarray}
\dcalm^{\rm exp}=\frac{\dcalm^{\rm exp}_{-2}}{\rmqp^2}
+\frac{\dcalm^{\rm exp}_{-1}}{\rmqp}+\dcalm^{\rm exp}_0+O(\rmqp) \,. \;\;
\label{eq_3_50}
\end{eqnarray}
Analogous to the unpolarized squared amplitude (\ref{eq:unpolsqramp}), 
the threshold coefficients $\dcalm^{\rm exp}_{-2}$, 
$\dcalm^{\rm exp}_{-1}$ are known due
to the LET. 
It was found in Refs.~\cite{Vdh97a,GuiVdh} that 
the polarized squared amplitude $\Delta \calm^{\rm exp}_0$ 
can be expressed in terms of three new structure functions
$P_{LT}^z(\rmq)$, $P_{LT}^{'z}(\rmq)$, and $P_{LT}^{'\perp}(\rmq)$. 
These new structure functions are related to the spin GPs according to
\cite{Vdh97a,GuiVdh}~: 
\begin{eqnarray}
&&P_{LT}^z \;=\; \frac{3 \, Q \,\rmq}{2 \, \qt0} \, G_M \, P^{(01,01)1}
\,-\, \frac{3\, M \, \rmq}{Q} \, G_E \, P^{(11,11)1} \;, \\
&&P_{LT}^{'z} \;=\; -\frac{3}{2}\, Q \, G_M \, P^{(01,01)1}
\,+\, \frac{3 \, M \, \rmq^2}{Q \, \qt0} \, G_E \, P^{(11,11)1} \;, \\
&&P_{LT}^{'\perp} \;=\; \frac{3 \,\rmq \, Q}{2 \, \qt0} \,G_M \, 
\left(P^{(01,01)1} \,-\, \sqrt{\frac{3}{2}} \, \qt0 \, P^{(11,02)1}\right) \;.
\label{eq:polobsgp}
\end{eqnarray}
\indent
While $P_{LT}^z$ and $P_{LT}^{'z}$ can be accessed by in-plane
kinematics ($\Phi = 0^o$), the measurement of $P_{LT}^{'\perp}$
requires an out-of-plane experiment. 
\newline
\indent
In Fig.~\ref{fig:vcs_mami_doublepol}, 
we show the dispersion results for the double polarization
observables, with polarized electron and by measuring the recoil
proton polarization either along the virtual photon direction
($z$-direction) or parallel to the reaction plane and perpendicular to
the virtual photon ($x$-direction). The double polarization
asymmetries are quite large (due to a non-vanishing
asymmetry for the BH + Born mechanism), but our DR calculations show
only small relative effects due to the spin GPs below pion threshold. 
Although these observables are tough to measure, 
a first test experiment is already planned at MAMI \cite{mamipol}.
\newline
\indent
When measuring double polarization observables above 
pion threshold, one can enhance the sensitivity to
the GPs, as we remarked before for the unpolarized observables. 
In Fig.~\ref{fig:vcs_mami_doublepolz_spectrum_thm50}, we show as an
example the double polarization asymmetry in MAMI kinematics 
for polarized beam and recoil proton polarization 
measured along the virtual photon direction as function of 
the outgoing photon energy through the $\Delta(1232)$ region. 
The $\Delta(1232)$ resonance excitation clearly shows up as a 
deviation from the LEX result above about $\rmqp$ = 100 MeV.
\newline
\indent
As discussed before, VCS polarization experiments below pion threshold, 
require the measurement of double polarization observables 
to get non-zero values, because the VCS amplitude is purely real below
pion threshold. 
However, when crossing the pion threshold, the VCS amplitude acquires
an imaginary part due to the coupling to the $\pi N$
channel. Therefore, single polarization observables become non-zero
above pion threshold. A particularly relevant observable is the
electron single spin asymmetry (SSA), which is obtained by flipping the
electron beam helicity \cite{GuiVdh}. 
For VCS, this observable is mainly due to the interference of the 
real BH + Born amplitude with the imaginary part of the VCS amplitude. 
In Fig.~\ref{fig:vcs_ssa}, the SSA is shown for two kinematics in the 
$\Delta(1232)$ region. As the SSA vanishes in-plane, its 
measurement requires an out-of-plane experiment, 
such as is accessible at MIT-Bates \cite{Kalos97}. 
Our calculation shows firstly that the SSA is quite
sizeable in the $\Delta(1232)$ region. 
Moreover, it displays
only a rather weak dependence on the GPs, because the SSA is mainly
sensitive to the imaginary part of the VCS amplitude. 
Therefore, it provides an excellent cross-check of the 
dispersion formalism for VCS, in particular by comparing 
at the same time the pion and photon electroproduction channels 
through the $\Delta$ region.

\section{Conclusions}
\label{conclusions}

In this work, we have presented a dispersion relation (DR) formalism for
VCS off a proton target. Such a formalism can serve as a
tool to extract generalized polarizabilities (GPs) from VCS observables over
a larger energy range. 
The way we evaluated our dispersive integrals using
$\pi N$ intermediate states, allows to apply the present formalism for
VCS observables through the $\Delta(1232)$-resonance region. 
\newline
\indent
The presented DR framework, when applied at a fixed value of $Q^2$,  
involves two free parameters which can be
expressed in terms of the electric and magnetic GPs, 
and which are to be extracted from a fit to VCS data. 
We proposed a parametrization of these two free parameters (asymptotic
terms to $\alpha$ and $\beta$) in terms of a dipole $Q^2$ dependence,
and investigated the sensitivity of VCS observables to the
corresponding dipole mass scales. 
\newline
\indent
We confronted our dispersive calculations with existing VCS data taken at
MAMI below pion threshold. Compared to the low energy expansion (LEX) analysis
which was previously applied to those data, 
we found only a modest additional energy dependence up to
photon energies of around 100 MeV, which supports such a LEX analysis.
When increasing the photon energy, 
our dispersive calculations show that the region between pion
threshold and the $\Delta$-resonance peak displays an enhanced
sensitivity to the GPs. It seems therefore very promising to measure
VCS observables in this energy region in order to extract GPs with
an enhanced precision.  
\newline
\indent
Furthermore, we showed our DR predictions for VCS data at higher
values of $Q^2$, in the range $Q^2$ = 1 - 2 GeV$^2$, where VCS data have been
taken at JLab which are presently under analysis. 
It was found for the JLab kinematics that the DR results show already
a noticeable deviation from the LEX result even for outgoing photon
energies as low as 75 MeV. Therefore, the DR analysis seems already to
be needed below pion threshold to extract GPs from the JLab data. 
We also showed predictions at $Q^2$ = 1 GeV$^2$ at higher outgoing
photon energies, through the
$\Delta(1232)$-resonance region, where data have also been taken at
JLab. At backward scattering angles, we found a very sizeable
sensitivity to the generalized electric polarizability. 
The two different JLab data sets, both below pion threshold and in
the $\Delta$-region, at the same value of $Q^2$ (in the range 
$Q^2$ = 1 - 2~GeV$^2$) will provide a very
interesting check on the presented DR formalism 
to demonstrate that a consistent value of the GPs can be extracted
by a fit in both energy regions. 
\newline
\indent
Besides unpolarized VCS experiments, which give access to a
combination of 3 (out of 6) GPs, we investigated the potential of
double polarization VCS observables. Although such double polarization
experiments with polarized beam and recoil proton polarization are
quite challenging, they are needed to access and quantify the remaining
three GPs. Using the DR formalism one can also analyze these
observables above pion threshold. 
\newline
\indent
Finally, above pion threshold also single
polarization observables are non-zero. In particular, the electron
single spin asymmetry, using a polarized electron beam,  
is sizeable in the $\Delta$-region and can
provide a very valuable cross-check of the VCS dispersion calculations,
as it is mainly sensitive to the imaginary part of the VCS amplitude,
which is linked through unitarity to the $\pi N$ channel.

\section*{acknowledgments}

The authors thank P. Bertin, N. d'Hose, H. Fonvieille, J.M. Friedrich,
S. Jaminion, S. Kamalov, G. Laveissiere, A. L'vov, H. Merkel, L. Tiator, 
R. Van de Vyver, and L. Van Hoorebeke for useful discussions. 
This work was supported by the ECT*, 
by the Deutsche Forschungsgemeinschaft (SFB 443), 
and by the European Commission IHP program under contract HPRN-CT-2000-00130.

\newpage

\begin{appendix}

\section{Gauge-invariant tensor basis for VCS}
\label{app:tensor}

\subsection{VCS tensor basis $\rho^{\mu \nu}_i$}

In writing down a gauge-invariant tensor basis for VCS, 
it will be useful to introduce the following 
symmetric combinations of the four-momenta 
(in the notations of Sec.~\ref{invampl})~: 
\begin{equation}        
P = {1 \over 2} \left( p + p' \right) \;, \hspace{1cm}
K = {1 \over 2} \left( q + q' \right)\;.
\label{eq:pk}
\end{equation}
                       
The 12 independent tensors $\rho^{\mu \nu}_i$ entering the VCS
amplitude of Eq.~(\ref{eq:nonborn}), that were  
introduced in \cite{Dre98}, are given by~:
\begin{eqnarray}
\rho_{1}^{\mu\nu} & = &
-\qqs g^{\mu\nu} + q^{\prime \mu} q^{\nu}
\vphantom{\frac{1}{1}} \, ,
\nonumber \\
\rho_{2}^{\mu\nu} & = &
- (2 M \nu)^2 g^{\mu\nu} - 4 \qqs P^{\mu} P^{\nu}
+ 4 M \nu \, \Big( P^{\mu} q^{\nu} + P^{\nu} q^{\prime \mu} \Big)
\vphantom{\frac{1}{1}} \, ,
\nonumber \\
\rho_{3}^{\mu\nu} & = &
- 2 M \nu Q^2 g^{\mu\nu} - 2 M \nu \, q^{\mu} q^{\nu} 
 + 2 Q^2 P^{\nu} q^{\prime \mu} + 2 \qqs \, P^{\nu} q^{\mu}
\vphantom{\frac{1}{1}} \, ,
\nonumber \\
\rho_{4}^{\mu\nu} & = &
8 P^{\mu} P^{\nu} \kdagger 
- 4 M \nu \, \Big( P^{\mu} \gamma^{\nu} + P^{\nu} \gamma^{\mu} \Big)
+ i \, 4 M \nu \, \gamma_5 \, \varepsilon^{\mu \nu \alpha \beta} 
 K_{\alpha} \gamma_{\beta}
\vphantom{\frac{1}{1}} \, ,
\nonumber \\
\rho_{5}^{\mu\nu} & = &
P^{\nu} q^{\mu} \kdagger 
- \frac{Q^2}{2} \, \Big( P^{\mu} \gamma^{\nu} 
 - P^{\nu} \gamma^{\mu} \Big )
- M \nu \, q^{\mu} \gamma^{\nu}
-\frac{i}{2} \, Q^2 \, \gamma_5 \, \varepsilon^{\mu\nu\alpha\beta} 
 K_{\alpha} \gamma_{\beta} \, ,
\nonumber \\
\rho_{6}^{\mu\nu} & = &
- 8 \qqs P^{\mu} P^{\nu} 
+ 4 M \nu \, \Big( P^{\mu} q^{\nu} + P^{\nu} q^{\prime \mu} \Big)
+ 4 M \qqs \, \Big( P^{\mu} \gamma^{\nu} + P^{\nu} \gamma^{\mu} \Big)
\vphantom{\frac{1}{1}}
\nonumber \\
& & - 4 M^2 \nu \, \Big( q^{\prime \mu} \gamma^{\nu} 
 + q^{\nu} \gamma^{\mu} \Big)
+ i \, 4 M \nu \, \Big( q^{\prime \mu} \sigma^{\nu\alpha} K_{\alpha}
   - q^{\nu} \sigma^{\mu\alpha} K_{\alpha} + \qqs \sigma^{\mu\nu} \Big)
\vphantom{\frac{1}{1}}
\nonumber \\
& & + i \, 4 M \qqs \, \gamma_5 \, \varepsilon^{\mu\nu\alpha\beta} 
  K_{\alpha} \gamma_{\beta}
\vphantom{\frac{1}{1}} \, ,
\nonumber \\
\rho_{7}^{\mu\nu} & = &
\Big( P^{\mu} q^{\nu} - P^{\nu} q^{\prime \mu} \Big) \kdagger
- \qqs \,\Big( P^{\mu} \gamma^{\nu} - P^{\nu} \gamma^{\mu} \Big)
+ M \nu \, \Big( q^{\prime \mu} \gamma^{\nu} 
  - q^{\nu} \gamma^{\mu} \Big)
\vphantom{\frac{1}{1}} \, ,
\nonumber \\
\rho_{8}^{\mu\nu} & = &
M \nu \, q^{\mu} q^{\nu} 
+ \frac{Q^2}{2} \Big( P^{\mu} q^{\nu} - P^{\nu} q^{\prime \mu} \Big)
- \qqs \, P^{\nu} q^{\mu}
- M q^{\mu} q^{\nu} \kdagger
+ M \qqs \, q^{\mu} \gamma^{\nu}
\nonumber \\
& & + \frac{M}{2} Q^2 \Big( q^{\prime \mu} \gamma^{\nu} 
  - q^{\nu} \gamma^{\mu} \Big)
-\frac{i}{2} \, Q^2 \Big( q^{\prime \mu} \sigma^{\nu\alpha} K_{\alpha}
   - q^{\nu} \sigma^{\mu\alpha} K_{\alpha} + \qqs \sigma^{\mu\nu} \Big) 
\, , \nonumber \\
\rho_{9}^{\mu\nu} & = &
2 M \nu \, \Big( P^{\mu} q^{\nu} - P^{\nu} q^{\prime \mu} \Big)
- 2 M \qqs \, \Big( P^{\mu} \gamma^{\nu} - P^{\nu} \gamma^{\mu} \Big)
+ 2 M^2 \nu \, \Big( q^{\prime \mu} \gamma ^{\nu} 
  - q^{\nu} \gamma^{\mu} \Big)
\vphantom{\frac{1}{1}}
\nonumber \\
& & + i \, 2 \qqs \, \Big( P^{\mu} \sigma^{\nu\alpha} K_{\alpha}
   + P^{\nu} \sigma^{\mu\alpha} K_{\alpha} \Big)
-i \,2 M \nu \, \Big( q^{\prime \mu} \sigma^{\nu\alpha} K_{\alpha}
   + q^{\nu} \sigma^{\mu\alpha} K_{\alpha} \Big)
\vphantom{\frac{1}{1}} \, ,
\nonumber \\
\rho_{10}^{\mu\nu} & = &
- 4 M \nu \, g^{\mu\nu}
+ 2 \, \Big( P^{\mu} q^{\nu} + P^{\nu} q^{\prime \mu} \Big)
+ 4 M \, g^{\mu\nu} \kdagger
- 2 M \, \Big( q^{\prime \mu} \gamma^{\nu} 
  + q^{\nu} \gamma^{\mu} \Big)
\vphantom{\frac{1}{1}}
\nonumber \\
& & -2 \, i \, \Big( q^{\prime \mu} \sigma^{\nu\alpha} K_{\alpha}
  - q^{\nu} \sigma^{\mu\alpha} K_{\alpha} + \qqs \sigma^{\mu\nu} \Big)
\vphantom{\frac{1}{1}} \, ,
\nonumber \\
\rho_{11}^{\mu\nu} & = &
4 \, \Big( P^{\mu} q^{\nu} + P^{\nu} q^{\prime \mu} \Big) \kdagger
- 4 M \nu \, \Big( q^{\prime \mu} \gamma^{\nu} 
  + q^{\nu} \gamma^{\mu} \Big)
+ i \, 4 \qqs \, \gamma_5 \, \varepsilon^{\mu\nu\alpha\beta} 
   K_{\alpha} \gamma_{\beta}
\vphantom{\frac{1}{1}} \, ,
\nonumber \\
\rho_{12}^{\mu\nu} & = &
2 Q^2 P^{\mu} P^{\nu} + 2 M \nu \, P^{\nu} q^{\mu}
- 2 M Q^2 P^{\mu} \gamma^{\nu}
-2 M^2 \nu \, q^{\mu} \gamma^{\nu}
+ i \, 2 M \nu \, q^{\mu} \sigma^{\nu\alpha} K_{\alpha}
\vphantom{\frac{1}{1}}
\nonumber \\
& & + i \, Q^2 \, \Big( P^{\mu} \sigma^{\nu\alpha} K_{\alpha}
  + P^{\nu} \sigma^{\mu\alpha} K_{\alpha}
  - M \nu \, \sigma^{\mu\nu} \Big)
- i \, M Q^2 \, \gamma_5 \, \varepsilon^{\mu\nu\alpha\beta} 
  K_{\alpha} \gamma_{\beta}
\vphantom{\frac{1}{1}} \, ,
\label{eq:vcstensors}
\end{eqnarray}
where we follow the conventions of Bjorken and Drell 
\cite{BD}, i.e. $\sigma^{\mu \nu} = i/2 \, [\gamma^{\mu}, \gamma^{\nu}]$ 
and in particular $\varepsilon_{0123}$ = +1. 

\subsection{VCS invariant amplitudes $B_i$ of Berg and Lindner}
\label{app:berglindner}

For further reference, it also turns out to be useful to work with 
an alternative tensor basis for VCS,  
introduced by Berg and Lindner \cite{Ber58}. 
\newline
\indent
One starts by defining, besides the four-vectors $P$ and $K$ 
of Eq.~(\ref{eq:pk}), the combination
\begin{equation}
L \;=\;{1\over 2}(q' \,-\,q ) \;, 
\end{equation}
and constructs from $K, P$, and $L$, the following four-vectors 
which are orthogonal to each other~:
\begin{eqnarray}
L'^\mu&\equiv&L^\mu\,-\,\frac{(L\cdot K)}{K^2} K^\mu\, ,
\nonumber\\
P'^\mu&\equiv&P^\mu\,-\,\frac{(P\cdot K)}{K^2} K^\mu
\,-\,\frac{(P\cdot L')}{L'^2} L'^\mu\, ,
\nonumber\\
N^\mu&\equiv&\varepsilon^{\mu\nu\alpha\beta}
P'_\nu L'_\alpha K_\beta\;.
\end{eqnarray}
One next constructs the combination of the four-vectors 
$K$ and $L'$ which is gauge-invariant with respect to the virtual
photon four-momentum $q$ as~:
\begin{equation}
K'^\mu\;\equiv\;K^\mu\,-\,\frac{q\cdot K}{q\cdot L'} L'^\mu \;,
\end{equation}
which satisfies $q\,\cdot K'\;=\;0$. In terms of these
four-vectors, the Lorentz- and gauge-invariant VCS tensor 
${\cal M}^{\mu\nu}$ can now be written as~:
\begin{eqnarray}
{\cal M}^{\mu\nu}&=& 
\frac{\psmu \psnu}{\pss} \Big( B_1 + B_2 \kdagger \, \Big )
+ \frac{N^{\mu} N^{\nu}}{N^2} \Big( B_3 + B_4 \kdagger \, \Big )
\nonumber \\
& & + \frac{\psmu N^{\nu} + \psnu N^{\mu}}{\pss N^2} 
 \Big( B_5 \, i \gamma_5 + B_6 \ndagger \, \Big ) 
+ \frac{\psmu N^{\nu} - \psnu N^{\mu}}{\pss N^2} 
 \Big( B_7 \, i \gamma_5 + B_8 \ndagger \, \Big )
\nonumber \\
& & + \frac{\ksmu \psnu}{K^2 \pss} \Big( B_9 + B_{10} \kdagger \, \Big)
+ \frac{\ksmu N^{\nu}}{K^2 N^2} 
 \Big( B_{11} \, i \gamma_5 + B_{12} \ndagger \, \Big) \;,
\label{eq:blbasis}
\end{eqnarray}
where $B_i (i = 1,...,12)$ are the VCS invariant amplitudes of Berg and
Lindner \cite{Ber58}. 
\newline
\indent
The invariant amplitudes $F_i$ defined in Eq.~(\ref{eq:fampl}) which
correspond to the tensor basis of Eq.~(\ref{eq:vcstensors}) can be
expressed in terms of the invariant amplitudes $B_i$ defined in 
Eq.~(\ref{eq:blbasis}). These expressions read~:
\begin{eqnarray}
F_1 & = &
\frac{2}{(t + Q^2)^3 \pss} \bigg \{
 2 M^2 \nu^2 (t - Q^2) \Big[ B_1 - B_3 + \nu (B_2 - B_4) \Big]
+ (t + Q^2)^2 \pss \Big[ B_3 + \nu B_4 \Big]
\nonumber \\ 
& & - \frac{M}{\pss} \nu^2 t (t - Q^2) B_6
+ 8 M \nu Q^2 \bigg[ B_9 + \nu B_{10} - \frac{t}{4M} B_{12} \bigg]
 \bigg \} \;,
\nonumber \\
F_2 & = &
\frac{1}{2 (t + Q^2) \pss}  \bigg\{ 
\Big[ B_1 - B_3 + \nu (B_2 - B_4) \Big]
- \frac{t}{2 M \pss} B_6 \bigg\} \;,
\nonumber \\
F_3 & = &
\frac{1}{(t + Q^2)^2 \pss} \bigg\{
- 2 M \Big[ B_1 - B_3 + \nu (B_2 - B_4) \Big]
+ \frac{t}{\pss} B_6
\nonumber \\
& & + \frac{4}{\nu} 
\bigg[ B_9 + \nu B_{10} - \frac{t}{4M} B_{12} \bigg] \bigg\} \;,  
\nonumber \\
F_4 & = &
\frac{1}{8 \nu \pss} \Big[ B_2 - B_4 \Big]
- \frac{M t}{(t + Q^2)^2 \psf} B_6 \;,
\nonumber \\
F_5 & = &
\frac{2}{(t + Q^2) \pss} \bigg \{
 M \nu \bigg[1 - \frac{2 \nu^2 (t - Q^2)}{(t + Q^2)^2} \bigg] 
 \Big[B_2 - B_4 \Big]
- \frac{\nu}{M} \pss B_4
\nonumber \\
& & - \frac{4 M}{(t + Q^2) \pss} 
\bigg[1 + \frac{4 \nu^2 Q^2}{(t + Q^2)^2} \bigg] B_5
- \bigg[ 1 - \frac{4 M^2 - t}{2 \pss}
 \bigg(1 - \frac{2 \nu^2 (t - Q^2)}{(t + Q^2)^2} \bigg) \bigg] B_6
\nonumber \\
& & + \frac{4}{M (t + Q^2)} B_7 - B_8
- 2 \bigg[1 + \frac{4 \nu^2 Q^2}{(t + Q^2)^2} \bigg] B_{10}
\nonumber \\
& & + \frac{32 \nu Q^2}{(t + Q^2)^3} B_{11}
+ \frac{8 \nu (t - Q^2)}{M (t + Q^2)^2}
\bigg[ M^2   
+ \frac{t Q^2}{4 (t - Q^2)} \bigg] B_{12} \bigg\} \;,
\nonumber \\
F_6 & = &
\frac{1}{(t + Q^2)^3 \pss} \bigg\{
- \frac{\nu}{4} (t^2 + t Q^2 + 2Q^4) \Big[B_2 - B_4 \Big]
+ \frac{Q^2 (t + Q^2)^2 \pss}{8 M^2 \nu} B_4
+ \frac{4 Q^2}{\pss} B_5
\nonumber \\
& & - \Big( M^2 - \frac{t}{4} \Big)
\frac{t^2 + t Q^2 + 2 Q^4}{2 M \pss} B_6
+ \frac{Q^4}{M} B_{10}
- \frac{4 Q^2}{M \nu} B_{11}
+ \frac{Q^4}{M^2 \nu} \Big(M^2 - \frac{t}{4} \Big) B_{12} \bigg\} \;,
\nonumber \\
F_7 & = &
\frac{1}{(t + Q^2)^2 \pss} \bigg\{
2 M \nu Q^2 \Big[ B_2 - B_4 \Big]
- \frac{8 M}{\pss} B_5
+ \frac{Q^2}{\pss} \Big((4 M^2 - t) - 2\pss \Big) B_6
\nonumber \\
& & + 2 t B_8 - 4 Q^2 B_{10}
+ \frac{16 M \nu t Q^2}{(t + Q^2)^2} B_{12} \bigg\} \;,
\nonumber \\
F_8 & = &
\frac{2}{(t + Q^2)^2 \pss} \bigg \{
- 2 M \nu \Big[ B_2 - B_4 \Big]
+ \frac{16 M}{(t + Q^2) \pss} B_5
- \frac{1}{\pss} (4 M^2 - t) B_6
\nonumber \\
& & + 4 B_{10}
- \frac{16}{\nu (t + Q^2)} B_{11}
+ \frac{4 M^2 - t}{M \nu} B_{12} \bigg\} \;,
\nonumber \\
F_9 & = &
\frac{1}{(t + Q^2)^3 \pss} \bigg\{
- \frac{\nu}{2} Q^2 (t - Q^2) \Big[ B_2 - B_4 \Big]
- \frac{Q^2}{4 M^2 \nu} (t + Q^2)^2 \pss B_4
+ \frac{4}{\pss} (t - Q^2) B_5
\nonumber \\ & & - \frac{Q^2}{M \pss} 
\Big( M^2 - \frac{t}{4} \Big)(t - Q^2) B_6
- \frac{2 Q^4}{M} B_{10}
+ \frac{8 Q^2}{M \nu} B_{11}
- \frac{2 Q^4}{M^2 \nu} \Big( M^2 - \frac{t}{4} \Big) B_{12} \bigg\} \;,
\nonumber \\
F_{10} & = &
\frac{1}{4M \nu} B_4 \;,
\nonumber \\
F_{11} & = &
\frac{1}{(t + Q^2)^2 \pss} \bigg\{
\frac{M \nu}{2} \bigg[ t + \frac{2 \nu^2 (t - Q^2)}{t + Q^2} \bigg] 
 \Big[ B_2 - B_4 \Big]
+ \frac{\nu}{2 M} (t + Q^2) \pss B_4
\nonumber \\
& & + \frac{8 M \nu^2 Q^2}{(t + Q^2)^2 \pss} B_5
+ \frac{1}{\pss} \bigg[ \Big( M^2 - \frac{t}{4} \Big)
 \bigg( t + \frac{2 \nu^2 (t - Q^2)}{t + Q^2} \bigg) 
 -\frac{t \pss}{2} \bigg] B_6
- \frac{2}{M} B_7 
\nonumber \\
& & + \frac{Q^2}{2} B_8
+ Q^2 \bigg[ 1 + \frac{4 \nu^2}{t + Q^2} \bigg] B_{10}
- \frac{16 \nu Q^2}{(t + Q^2)^2} B_{11}
+ \frac{4 M \nu Q^2}{t + Q^2}
 \bigg[ \frac{Q^2}{t + Q^2} - \frac{t}{4 M^2} \bigg] B_{12} \bigg\} \;
 , \nonumber \\
F_{12} & = &
\frac{1}{(t + Q^2)^2 \pss} \bigg\{
- \frac{\nu}{2} (t - Q^2) \Big[ B_2 - B_4 \Big]
- \frac{1}{4 M^2 \nu} (t + Q^2)^2 \pss B_4
- \frac{4}{\pss} B_5
\nonumber \\
& & - \frac{1}{M \pss} \Big( M^2 - \frac{t}{4} \Big)(t - Q^2) B_6
- \frac{2 Q^2}{M} B_{10}
+ \frac{8}{M \nu} B_{11}
- \frac{2 Q^2}{M^2 \nu} \Big( M^2 - \frac{t}{4} \Big) B_{12} \bigg\} \;.
\label{eq:fafob}
\end{eqnarray}

\section{Born contributions to invariant amplitudes}
\label{app:born}

For the invariant amplitudes $F_i$, defined through Eq.~(\ref{eq:fampl}), 
one finds the following Born contributions $F_i^B$, 
corresponding to a nucleon intermediate state in the $s$- and
$u$-channel of the $\gamma^* p \to \gamma p$ process~:
\begin{eqnarray}
F_{1}^{B} & = &
\frac{1}{M (s - M^2)(u - M^2)} \bigg\{ \frac{t + Q^2}{2}
 \bigg [ \kappa F_1(Q^2) + (1 + \kappa) F_2(Q^2) \bigg ]
 - \nu^2 \kappa F_2(Q^2) \bigg\}\;,
\nonumber \\
F_{2}^{B} & = &
- \frac{1}{M (s - M^2)(u - M^2)} \bigg[ F_1(Q^2)
 + \frac{t + Q^2}{8 M^2} \kappa F_2(Q^2) \bigg] \;,
\nonumber \\
F_{3}^{B} & = &
0 \vphantom{\frac{1}{1}} \;,
\nonumber \\
F_{4}^{B} & = &
- \frac{1}{2 M (s - M^2)(u - M^2)} \kappa F_2(Q^2) \;,
\nonumber \\
F_{5}^{B} & = &
\frac{1}{M^2 (s - M^2)(u - M^2)} \bigg\{ - \frac{t + Q^2}{4} 
 \bigg[ \kappa F_1(Q^2) + (1 + 2\kappa) F_2(Q^2) \bigg]
 + \nu^2 \kappa F_2(Q^2) \bigg\} \;,
\nonumber \\
F_{6}^{B} & = &
\frac{1}{4 M (s - M^2)(u - M^2)} \bigg[ (2 + \kappa) F_1(Q^2) 
 + F_2(Q^2) + \frac{t}{4 M^2} \kappa F_2(Q^2) \bigg] \;,
\nonumber \\
F_{7}^{B} & = &
0 \vphantom{\frac{1}{1}} \;,
\nonumber \\
F_{8}^{B} & = &
0 \vphantom{\frac{1}{1}} \;,
\nonumber \\
F_{9}^{B} & = &
\frac{1}{2 M (s - M^2)(u - M^2)} \bigg[ - \kappa F_1(Q^2) + F_2(Q^2)
 + \frac{Q^2}{4 M^2} \kappa F_2(Q^2) \bigg] \;,
\nonumber \\
F_{10}^{B} & = &
\frac{1}{(s - M^2)(u - M^2)} (1 + \kappa) 
 \Big( F_1(Q^2) + F_2(Q^2) \Big) \;,
\nonumber \\
F_{11}^{B} & = &
\frac{1}{4 M^2 (s - M^2)(u - M^2)} \bigg[ 
 \frac{t + Q^2}{4} \Big(\kappa F_1(Q^2) + F_2(Q^2)\Big)
 - \nu^2 \kappa F_2(Q^2) \bigg] \;,
\nonumber \\
F_{12}^{B} & = &
\frac{t + Q^2}{8 M^3 (s - M^2) (u - M^2)} \kappa F_2(Q^2) \;,
\end{eqnarray}    
where $F_1(Q^2)$ and $F_2(Q^2)$ are the Dirac and Pauli nucleon form
factors respectively. 

\section{s-channel helicity amplitudes for VCS}
\label{app:shel}

\subsection{Definitions and conventions}
\label{app:sheldef}

The $s$-channel helicity amplitudes for virtual Compton scattering 
are denoted by $T^s_{\lambda' \, \lambda'_N \, , \, \lambda \,
  \lambda_N}$, and were defined in Eq.~(\ref{eq:matrixele}).
In this appendix, we express the invariant amplitudes $F_i$ in terms of 
these s-channel helicity amplitudes. In addition, we quote the explicit
results for the imaginary parts of the helicity amplitudes in the case of
$\pi N$ intermediate states.
\newline
\indent
We work in the c.m. system of the $s$-channel process $\gamma^* N
\rightarrow \gamma N$, and all kinematical quantities are understood
in this system. 
The energies of the incoming (outgoing) nucleon are denoted by $E$
($E'$) respectively.  
The incoming photon has energy $q_0$ and its momentum $\vec q$
is chosen to point in the $z$-direction. The outgoing photon 
momentum $\vec q \,'$ is chosen to lie in the $xz$ plane and makes
an angle $\theta$ with the $z$-axis. We use the Lorentz gauge for the
photon polarization vectors. For the initial (virtual) photon, the transverse
and longitudinal polarization vectors are given by~:
\begin{eqnarray}
&&\varepsilon^\mu \left(q, \lambda = \pm 1 \right) \,=\,
\left(0, \mp {1 \over {\sqrt{2}}}, - {i \over {\sqrt{2}}}, 0 \right)
\, , \nonumber \\
&&\varepsilon^\mu \left(q, \lambda = 0 \right) \,=\,
\left({{| \vec q \, |} \over Q}, 0, 0, {{q_0} \over Q} \right) \, ,
\end{eqnarray}
whereas for the final (real) photon, the polarization
vectors are given by~:
\begin{equation}
\varepsilon^{' \mu} \left(q', \lambda^{'} = \pm 1 \right) \,=\,
\left(0, \mp {1 \over {\sqrt{2}}} \cos \theta, 
- {i \over {\sqrt{2}}}, \pm {1 \over {\sqrt{2}}} \sin \theta \right) \,.
\end{equation}
\newline
\indent
The initial nucleon, characterized by the momentum $\vec{p}$ and the
 polarization $\lambda_{N}$, is propagating in the negative $z$-direction. 
The final nucleon, with momentum $\vec{p}^{\, '}$ 
and polarization $\lambda'_{N}$,
makes an angle $180^o - \theta$ with respect to
the virtual photon, and has the azimuthal 
angle $180^o + \phi_{\gamma^* \gamma}$. 
This leads to the following spinor conventions 
for the incoming and outgoing nucleons~:
\begin{eqnarray}
u(\vec{p},\,\lambda_N)&=&\sqrt{E+M}
\left[
\begin{array}{c}
\chi_{\lambda_N} \\
\\
2\lambda_N {{|\vec{p}|}\over{E+M}}\,\chi_{\lambda_N} 
\end{array}
\right] \;, \nonumber\\
\nonumber\\
u(\vec{p}^{\, '},\,\lambda'_N)&=&\sqrt{E'+M}
\left[
\begin{array}{c}
\chi'_{\lambda'_N} \\
\\
2\lambda'_N {{|\vec{p}^{\, '}|}\over{E'+M}}\,{\chi'_{\lambda'_N}}
\end{array}
\right] \;,
\end{eqnarray}
where
\begin{eqnarray}
\chi_{\frac{1}{2}}&=&\left[
\begin{array}{c}
0\\
\\
1
\end{array}
\right],
\;\;\;\;\;\;\;\;\;
\chi_{-\frac{1}{2}}\,=\,\left[
\begin{array}{c}
-1\\
\\
0
\end{array}
\right],\nonumber\\
\chi'_{\frac{1}{2}}&=&\left[
\begin{array}{c}
\sin{\frac{\theta}{2}}\\
\\
-\cos{\frac{\theta}{2}}
\end{array}
\right],
\;\;\;\;\;\;\;\;\;
\chi'_{-\frac{1}{2}}\,=\,\left[
\begin{array}{c}
\cos{\frac{\theta}{2}}\\
\\
\sin{\frac{\theta}{2}}
\end{array}
\right] \;.
\end{eqnarray}

\subsection{VCS reduced helicity amplitudes}
\label{app:reduced_ampl}

The reduced helicity amplitudes $\tau_i$ are defined by factorizing
out from the helicity amplitudes 
$T^s_{\lambda'\lambda'_N;\lambda \lambda_N}$ the kinematical factors
in  $\left(\cos\frac{\theta}{2}\right)^{|\Lambda+\Lambda'|}$ and
$\left(\sin\frac{\theta}{2}\right)^{|\Lambda-\Lambda'|},$
with $\Lambda=\lambda-\lambda_N$ and 
$\Lambda'=\lambda'-\lambda'_N$.
The relations between the 12 independent VCS helicity amplitudes
and the reduced helicity amplitudes $\tau_i$ ($i=1,..,12$) read~: 
\begin{eqnarray}
T^s_{1\frac{1}{2};1\frac{1}{2}}=\cos\frac{\theta}{2}\tau_1 , 
& &\quad\quad
T^s_{-1\frac{1}{2};-1\frac{1}{2}}=\cos^3\frac{\theta}{2}\tau_2 , \nonumber\\
T^s_{1-\frac{1}{2};1\frac{1}{2}}=
\cos^2\frac{\theta}{2}\sin\frac{\theta}{2}\tau_3 , & &\quad\quad
T^s_{1\frac{1}{2};-1\frac{1}{2}}=
\cos\frac{\theta}{2}\sin^2\frac{\theta}{2}\tau_4 , \nonumber\\
T^s_{-1-\frac{1}{2};1\frac{1}{2}}=
\sin\frac{\theta}{2}\tau_5, & &\quad\quad
T^s_{1-\frac{1}{2};-1\frac{1}{2}}=
\sin^3\frac{\theta}{2}\tau_6, \nonumber\\
T^s_{-1\frac{1}{2};1\frac{1}{2}}=
\cos\frac{\theta}{2}\sin^2\frac{\theta}{2}\tau_7, & &\quad\quad
T^s_{-1-\frac{1}{2};-1\frac{1}{2}}=
\cos^2\frac{\theta}{2}\sin\frac{\theta}{2}\tau_8, \nonumber\\
T^s_{1\frac{1}{2};0\frac{1}{2}}=
\sin\frac{\theta}{2}\tau_9, & &\quad\quad
T^s_{-1-\frac{1}{2};0\frac{1}{2}}=
\cos\frac{\theta}{2}\tau_{10}, \nonumber\\ 
T^s_{-1\frac{1}{2};0\frac{1}{2}}=
\sin\frac{\theta}{2}\cos^2\frac{\theta}{2}\tau_{11}, & &\quad\quad
T^s_{1-\frac{1}{2};0\frac{1}{2}}=
\cos\frac{\theta}{2}\sin^2\frac{\theta}{2}\tau_{12} . 
\label{eq:vcsredhel}
\end{eqnarray}

\subsection{Relations between the invariant amplitudes of Berg and Lindner
  and the VCS reduced helicity amplitudes}
\label{app:relblhel}

The imaginary parts of the invariant amplitudes $F_i$, which enter the
dispersion integrals of Eq.~(\ref{eq:unsub}), are constructed from the
VCS reduced helicity amplitudes $\tau_i$, 
which were defined in~(\ref{eq:vcsredhel}). 
To avoid too lengthy formulas, we display here the relations between the
amplitudes $B_i$ and the $\tau_i$. The relations between the $F_i$ and
the $\tau_i$ are then obtained from those relations, and by using 
Eq.~(\ref{eq:fafob}), which expresses the $F_i$ in terms of the $B_i$. 
\newline
\indent
For convenience we define the following abbreviations for kinematical factors~:
\begin{eqnarray}
& & C_1 = 1 + \frac{|\vec{q}\,|}{E + M} \;, \hspace{3.15cm} 
 C_2 = 1 - \frac{|\vec{q}\,|}{E + M} \;, 
\nonumber \\
& & C_3 = 1 + \frac{\sqrt{s} - M}{\sqrt{s} + M} \,
 \frac{|\vec{q}\,|}{E + M} \;,
 \hspace{1.5cm}
 C_4 = 1 - \frac{\sqrt{s} - M}{\sqrt{s} + M} \, \frac{|\vec{q}\,|}{E + M} \;,
\nonumber \\
& & C_5 = \frac{|\vec{q}\,|}{E + M} + \frac{|\vec{q}\,'|}{E' + M} \;,
 \hspace{1.9cm}
 C_6 = \frac{|\vec{q}\,|}{E + M} - \frac{|\vec{q}\,'|}{E' + M} \;.
 \nonumber 
\end{eqnarray}

With these definitions, 
the relations between the amplitudes $B_i$ of Berg and Lindner  
and the reduced helicity amplitudes $\tau_i$ are given by~:
\begin{eqnarray}
\label{eq:B1}
(- e^2) \, B_1 && = 
 - \; \frac{\sqrt{(E + M)(E'+ M)}}{4 \, \sqrt{s}\,|\vec{q}\,|} \, 
(\sqrt{s} - M) \, \frac{|\vec{q}\,| - q_0 \cos{\theta}}{t + Q^2}
 \\
& & \Bigg\{ \;\; C_2 \Bigg[ \, \tau_1 + \cos^2 \frac{\theta}{2} \, \tau_2
 -\sin^2 \frac{\theta}{2} \Big( \tau_4 + \tau_7 \Big)
 - \frac{2 \sqrt{2} \, Q}{|\vec{q}\,| - q_0 \cos{\theta}} \sin^2 
\frac{\theta}{2}
 \bigg( \tau_9 - \cos^2 \frac{\theta}{2} \, \tau_{11} \bigg) \Bigg]
 \nonumber\\
& & 
 + \; C_1 \Bigg[ - \tau_5 - \sin^2 \frac{\theta}{2} \, \tau_6
 +\cos^2 \frac{\theta}{2} \Big( \tau_3 + \tau_8 \Big)
 + \frac{2 \sqrt{2} \, Q}{|\vec{q}\,| - q_0 \cos{\theta}} \cos^2 
\frac{\theta}{2} \bigg( \tau_{10} - \sin^2 \frac{\theta}{2} \, 
\tau_{12} \bigg) \Bigg] \Bigg\} \,,
 \nonumber\\
& & \vphantom{\frac{1}{1}}
 \nonumber \\
(- e^2) \, B_2 && = 
 \frac{\sqrt{(E + M)(E'+ M)}}{4 \, \sqrt{s}\,|\vec{q}\,|} \,
 \frac{|\vec{q}\,| - q_0 \cos{\theta}}{t + Q^2}
 \\
& & \Bigg\{ \;\; C_3 \Bigg[ \, \tau_1 + \cos^2 \frac{\theta}{2} \, \tau_2
 -\sin^2 \frac{\theta}{2} \Big( \tau_4 + \tau_7 \Big)
 - \frac{2 \sqrt{2} \, Q}{|\vec{q}\,| - q_0 \cos{\theta}} \sin^2 
\frac{\theta}{2}
 \bigg( \tau_9 - \cos^2 \frac{\theta}{2} \, \tau_{11} \bigg) \Bigg]
 \nonumber\\
& & 
 + \; C_4 \Bigg[ - \tau_5 - \sin^2 \frac{\theta}{2} \, \tau_6
 +\cos^2 \frac{\theta}{2} \Big( \tau_3 + \tau_8 \Big)
 + \frac{2 \sqrt{2} \, Q}{|\vec{q}\,| - q_0 \cos{\theta}} \cos^2 
\frac{\theta}{2}
\bigg( \tau_{10} - \sin^2 \frac{\theta}{2} \, \tau_{12} \bigg) \Bigg]
 \Bigg\} \,,
 \nonumber \\
& & \vphantom{\frac{1}{1}}
 \nonumber \\
(- e^2) \, B_3 & = &
 - \, \frac{\sqrt{(E + M)(E'+ M)}}{8 \, \sqrt{s}\,|\vec{q}\,||\vec{q}\,'|} \, 
 (\sqrt{s} - M)
\\
& & \Bigg\{ \;\; C_2 \Bigg[ \, \tau_1 + \cos^2 \frac{\theta}{2} \, \tau_2
 + \sin^2 \frac{\theta}{2} \Big( \tau_4 + \tau_7 \Big) \Bigg]
 \nonumber \\
& & + \; C_1 \Bigg[ \, \tau_5 + \sin^2 \frac{\theta}{2} \, \tau_6
 +\cos^2 \frac{\theta}{2} \Big( \tau_3 + \tau_8 \Big) \Bigg] \Bigg\} \,,
 \nonumber \\
& & \vphantom{\frac{1}{1}}
 \nonumber \\
(- e^2) \, B_4 & = &
 \frac{\sqrt{(E + M)(E'+ M)}}{8 \, \sqrt{s}\,|\vec{q}\,||\vec{q}\,'|} 
 \\
& & \Bigg\{ \;\; C_3 \Bigg[ \, \tau_1 + \cos^2 \frac{\theta}{2} \, \tau_2
 + \sin^2 \frac{\theta}{2} \Big( \tau_4 + \tau_7 \Big) \Bigg]
 \nonumber \\
& & + \; C_4 \Bigg[ \, \tau_5 + \sin^2 \frac{\theta}{2} \, \tau_6
 +\cos^2 \frac{\theta}{2} \Big( \tau_3 + \tau_8 \Big) \Bigg] \Bigg\} \,,
 \nonumber \\
& & \vphantom{\frac{1}{1}}
 \nonumber \\
(- e^2) \, B_5 & = &
 - \, \frac{\sqrt{(E + M)(E'+ M)}}{4 \, t} \,
 \frac{s \, |\vec{q}\,|^2 \, |\vec{q}\,'|^3 \sin^4 {\theta}}{(t + Q^2)^2}
 \\
& & \Bigg\{ \;\; C_5 \Bigg[ \, (|\vec{q}\,| - q_0) 
 \bigg( \tau_5 - \sin^2 \frac{\theta}{2} \, \tau_6 \bigg)
 + (|\vec{q}\,| + q_0) \sin^2 \frac{\theta}{2} \Big( \tau_3 - \tau_8 \Big)
 \nonumber \\
& & \hspace{1.5cm}
 - \sqrt{2} \, Q \bigg( \tau_{10} 
   + \sin^{2} \frac{\theta}{2} \, \tau_{12} \bigg) \Bigg]
 \nonumber \\
& & - \; C_6 \Bigg[ \, (|\vec{q}\,| + q_0) 
 \bigg( \tau_1 - \cos^2 \frac{\theta}{2} \, \tau_2 \bigg)
 - (|\vec{q}\,| - q_0) \cos^2 \frac{\theta}{2} \Big( \tau_4 - \tau_7 \Big)
 \nonumber \\
& & \hspace{1.5cm}
 - \sqrt{2} \, Q \bigg( \tau_{9} 
   + \cos^{2} \frac{\theta}{2} \, \tau_{11} \bigg) \Bigg] \Bigg\} \,,
 \nonumber \\
& & \vphantom{\frac{1}{1}}
 \nonumber \\
(- e^2) \, B_6 & = &
 \frac{\sqrt{(E + M)(E'+ M)}}{t} \,
 \frac{\sqrt{s} \, |\vec{q}\,| \, |\vec{q}\,'|^2 \sin^2 {\theta}}{(t + Q^2)^2}
 \\
& & \Bigg\{ \;\; C_5 \, \sin^2 \frac{\theta}{2} \Bigg[ \, (|\vec{q}\,| + q_0) 
 \bigg( \tau_1 - \cos^2 \frac{\theta}{2} \, \tau_2 \bigg)
 - (|\vec{q}\,| - q_0) \cos^2 \frac{\theta}{2} \Big( \tau_4 - \tau_7 \Big)
 \nonumber \\
& & \hspace{2.5cm}
 - \sqrt{2} \, Q \bigg( \tau_{9} 
   + \cos^{2} \frac{\theta}{2} \, \tau_{11} \bigg) \Bigg]
 \nonumber \\
& & \; + \; C_6 \, \cos^2 \frac{\theta}{2} \Bigg[ \, (|\vec{q}\,| - q_0) 
 \bigg( \tau_5 - \sin^2 \frac{\theta}{2} \, \tau_6 \bigg)
 + (|\vec{q}\,| + q_0) \sin^2 \frac{\theta}{2} \Big( \tau_3 - \tau_8 \Big)
 \nonumber \\
& & \hspace{2.5cm}
 - \sqrt{2} \, Q \bigg( \tau_{10} 
   + \sin^{2} \frac{\theta}{2} \, \tau_{12} \bigg) \Bigg]  \Bigg\} \,,
 \nonumber \\
& & \vphantom{\frac{1}{1}}
 \nonumber \\
(- e^2) \, B_7 & = &
 - \, \frac{\sqrt{(E + M)(E'+ M)}}{t} \,
 \frac{s \, |\vec{q}\,|^2 \, |\vec{q}\,'|^3 \sin^2 {\theta}}{(t + Q^2)^2}
 \\
& & \Bigg\{ \;\; C_5 \, \sin^2 \frac{\theta}{2} \Bigg[ \, (|\vec{q}\,| + q_0) 
 \sin^2 \frac{\theta}{2} \bigg( \tau_5 - \sin^2 \frac{\theta}{2} \, 
\tau_6 \bigg)
 + (|\vec{q}\,| - q_0) \cos^4 \frac{\theta}{2} \Big( \tau_3 - \tau_8 \Big)
 \nonumber \\
& & \hspace{2.5cm}
 - \sqrt{2} \, Q \, \cos^2 \frac{\theta}{2} \bigg( \tau_{10} 
   + \sin^{2} \frac{\theta}{2} \, \tau_{12} \bigg) \Bigg]
 \nonumber \\
& & \; - \; C_6 \, \cos^2 \frac{\theta}{2} \Bigg[ \, (|\vec{q}\,| - q_0) 
 \cos^2 \frac{\theta}{2} \bigg( \tau_1 - \cos^2 \frac{\theta}{2} \, 
\tau_2 \bigg)
 - (|\vec{q}\,| + q_0) \sin^4 \frac{\theta}{2} \Big( \tau_4 - \tau_7 \Big)
 \nonumber \\
& & \hspace{2.5cm}
 - \sqrt{2} \, Q \, \sin^2 \frac{\theta}{2} \bigg( \tau_{9} 
   + \cos^{2} \frac{\theta}{2} \, \tau_{11} \bigg) \Bigg] \Bigg\} \,,
 \nonumber \\
& & \vphantom{\frac{1}{1}}
 \nonumber \\
(- e^2) \, B_8 & = &
 \frac{\sqrt{(E + M)(E'+ M)}}{t} \,
 \frac{\sqrt{s} \, |\vec{q}\,| \, |\vec{q}\,'|^2 \sin^2 {\theta}}{(t + Q^2)^2}
 \\
& & \Bigg\{ \;\; C_5 \Bigg[ \, (|\vec{q}\,| - q_0) 
 \cos^2 \frac{\theta}{2} \bigg( \tau_1 - \cos^2 \frac{\theta}{2} \, 
\tau_2 \bigg)
 - (|\vec{q}\,| + q_0) \sin^4 \frac{\theta}{2} \Big( \tau_4 - \tau_7 \Big)
 \nonumber \\
& & \hspace{2.5cm}
 - \sqrt{2} \, Q \, \sin^2 \frac{\theta}{2} \bigg( \tau_{9} 
   + \cos^{2} \frac{\theta}{2} \, \tau_{11} \bigg) \Bigg]
 \nonumber \\
& & \; + \; C_6 \Bigg[ \, (|\vec{q}\,| + q_0) 
 \sin^2 \frac{\theta}{2} \bigg( \tau_5 - \sin^2 \frac{\theta}{2} \, 
\tau_6 \bigg)
 + (|\vec{q}\,| - q_0) \cos^4 \frac{\theta}{2} \Big( \tau_3 - \tau_8 \Big)
 \nonumber \\
& & \hspace{2.5cm}
 - \sqrt{2} \, Q \, \cos^2 \frac{\theta}{2} \bigg( \tau_{10} 
   + \sin^{2} \frac{\theta}{2} \, \tau_{12} \bigg) \Bigg]  \Bigg\} \,,
 \nonumber \\
& & \vphantom{\frac{1}{1}}
 \nonumber \\
(- e^2) \, B_9 & = &
 - \, \frac{\sqrt{(E + M)(E'+ M)}}{8 \, (t + Q^2)} \,
 (\sqrt{s} - M)\, | \vec{q}\,'| \,{\sin^2{\theta}}
\\
& & \Bigg\{ \;\; C_2 \Bigg[ \, \tau_1 + \cos^2 \frac{\theta}{2} \, \tau_2
 -\sin^2 \frac{\theta}{2} \Big( \tau_4 + \tau_7 \Big)
 - \frac{|\vec{q}\,| - q_0 \cos{\theta}}{\sqrt{2} \, Q \cos^2 \frac{\theta}{2}}
 \bigg( \tau_9 - \cos^2 \frac{\theta}{2} \, \tau_{11} \bigg) \Bigg]
 \nonumber\\
& & 
\; + \; C_1 \Bigg[ - \tau_5 - \sin^2 \frac{\theta}{2} \, \tau_6
 +\cos^2 \frac{\theta}{2} \Big( \tau_3 + \tau_8 \Big)
 + \frac{|\vec{q}\,| - q_0 \cos{\theta}}{\sqrt{2} \, Q \sin^2 \frac{\theta}{2}}
 \bigg( \tau_{10} - \sin^2 \frac{\theta}{2} \, \tau_{12} \bigg) \Bigg]
 \Bigg\} \,,
 \nonumber\\
& & \vphantom{\frac{1}{1}}
 \nonumber \\
(- e^2) \, B_{10} & = &
 \frac{\sqrt{(E + M)(E'+ M)}}{8 \, (t + Q^2)} \,
 | \vec{q}\,'| \,{\sin^2{\theta}}
\\
& & \Bigg\{ \;\; C_3 \Bigg[ \, \tau_1 + \cos^2 \frac{\theta}{2} \, \tau_2
 -\sin^2 \frac{\theta}{2} \Big( \tau_4 + \tau_7 \Big)
 - \frac{|\vec{q}\,| - q_0 \cos{\theta}}{\sqrt{2} \, Q \cos^2 \frac{\theta}{2}}
 \bigg( \tau_9 - \cos^2 \frac{\theta}{2} \, \tau_{11} \bigg) \Bigg]
 \nonumber\\
& & 
\; + \; C_4 \Bigg[ - \tau_5 - \sin^2 \frac{\theta}{2} \, \tau_6
 +\cos^2 \frac{\theta}{2} \Big( \tau_3 + \tau_8 \Big)
 + \frac{|\vec{q}\,| - q_0 \cos{\theta}}{\sqrt{2} \, Q \sin^2 \frac{\theta}{2}}
 \bigg( \tau_{10} - \sin^2 \frac{\theta}{2} \, \tau_{12} \bigg) \Bigg]
 \Bigg\} \,,
 \nonumber \\
& & \vphantom{\frac{1}{1}}
 \nonumber \\
(- e^2) \, B_{11} && = 
 \frac{\sqrt{(E + M)(E'+ M)}}{8 \, t}\,
 \sqrt{s} \, |\vec{q}\,| \, |\vec{q}\,'|^2 \sin^2 {\theta}
 \\
& & \Bigg\{ \;\; C_5 \, \sin^2 \frac{\theta}{2} \Bigg[ \,
 \tau_5 - \sin^2 \frac{\theta}{2} \, \tau_6
 +\cos^2 \frac{\theta}{2} \Big( \tau_3 - \tau_8 \Big)
 - \frac{|\vec{q}\,| - q_0 \cos{\theta}}{\sqrt{2} \, Q \sin^2 \frac{\theta}{2}}
 \bigg( \tau_{10} + \sin^2 \frac{\theta}{2} \, \tau_{12} \bigg) \Bigg]
 \nonumber \\
& & - \; C_6 \, \cos^2 \frac{\theta}{2} \Bigg[ 
 \, \tau_1 - \cos^2 \frac{\theta}{2} \, \tau_2
 - \sin^2 \frac{\theta}{2} \Big( \tau_4 - \tau_7 \Big)
 - \frac{|\vec{q}\,| - q_0 \cos{\theta}}{\sqrt{2} \, Q \cos^2 \frac{\theta}{2}}
 \bigg( \tau_9 + \cos^2 \frac{\theta}{2} \, \tau_{11} \bigg) \Bigg] \Bigg\} \,,
 \nonumber \\
& & \vphantom{\frac{1}{1}}
 \nonumber \\
\label{eq:B12}
(- e^2) \, B_{12} && = 
 - \, \frac{\sqrt{(E + M)(E'+ M)}}{8 \, t} \, |\vec{q}\,'| \, \sin^2 {\theta}
 \\
& & \Bigg\{ \;\; C_5 \, \Bigg[ \,
 \tau_1 - \cos^2 \frac{\theta}{2} \, \tau_2
 - \sin^2 \frac{\theta}{2} \Big( \tau_4 - \tau_7 \Big)
 - \frac{|\vec{q}\,| - q_0 \cos{\theta}}{\sqrt{2} \, Q \cos^2 \frac{\theta}{2}}
 \bigg( \tau_9 + \cos^2 \frac{\theta}{2} \, \tau_{11} \bigg) \Bigg]
 \nonumber \\
& & + \; C_6 \, \Bigg[
 \tau_5 - \sin^2 \frac{\theta}{2} \, \tau_6
 +\cos^2 \frac{\theta}{2} \Big( \tau_3 - \tau_8 \Big)
 - \frac{|\vec{q}\,| - q_0 \cos{\theta}}{\sqrt{2} \, Q \sin^2 \frac{\theta}{2}}
 \bigg( \tau_{10} + \sin^2 \frac{\theta}{2} \, \tau_{12} \bigg) \Bigg]
 \Bigg\} \,.
\nonumber
\end{eqnarray}

Eqs.~(\ref{eq:B1} - \ref{eq:B12}) 
together with the equations in (\ref{eq:fafob}) eventually allow 
to express the $F_i$ in terms of the VCS helicity amplitudes.

\subsection{Unitarity relations between the VCS reduced helicity amplitudes 
and pion photo- and electroproduction multipoles}
\label{app:pionprod}

If we write down the unitarity equations for the VCS helicity
amplitudes and consider only $\pi N$ intermediate states, then the 
imaginary parts of the VCS helicity amplitudes can be expressed in terms of 
the $\gamma^* N \rightarrow \pi N$ times $\gamma N \rightarrow
\pi N$ multipoles~:

\begin{eqnarray}
\label{eq:unitarity}
\mbox{Im}\tau_1&=&-8\pi q_\pi\sqrt{s}\sum_l(2l+2)\left[A_{l+}(Q^2)A^*_{l+}(0)
+A_{(l+1)-}(Q^2)A^*_{(l+1)-}(0)\right]\nonumber\\
& &\times
F(-l;l+2;1;\sin^2\frac{\theta}{2}),\nonumber\\
\mbox{Im}\tau_2&=&-8\pi q_\pi\sqrt{s}\sum_l\frac{l(l+1)(l+2)}{2}
\left[B_{l+}(Q^2)B^*_{l+}(0)
+B_{(l+1)-}(Q^2)B^*_{(l+1)-}(0)\right]\nonumber\\
& &\times
F(-l+1;l+3;1;\sin^2\frac{\theta}{2}),\nonumber\\
\mbox{Im}\tau_3&=&-8\pi q_\pi\sqrt{s}\sum_l l(l+1)(l+2)
\left[A_{l+}(Q^2)B^*_{l+}(0)
+A_{(l+1)-}(Q^2)B^*_{(l+1)-}(0)\right]\nonumber\\
& &\times
F(-l+1;l+3;2;\sin^2\frac{\theta}{2}),\nonumber\\
\mbox{Im}\tau_4&=&-8\pi q_\pi\sqrt{s}\sum_l \frac{l(l+1)^2(l+2)}{2}
\left[B_{l+}(Q^2)A^*_{l+}(0)
-B_{(l+1)-}(Q^2)A^*_{(l+1)-}(0)\right]\nonumber\\
& &\times
F(-l+1;l+3;3;\sin^2\frac{\theta}{2}),\nonumber\\
\mbox{Im}\tau_5&=&8\pi q_\pi\sqrt{s}\sum_l 2(l+1)^2
\left[A_{l+}(Q^2)A^*_{l+}(0)
-A_{(l+1)-}(Q^2)A^*_{(l+1)-}(0)\right]\nonumber\\
& &\times
F(-l;l+2;2;\sin^2\frac{\theta}{2}),\nonumber\\
\mbox{Im}\tau_6&=&-8\pi q_\pi\sqrt{s}\sum_l \frac{l^2(l+1)^2(l+2)^2}{12}
\left[B_{l+}(Q^2)B^*_{l+}(0)
-B_{(l+1)-}(Q^2)B^*_{(l+1)-}(0)\right]\nonumber\\
& &\times
F(-l+1;l+3;4;\sin^2\frac{\theta}{2}),\nonumber\\
\mbox{Im}\tau_7&=&-8\pi q_\pi\sqrt{s}\sum_l \frac{l(l+1)^2(l+2)}{2}
\left[A_{l+}(Q^2)B^*_{l+}(0)
-A_{(l+1)-}(Q^2)B^*_{(l+1)-}(0)\right]\nonumber\\
& &\times
F(-l+1;l+3;3;\sin^2\frac{\theta}{2}),\nonumber\\
\mbox{Im}\tau_8&=&-8\pi q_\pi\sqrt{s}\sum_l l(l+1)(l+2)
\left[B_{l+}(Q^2)A^*_{l+}(0)
+B_{(l+1)-}(Q^2)A^*_{(l+1)-}(0)\right]\nonumber\\
& &\times
F(-l+1;l+3;2;\sin^2\frac{\theta}{2}),\nonumber\\
\mbox{Im}\tau_9&=&8\pi q_\pi\sqrt{s}\frac{\sqrt{2Q^2}}{q_0}
\sum_l (l+1)^3
\left[L_{l+}(Q^2)A^*_{l+}(0)
+L_{(l+1)-}(Q^2)A^*_{(l+1)-}(0)\right]\nonumber\\
& &\times
F(-l;l+2;2;\sin^2\frac{\theta}{2}),\nonumber\\
\mbox{Im}\tau_{10}&=&8\pi q_\pi\sqrt{s}\frac{\sqrt{2Q^2}}{q_0}
\sum_l (l+1)^2
\left[L_{l+}(Q^2)A^*_{l+}(0)
-L_{(l+1)-}(Q^2)A^*_{(l+1)-}(0)\right]\nonumber\\
& &\times
F(-l;l+2;1;\sin^2\frac{\theta}{2}),\nonumber\\
\mbox{Im}\tau_{11}&=&-8\pi q_\pi\sqrt{s}\frac{\sqrt{2Q^2}}{q_0}
\sum_l \frac{l(l+1)^2(l+2)}{2}
\left[L_{l+}(Q^2)B^*_{l+}(0)
-L_{(l+1)-}(Q^2)B^*_{(l+1)-}(0)\right]\nonumber\\
& &\times
F(-l+1;l+3;2;\sin^2\frac{\theta}{2}),\nonumber\\
\mbox{Im}\tau_{12}&=&8\pi q_\pi\sqrt{s}\frac{\sqrt{2Q^2}}{q_0}
\sum_l \frac{l(l+1)^3(l+2)}{4}
\left[L_{l+}(Q^2)B^*_{l+}(0)
+L_{(l+1)-}(Q^2)B^*_{(l+1)-}(0)\right]\nonumber\\
& &\times
F(-l+1;l+3;3;\sin^2\frac{\theta}{2}), 
\label{eq:imtau}
\end{eqnarray}
where $\tau_i$ are the reduced helicity amplitudes defined in 
Eq.~(\ref{eq:vcsredhel}).
In Eqs. (\ref{eq:unitarity}), $q_\pi$ is the pion c.m. momentum 
in the intermediate state, and $F$ is a hypergeometric polynomial defined as~:
\begin{equation}
F(a;b;c;x)=1+\frac{a b}{c}\frac{x}{1!}
+\frac{a(a+1)b(b+1)}{c(c+1)}\frac{x^2}{2!}+ ...
\end{equation}
In Eq.~(\ref{eq:imtau}), the transverse multipoles $A_{l \pm}$, $B_{l \pm}$, 
and the longitudinal multipoles $L_{l \pm}$ 
are defined as in Ref.~\cite{Walker69}.

\end{appendix}

\newpage

\newpage

\newpage

\newpage

\newpage

\newpage

\newpage

\newpage

\newpage

\newpage

\newpage

\newpage

\newpage

\newpage

\newpage

\newpage

\begin{references}

\bibitem{GuiVdh}
P.A.M. Guichon and M. Vanderhaeghen, Prog. Part. Nucl. Phys. {\bf 41},
125 (1998).
\bibitem{Vdh00}
M. Vanderhaeghen, Eur. Phys. J. A {\bf 8}, 455 (2000).
\bibitem{Gui95}
P.A.M. Guichon, G.Q. Liu, and A.W. Thomas, Nucl. Phys. {\bf A591}, 606 (1995).

\bibitem{Roc00}
J. Roche et al., Phys. Rev. Lett. {\bf 85}, 708 (2000).
\bibitem{JLab}
P.Y. Bertin, P.A.M. Guichon, and C. Hyde-Wright, spokespersons 
JLab experiment, E-93-050.
\bibitem{Bates}
J. Shaw and R. Miskimen, spokespersons MIT-Bates experiment, 97-03.
\bibitem{mamipol}
N. d'Hose and H. Merkel, spokespersons MAMI experiment, (2001).

\bibitem{lvov97}      
A. L'vov, V.A. Petrun'kin, and M. Schumacher, 
Phys. Rev. C {\bf 55}, 359 (1997).
\bibitem{Dre99}
D. Drechsel, M. Gorchtein, B. Pasquini, and M. Vanderhaeghen, 
Phys. Rev. C {\bf 61}, 015204 (1999).
\bibitem{Pas00}
B. Pasquini, D. Drechsel, M. Gorchtein, A. Metz, and M. Vanderhaeghen, 
Phys. Rev. C {\bf 62}, 052201 (2000).


\bibitem{Dre98}
D. Drechsel, G. Kn\"ochlein, A.Yu. Korchin, A. Metz, and S. Scherer, 
Phys. Rev. C {\bf 57}, 941 (1998); Phys. Rev. C {\bf 58}, 1751 (1998).
\bibitem{Vdh97a}
M. Vanderhaeghen, Phys. Lett. B {\bf 402}, 243 (1997).

\bibitem{Bj65}
J.D. Bjorken and S.D. Drell, 
{\it Relativistic Quantum Fields}, McGraw-Hill, New York (1965).
\bibitem{Pil79}
H. Pilkuhn, 
{\it Relativistic Particle Physics}, Springer Verlag, Heidelberg (1979).
\bibitem{Jaf92}
R.L. Jaffe and P.F. Mende, Nucl. Phys. {\bf B369}, 189 (1992).
\bibitem{Oeh95}
R. Oehme, Int. J. Phys. {\bf A 10}, 1995 (1995). 

\bibitem{Ber58}
R.A. Berg and C.N. Lindner, Nucl. Phys. {\bf 26}, 259 (1961).
\bibitem{Tarrach}
R. Tarrach, Nuovo Cimento A {\bf 28}, 409 (1975).
\bibitem{Dre97}
D. Drechsel, G. Kn\"ochlein, A. Metz, and S. Scherer, 
Phys. Rev. C {\bf 55}, 424 (1997). 

\bibitem{maid00}
D. Drechsel, O. Hanstein, S. Kamalov, and L. Tiator, 
Nucl. Phys. {\bf A645}, 145 (1999).

\bibitem{vpi99}
M.M. Pavan, R.A. Arndt, I.I. Strakovsky, and R.L. Workman, 
PiN Newslett. {\bf 15}, 171 (1999).
\bibitem{bl81}
S.J. Brodsky and G.P. Lepage, Phys. Rev. D {\bf 24}, 1808 (1981).

\bibitem{Hemmert00}
T.R. Hemmert, B.R. Holstein, G. Kn\"ochlein, and D. Drechsel,
Phys. Rev. D {\bf 62}, 014013 (2000).
\bibitem{Metz96}
A. Metz and D. Drechsel, 
Z. Phys. {\bf A356}, 351 (1996); Z. Phys. {\bf A359}, 165 (1997).
\bibitem{PSD00}
B. Pasquini, S. Scherer, and D. Drechsel,
Phys. Rev. C {\bf 63}, 025205 (2001).

\bibitem{Bab98}  
D. Babusci, G. Giordano, and G. Matone, Phys. Rev. C {\bf 57}, 291 (1998).
      
\bibitem{Wiss00}  
V. Olmos de Le\'on, et al., submitted to Eur. Phys. J. A.

\bibitem{Hohler}
G. H\"ohler, Pion-Nucleon Scattering, Landolt-B\"ornstein, Vol. I/9b2,
edited by H. Schopper (Springer, Berlin, 1983).
\bibitem{Hol94}
B.R. Holstein and A.M. Nathan, Phys. Rev. D {\bf 49}, 6101 (1994).

\bibitem{Hoe76}
G. H\"ohler, E. Pietarinen, I. Sabba-Stefanescu, F. Borkowski,
G.G. Simon, V.H. Walther, and R.D. Wendling, 
Nucl. Phys. {\bf B 114}, 505 (1976).

\bibitem{natalie}
N. Degrande, Ph.D. thesis, University Gent, (2001); 
Luc Van Hoorebeke, private communication.
\bibitem{helene}
S. Jaminion, Ph.D. thesis, Universit\'e Blaise Pascal, Clermont Ferrand, (2000);
H. Fonvieille, private communication. 
\bibitem{geraud}
G. Laveissiere, private communication.
\bibitem{Bost95}
P.E. Bosted, Phys. Rev. C {\bf 51}, 409 (1995).
\bibitem{Jon00}
M.K. Jones et al., Phys. Rev. Lett. {\bf 84}, 1398 (2000).

\bibitem{Kalos97}
N.I. Kaloskamis and C.N. Papanicolas, 
spokespersons MIT-Bates experiment, (1997).

\bibitem{BD}
J.D. Bjorken and S.D. Drell, 
{\it Relativistic Quantum Mechanics}, McGraw-Hill, New York (1964).
\bibitem{Walker69}
R.L. Walker, Phys. Rev. {\bf 182}, 1729 (1969).

\end{references}
\end{document}